\documentclass[aps,pra,showpacs,showkeys,twocolumn]{revtex4}
\usepackage{graphicx}
\usepackage{amsmath, amsthm, amssymb}
\newcommand{\vect}[1] {\mathbf{#1}}
\newcommand{\dif} {\mathrm{d}}
\newcommand{\I} {\mathrm{i}}
\newcommand{\E} {\mathrm{e}}
\newcommand{\up} {\uparrow}
\newcommand{\down} {\downarrow}

\newcommand{\smomentum} {spin-momentum }
\newcommand{\smomenta}  {spin-momenta }

\newcommand{\Ht} {\widetilde{H}}
\newcommand{\Htccq} {\Ht^{(22)\vect q}}
\newcommand{\Htcq} {\Ht^{(2)\vect q}}
\newcommand{\A} {F}
\newcommand{\devp} {d'}
\newcommand{\dev}{d}
\newcommand{\proj}{P^\perp}
\newcommand{\deltamu} {\widetilde{\mu}}
\newcommand{\Cq} {C^{\mspace{2.0mu}\vect q}}
\newcommand{\norms}[1] {\lvert #1 \rvert^2}

\newcommand{\brac}[1] {\langle #1\rvert}
\newcommand{\ket}[1] {\lvert #1\rangle}

\begin{document}

\title{A new perturbation theory for the superfluid Fermi gas in the molecular Bose-Einstein condensed state}
\date{\today}
\author{Shina~Tan}
\affiliation{James Franck Institute and Department of Physics,
  University of Chicago, Chicago, Illinois 60637}

\begin{abstract}
We demonstrate how solutions to quantum few-fermion scattering 
problems can be the point-of-departure of a 
new treatment of a generalized many-body wave function. Our
focus is on a particular ansatz for the ground state wavefunction
of a superfluid Fermi gas introduced
earlier (cond-mat/0506293). 
Our method is perturbative in the sense that
the probability amplitudes for few-fermion scattering processes 
are treated as small quantities; it is also \emph{nonperturbative} in the sense
that whenever such scattering events occur, nonperturbative quantum few-fermion scattering physics dominates.
This approach can be viewed as a
new diagrammatic methodology, based on a wave function as
distinct from a perturbation series in the interparticle interactions.
Some generic properties of the wave function are studied,
     and its parameters in the Bose-Einstein condensed limit are computed beyond mean-field.
These results enable us to predict many observable properties of this Fermi gas with well-controlled accuracy, such
as the superfluid pairing function, the four-fermion and six-fermion  correlation functions, the momentum distribution, and
the two-body reduced density matrix, etc.
\end{abstract}


\maketitle

\section{Introduction\label{sec:intro}}
In this paper we consider
a superfluid Fermi gas with \emph{strong enough} attractive interactions such that molecules of pairs
of fermions are formed and Bose-Einstein condensed. 
A physical example of the system we have in mind
is a two-component Fermi gas near a Feshbach resonance,
with positive inter-fermionic scattering length.
Our goal is to develop a new form of perturbation theory which will
enable us to address fundamental and, now, experimentally accessible
properties of this gas. 
One can not determine many observables of this system
by simply resorting to the common wisdom of conventional Bose gases.
In many instances, the underlying fermionic character must play
a profound role.
The starting point for the present paper
is a wavefunction ansatz introduced earlier
\cite{_TanLevin} for the ground state of a superfluid Fermi gas. 
In this earlier work we explored the equation of state
and its relation to breathing mode experiments. Here we present
in considerable detail our theoretical methodolgy. To demonstrate the predictive
power of our theory, we also compute the superfluid pairing function and the four-fermion correlation function.
In future work we will address 
the fermionic momentum distribution, the two-body reduced density matrix,
the higher correlation functions as well as other properties.

Because the attractive interactions are strong
enough to form bound states, whenever  
two or more molecules collide, the few-body physics is highly 
nonperturbative. In this paper we build on the
observation that there is, nevertheless,
an underlying perturbative component:
the interactions are weak in the sense that the \emph{probability amplitudes} 
(or component terms of the many-body
wave function) associated with collisions are small, 
provided that the density of the system is relatively low.
This is precisely
the situation in which standard many-body perturbation
techniques, based on weak inter-particle interactions are inappropriate.
By contrast, here we develop a new kind of perturbation method, 
taking full advantage of these small amplitudes.
Whenever molecules approach each other, however, the behavior of the constituent particles is  
treated nonperturbatively. In this way a complete many-body theory fully compatible with the nonperturbative
few-body physics will be presented.

Closely related ideas have been used in the companion paper
\cite{_Tan_BEC}
to address the problem of a gas of point-like particles. This is,
in some sense, 
the simpler analog of the bound states problem.
In this companion paper
\cite{_Tan_BEC} the ground state of the dilute Bose
gas of structureless particles with quite arbitrary interactions 
is studied, starting from
a general many-body wave function:
\begin{equation*}
\ket{\psi_\text{boson}}=\exp\biggl(\sum_{p=1}^{\infty}
\frac{1}{p\,!}\sum_{\vect k_1\dots\vect k_p}
\alpha^{(p)}_{\vect k_1\dots\vect k_p}
b_{\vect k_1}^\dagger\dots b_{\vect k_p}^\dagger
\biggr)\ket{0},
\end{equation*}
where the $b$'s are boson annihilation operators, the $\vect k$'s are momenta, and $\ket{0}$
is the particle vacuum. To our surprise, this familiar system has some fundamental properties
unknown to people, because the prevailing theories to date (Bogoliubov theory, pseudopotential method,
effective field theory, etc) are all low-energy effective theories, capable primarily of describing the physics at 
the length scale of the superfluid healing length. In the companion paper \cite{_Tan_BEC} such a limitation
is removed, and the \emph{short-distance} structure is clarified. As a byproduct, some fundamental theorems
derived initially for a Fermi system \cite{_Tan_EnergyTheorem, _Tan_Adiabatic}
are successfully extended to a dilute Bose system \cite{_Tan_BEC}.

In this paper we apply the same approach to a Bose-Einstein condensate of molecules formed out of strongly attractive fermions.
Each molecule is a bound state of two fermions
in two spin states (or any two internal states) $\up\equiv+1$ and $\down\equiv-1$. 
In Ref.~\cite{_TanLevin} we presented a generalized many-body wavefunction
of the form
\begin{equation}\label{eq:m_psi}
\ket{\psi_\text{fermion}}=\exp\biggl\{\sum_{p=2}^{\infty}\frac{1}{p\,!}
\sum_{\vect K\text{'s}}
\alpha^{(p)}_{\vect K_1\vect K_2\cdots\vect K_{p}}
c_{\vect K_1}^\dagger \dots c_{\vect K_{p}}^\dagger\biggr\}\ket{0},
\end{equation}
where $p=2,4,6,\cdots$ must be \emph{even} \cite{_odd_p}, the subscript $\vect K_i$ is the shorthand for 
$\vect k_i\sigma_i$, and $\sigma$ labels the spin state. The coefficient
$\alpha^{(p)}_{\vect K_1\cdots\vect K_p}$ has some elementary properties:
1) it is antisymmetric under the interchange of any two subscripts,
2) it is zero whenever the sum of all its subscripts is nonzero, and
3) it is zero whenever the sum of a \emph{nontrivial subset} (Appendix~\ref{a:terms}) of its subscripts
is zero (see Sec.~\ref{subsec:m_thermodynamic}).
(Here the sum of subscripts is defined by summing the momenta and the associated
$\sigma$'s separately, and when both sums are zero, the sum of these subscripts is called zero.)

The well-known
BCS or Eagles-Leggett (EL) wave function of fermionic superfluids
is the lowest order approximation of
Eq.~\eqref{eq:m_psi}. This wavefunction and its implications
for experiment have been reviewed
rather extensively elsewhere \cite{ChenRev}. 
Retaining only the $p=2$ term, one can verify that $\ket{\psi}$
becomes equivalent to a more familiar form of the EL wave function:
\begin{equation*}
\ket{\psi_\text{EL}}=\prod_{\vect k}
\bigl(u_\vect k+v_\vect k c_{\vect k\up}^\dagger c_{-\vect k\down}^\dagger\bigr)\ket{0}.
\end{equation*}

In Secs.~\ref{sec:m_general}, \ref{sec:m_fewbody}, and \ref{sec:m_special},
the ideas and methods in Ref.~\cite{_Tan_BEC}
are extended to bosonic molecules.
Here we present details to support our
earlier conclusions \cite{_TanLevin}
for the ground state energy density calculated to the 2.5-th order
in the molecular density. As discussed
in Ref.~\cite{_TanLevin}, the result indicates that the Lee-Yang formula \cite{LeeYang1957} is valid
not only for structureless bosons, but also for composite ones.

In Sec.~\ref{sec:m_general} we define the system, discuss those
properties of the many-body wave function that are valid for \emph{all} densities, and introduce
a new diagrammatic method.

In Sec.~\ref{sec:m_fewbody} we discuss the few-body functions and the properties that will be
referred to in the subsequent many-body low-density expansions.

In Sec.~\ref{sec:m_special} we introduce our low-density expansion method, derive the power-counting formula
for such expansions, and compute the parameters of the wave function beyond mean-field. The equation of state
is also determined beyond mean-field. Our method is variational, but \emph{unlike} the traditional sense of this term,
we identify \emph{all} the significant contributions up to any given
order in the density, ensuring that our results are asymptotically \emph{exact}, order by order.

In Sec.~\ref{sec:m_predictions} we compute two other physical observables, using the established knowledge
of the wave function, and discuss  the motivation and the feasibility of computing some other fundamental observables.

In Sec.~\ref{sec:m_conclusion}, we outline the extension of the calculation
to third order in the molecular density, and predict
that the Wu term \cite{Wu1959PR} must also be present, and contain the same scattering length as the Lee-Yang
term \cite{LeeYang1957}.

With each increasing order in the density expansion of
the energy, the \emph{many-body wave function} itself
is determined more and more accurately. In this way, 
we can determine \emph{all other physical
observables} with a high degree of
accuracy by evaluating their expectation values within
the many-body wave function. This approach also
makes it possible to address quantitative properties
slightly \emph{above} zero temperature.

Some trends in the system, upon increasing density, are clearly visible in the low-density expansion.
This opens a new avenue towards the many-body physics in the unitary scattering regime.

\section{\label{sec:m_general}Fermionic superfluid: some general analyses}
All results in this section are \emph{completely independent} of the density, and equally valid in regimes such
as the unitarity limit of a Fermi gas with large scattering length.
They are however a prerequisite of the low-density expansion in Sec.~\ref{sec:m_special}.
In light of the possibility of more concrete progress in the high density regime, it is desirable to separate the generally
applicable results from those that are restricted to low density.

\subsection{\label{subsec:m_hamiltonian}Hamiltonian}
We study a system of fermions with identical mass $m$ (let
$\hbar=m=1$) in two internal states (hereafter simply called spin states).
 For simplicity we restrict ourselves to a system in which there is only interaction between fermions
in the opposite spin states ($\up$ and $\down$). The hamiltonian $H$
and the operator $\hat{N}$ for the total number of fermions satisfy
\begin{align}\label{eq:m_H1}
&H-\mu \hat{N}=\sum_{\sigma}\int\dif^3rc_\sigma^\dagger(\vect r)
(-\nabla^2/2-\mu) c_\sigma^{}(\vect r)+\notag\\
&\int\dif^3r_1\cdots\dif^3r_4
U(\vect r_1\vect r_2\vect r_3\vect r_4)c_{\up}^\dagger(\vect r_1)
c_{\down}^\dagger(\vect r_2)c_{\down}^{}(\vect r_4)c_{\up}^{}(\vect r_3),
\end{align}
where $\mu$ is the fermionic chemical potential, $\sigma=\up$, $\down$,
and $c_\sigma(\vect r)$ is the standard fermion annihilation operator in the coordinate
representation, satisfying $\{c_\sigma(\vect r),c_{\sigma'}^\dagger(\vect r')\}=\delta_
{\sigma\sigma'}\delta(\vect r-\vect r')$.

We restrict ourselves to an interaction $U$ with the following properties:
\begin{itemize}
\item translational symmetry, $U(\vect r_1+\vect r, \vect r_2+\vect r, \vect r_3+\vect r,
\vect r_4+\vect r)=U(\vect r_1\vect r_2\vect r_3\vect r_4)$, for any $\vect r$.
\item rotational symmetry, $U(\mathcal{R}\vect r_1,\mathcal{R}\vect r_2,\mathcal{R}\vect r_3,
\mathcal{R}\vect r_4)=U(\vect r_1\vect r_2\vect r_3\vect r_4)$, for any
rotational transformation $\mathcal{R}$.
\item \emph{Galilean symmetry}, $U(\vect r_1\vect r_2\vect r_3\vect r_4)=0$
whenever $\vect r_1+\vect r_2\neq\vect r_3+\vect r_4$.
\item finite-rangedness: there is a finite length scale $l$ such that
$U(\vect r_1\vect r_2\vect r_3\vect r_4)$ approaches zero sufficiently fast whenever
any $\lvert\vect r_i-\vect r_j\rvert\gg l$, where $1\leq i,j\leq 4$.
\item the interaction supports a nondegenerate s-wave bound state (the molecule) of two fermions,
whose binding energy in the absence of any other particles will be denoted by $E_m<0$
(independent of the center-of-mass velocity of the molecule,
due to the Galilean symmetry); the typical distance between the two fermions in the molecule will
be denoted by $r_m$; the zero-velocity scattering length between two molecules, $a_m$, is positive and finite.
\item lower bound states do not exist, or they do exist but their formation rates are
so low that we can omit them \cite{_unique_bound_state}.
\end{itemize}
Also, $H=H^\dagger\Rightarrow U(\vect r_1\vect r_2\vect r_3\vect r_4)^*
=U(\vect r_3\vect r_4\vect r_1\vect r_2)$.

For later convenience, we use a symbol $\vect R\equiv(\vect r\sigma)$, called
``spin-spatial vector''; and similarly $\vect K\equiv(\vect k\sigma)$, called ``\smomentum''.
We also define some operations involving them.
The ``norm-square'': $\vect K^2\equiv k^2$.
The ``addition'': $\vect K_1+\vect K_2\equiv(\vect k_1+\vect k_2,\sigma_1+\sigma_2)$,
where each $\up$ is counted as $+1$ and each $\down$ is counted as $-1$;
 for any even number of spin-momenta $\vect K_1\cdots\vect K_p$,
if \emph{both} $\sigma_1+\cdots+\sigma_p=0$ \emph{and} $\vect k_1+\cdots+\vect k_p=0$,
we say that $\vect K_1+\cdots+\vect K_p=0$. The sum of a spin-momentum and a momentum:
$\vect K+\vect q\equiv(\vect k+\vect q,\sigma)$. The ``opposite spin-momentum'':
$-\vect K\equiv(-\vect k,-\sigma)$. The ``integration'': $\int\dif^3R\equiv\sum_\sigma\int\dif^3r$, and
$\int\dif^3K/(2\pi)^3\equiv\sum_\sigma\int\dif^3k/(2\pi)^3$;
and similarly a summation: $\sum_\vect K\equiv\sum_{\sigma\vect k}=\Omega\int\dif^3K/(2\pi)^3$,
where $\Omega$ is the system's volume.
Finally, if \emph{both} $\sigma_1+\cdots+\sigma_p=0$
\emph{and} $\vect k_1+\cdots+\vect k_p\sim q$, we say that $\vect K_1+\cdots+\vect K_p\sim q$,
where $q$ is some small momentum scale.

Now we define the function $U(\vect R_1\vect R_2\vect R_3\vect R_4)$, satisfying:
$U(\vect R_1\vect R_2\vect R_3\vect R_4)=-U(\vect R_2\vect R_1\vect R_3\vect R_4)
=-U(\vect R_1\vect R_2\vect R_4\vect R_3)$; it is zero whenever $\sigma_1+\sigma_2\neq0$
or $\sigma_3+\sigma_4\neq0$; $U(\vect r_1\up,\vect r_2\down,\vect r_3\up,\vect r_4\down)
\equiv U(\vect r_1\vect r_2\vect r_3\vect r_4)$. Equation~\eqref{eq:m_H1} is
thus rewritten as
\begin{align*}
H-\mu \hat{N}=&\int\dif^3Rc^\dagger(\vect R)(-\nabla^2/2-\mu)c(\vect R)\notag\\
&+\frac{1}{4}\int\notag\dif^3R_1\cdots\dif^3R_4U(\vect R_1\vect R_2\vect R_3\vect R_4)\notag\\
&\times c^\dagger(\vect R_1)c^\dagger(\vect R_2)c(\vect R_4)c(\vect R_3).
\end{align*}

Fourier transforms:
\begin{equation*}
c_{\vect k\sigma}\equiv\Omega^{-1/2}\int\dif^3r\:c(\vect r\sigma)\exp(-\I\vect k\cdot\vect r),
\end{equation*}
\begin{align*}
&U_{\vect k_1\sigma_1\vect k_2\sigma_2\vect k_3\sigma_3\vect k_4\sigma_4}
\equiv\Omega^{-2}\int\dif^3r_1\cdots\dif^3r_4\notag\\
&\times U(\vect r_1\sigma_1\vect r_2\sigma_2\vect r_3\sigma_3
\vect r_4\sigma_4)\notag\\
&\times\exp(-\I\vect k_1\cdot\vect r_1-\I\vect k_2\cdot\vect r_2
+\I\vect k_3\cdot\vect r_3+\I\vect k_4\cdot\vect r_4),
\end{align*}
\begin{align}\label{eq:m_H3}
H-\mu \hat{N}=&\sum_\vect K(k^2/2-\mu)c_{\vect K}^\dagger c_{\vect K}^{}\notag\\
&+\frac{1}{4}\sum_{\vect K\text{'s}}U_{\vect K_1\vect K_2\vect K_3\vect k_4}
c_{\vect K_1}^\dagger c_{\vect K_2}^\dagger c_{\vect K_4}^{} c_{\vect K_3}^{}.
\end{align}
$\{c_{\vect K}^{},c_{\vect K'}^\dagger\}=\delta_{\vect K\vect K'}$
and $\{c_{\vect K}^{},c_{\vect K'}^{}\}=0$.

$U_{\vect K_1\vect K_2\vect K_3\vect K_4}=0$ unless $\vect K_1+\vect K_2=\vect K_3+\vect K_4$.
When this \smomentum conservation condition is satisfied,
$U$ is a \emph{smooth} function of the remaining three independent momenta, as a consequence of
the condition that the interaction is finite-ranged. Third,
\begin{align*}
U_{\vect K_1\vect K_2\vect K_3\vect K_4}&=-U_{\vect K_2\vect K_1\vect K_3\vect K_4}
=-U_{\vect K_1\vect K_2\vect K_4\vect K_3}\\
&=U_{\vect K_3\vect K_4\vect K_1\vect K_2}^*=U_{\vect K_1+\vect k,\vect K_2+\vect k,
\vect K_3+\vect k,\vect K_4+\vect k},
\end{align*}
where the last equality is due to the Galilean symmetry ($\vect k$ is an arbitrary momentum).
Fourth, in the thermodynamic limit, any
nonzero $U_{\vect K_1\vect K_2\vect K_3\vect K_4}$ scales like \cite{_dimension_Q}
\begin{equation*}
U_{\vect K_1\vect K_2\vect K_3\vect K_4}\sim\Omega^{-1}.
\end{equation*}

In this paper, an expression like $x\sim\Omega^Q$ only refers to the behavior of $x$ for
$\Omega\rightarrow\Omega'$;
since the often dimensionful coefficient (independent of $\Omega$) is suppressed, one
should never use such expression to infer the \emph{dimension} of $x$.

\subsection{\label{subsec:m_wavefunction}Wave function}
The ground state wave function we have adopted is
\begin{equation*}
\ket{\psi}=\exp\Biggl(\sum_{p=2}^{\infty}\alpha^{(p)}\Biggr)\ket{0},
\end{equation*}
where $p=2,4,6,\cdots$ must be \emph{even} \cite{_odd_p}, and
\begin{equation*}\label{eq:m_alpha}
\alpha^{(p)}
\equiv\frac{1}{p\,!}\sum_{\vect K_1\cdots\vect K_p}\alpha^{(p)}_{\vect K_1\cdots\vect K_p}
c_{\vect K_1}^\dagger\cdots c_{\vect K_p}^\dagger
\end{equation*}
is called the $(p/2)$-th order \emph{generator};
\begin{equation*}\alpha^{(p)\dagger}=\frac{1}{p\,!}\sum_{\vect K_1\cdots\vect K_p}\alpha^{(p)*}_{\vect K_1\cdots\vect K_p}
c_{\vect K_p}^{}c_{\vect K_{p-1}}^{}\cdots c_{\vect K_2}^{}c_{\vect K_1}^{}\end{equation*}
is called the $(p/2)$-th order \emph{degenerator}. The prefactor $1/p\,!$ is for the
antisymmetry of the $c^\dagger$'s (or $c$'s).

In a very similar manner as in \cite{_Tan_BEC}, we can establish
the following properties of $\alpha^{(p)}_{\vect K_1\cdots\vect K_p}$.

Firstly, we assume that the ground state is translationally invariant, \emph{and}
that the superfluid pairing occurs in the spin-singlet channel only. So
$\alpha^{(p)}_{\vect K_1\cdots\vect K_p}=0$ whenever $\vect K_1+\cdots+\vect K_p\neq0$.

Secondly, in the thermodynamic limit, in which the density of fermions
\begin{equation*}
n\equiv N/\Omega
\end{equation*}
is kept constant, any nonzero coefficient $\alpha^{(p)}_{\vect K_1\cdots\vect K_p}$
scales with the system's volume as
\begin{equation}\label{eq:m_alpha_th_order}
\alpha^{(p)}_{\vect K_1\cdots\vect K_p}\sim\Omega^{1-p/2}.
\end{equation}

Thirdly, $\alpha^{(p)}_{\vect K_1\cdots\vect K_p}$ is \emph{antisymmetric}
under the interchange of $\vect K_i$ and $\vect K_j$, where $1\leq i<j\leq p$.

\subsection{\label{subsec:m_diagrams}Diagrams}
This subsection is largely parallel to a similar subsection in \cite{_Tan_BEC}.
The only essential additional feature is due to the antisymmetry of the
fermionic wave function.

The evaluation of \emph{any} physical observables can be reduced to the following general problem.

Consider an arbitrary product $O=o_1o_2\dots o_L$, where each
 $o_i$ ($1\leqslant i\leqslant L$) is a basic fermionic operator
(creation or annihilation) in the \smomentum space. The ground state expectation value
\begin{equation}\label{eq:m_expectation_old}
\langle O\rangle=
\frac
{ \brac{0}\exp\Bigl(\sum_{p=2}^{\infty}{\alpha^{(p)}}^\dagger\Bigr) O
  \exp\Bigl(\sum_{p'=2}^{\infty}\alpha^{(p')}\Bigr)\ket{0}
}
{ \brac{0}\exp\Bigl(\sum_{p=2}^{\infty}{\alpha^{(p)}}^\dagger\Bigr)
  \exp\Bigl(\sum_{p'=2}^{\infty}\alpha^{(p')}\Bigr)\ket{0}
}
\end{equation}
is calculated using the Wick's theorem.
We assign each fermion operator in the above expression a ``time": the creation
operators in the generators are assigned an earliest time $t_g$,
 the annihilation operators
in the degenerators a latest time $t_d$,
 and $o_1$ through $o_L$ are assigned some
intermediate times, $t_1$ through $t_L$, such that
$t_d>t_1>\dots > t_L>t_g$. The actual magnitudes of these ``times" do not matter;
only the ``causal order" matters. 

Equation~\eqref{eq:m_expectation_old} is thus rewritten as
\begin{equation}\label{eq:m_expectation}
\langle O\rangle=
\frac
{ \brac{0}\mathcal{T}O
  \exp\Bigl(\sum_{p=2}^{\infty}{\alpha^{(p)}}^\dagger+
\sum_{p'=2}^{\infty}\alpha^{(p')}\Bigr)\ket{0}
}
{ \brac{0}\mathcal{T}
  \exp\Bigl(\sum_{p=2}^{\infty}{\alpha^{(p)}}^\dagger+
\sum_{p'=2}^{\infty}\alpha^{(p')}\Bigr)\ket{0}
},
\end{equation}
where $\mathcal{T}$ is the ``time'' ordering operator.

The basic formulas to be used in the evaluation of Eq.~\eqref{eq:m_expectation} are
\begin{subequations}\label{eq:m_algebra}
\begin{equation}
\brac{0}\mathcal{T}c_{\vect Kt}^{}c_{\vect K't'}^{}\ket{0} 
=\brac{0}\mathcal{T}c_{\vect Kt}^\dagger c_{\vect K't'}^\dagger\ket{0}
=0,\label{eq:m_algebra_a}
\end{equation}
\begin{equation}
\brac{0}\mathcal{T}c_{\vect Kt}^{}c_{\vect K't'}^\dagger\ket{0}
=-\brac{0}\mathcal{T}c_{\vect K't'}^\dagger c_{\vect Kt}^{}\ket{0}
=\delta_{\vect K\vect K'}\theta(t-t').
\label{eq:m_algebra_b}
\end{equation}
\end{subequations}

Equation~\eqref{eq:m_expectation} is then expressed as an infinite sum of all the topologically
distinct diagrams with the following rules.
\begin{itemize}
\item Each $o_i$ is represented a small \emph{solid} circle if it is a fermion \emph{creation} operator, 
or a small \emph{hollow} one if it is an \emph{annihilation} operator, linked by a single solid line, with the designated spin-momentum.
\item Each $p/2$-th order generator (degenerator) is represented by a large \emph{solid} (\emph{hollow})
circle, linked by $p$ solid lines; it contributes a factor $\alpha^{(p)}_{\vect K_1\cdots\vect K_p}$ 
(or its complex conjugate).
\item Each solid line is associated with a \emph{single} spin-momentum, and it
 can only connect an \emph{earlier solid} object with a \emph{later
hollow} one, in accordance with Eq.~\eqref{eq:m_algebra}. The
diagram is thus bipartite.
\item Impose \smomentum conservation on all the internal vertices, that is, the sum of all the \smomenta
leaving/entering any generator/degenerator must equal zero. Sum over all the independent loop spin-momenta.
\item Each diagram has an overall symmetry factor $1/S$, and $S$ is the number of ways
of interchanging components without changing the diagram.
\item Vacuum bubbles (disconnected parts which are not connected to any external
vertex) are all canceled by the denominator in Eq.~\eqref{eq:m_expectation}.
\end{itemize}

The fermionic operators \emph{anticommute} under the ``time''-ordering operator; the subscripts
of $\alpha^{(p)}_{\vect K_1\cdots\vect K_p}$ also anticommute. It is easy to show that
there is a simple method to determine the \emph{sign} of a diagram.

Step 1. Divide all the solid lines of the diagram into a collection of continuous
trajectories. Each solid line can only be contained by one trajectory, and it can appear
in this trajectory only once. Each trajectory
is either an external trajectory, or an internal one. Each external trajectory
starts from an external vertex $o^{s}$, ends at a different external vertex $o^{e}$, and is of the
form $v_0\vect K_1v_1\vect K_2v_2\cdots \vect K_M v_M$, where $M\geq1$, $v_0=o^{s}$, $v_M=o^e$,
and the remaining $v$'s
are internal vertices. $\vect K$'s are spin-momenta, that is, the solid lines.
Each internal trajectory starts from an internal vertex $v_0$, and must come back to this same
vertex; it is of the form $v_0\vect K_1v_1\vect K_2v_2\cdots \vect K_Mv_M$,
 where $v_M=v_0$, and $M$ is necessarily
a positive \emph{even} number (recall that the diagram is bipartite); an internal trajectory
can never pass an external vertex. The total number of external trajectories is necessarily $L/2$.
The $L$ external vertices is thus divided into $L/2$ pairs, and can be written as
$o^s_1 o^e_1 o^s_2 o^e_2\cdots o^s_{L/2} o^e_{L/2}$; suppose that it takes $P$
interchanges of pairs of fermionic operators to transform the sequence
$o_1\cdots o_L$ to this new one. Suppose that $X$ of the $L/2$ external \emph{end} points
($o^e$'s) are fermion annihilation operators, that is, hollow circles.

Step 2. For every p-line internal vertex $v$, write a separate symbol $\alpha^{(p)}$
if it is solid, or $\alpha^{(p)*}$ if it is hollow; leave $p$ blank positions for its subscripts.

Step 3. To each internal vertex $v$ passed through by each trajectory, assign a pair of
ordered subscripts $\vect K\vect K'$, where the 3-tuple $\vect K v\vect K'$
(exactly in this order) is a segment of the trajectory. If $v$ is the starting-ending point of
an internal trajectory (containing $M$ solid lines), assign to it a pair of ordered subscripts
$\vect K_1\vect K_M$. Multiply the final expression by an overall factor $(-1)^{P+X}$.

There are different ways to divide the same diagram into trajectories, but the results
are equal.

\begin{figure}
\includegraphics{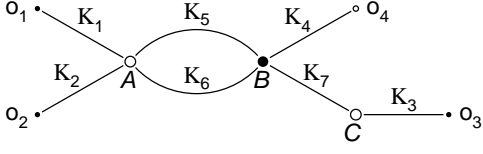}
\caption{\label{fig:m_example}A diagram in the expansion of $\langle c_{\vect K_1}^\dagger
c_{\vect K_2}^\dagger c_{\vect K_3}^\dagger c^{}_{\vect K_4}\rangle$.}
\end{figure}

We illustrate these rules with a term in the expansion of $\langle c_{\vect K_1}^\dagger
c_{\vect K_2}^\dagger c_{\vect K_3}^\dagger c^{}_{\vect K_4}\rangle$, shown in
Fig.~\ref{fig:m_example}. The diagram can be divided into two trajectories:
$o_1\vect K_1A\vect K_5B\vect K_4o_4$ and $o_2\vect K_2A\vect K_6B\vect K_7C\vect K_3o_3$,
so $(-1)^P=+1$, $X=1$, and $(-1)^{P+X}=-1$. The expression with the correct sign is therefore
$-A_{\vect K_1\vect K_5\vect K_2\vect K_6}B_{\vect K_5\vect K_4\vect K_6\vect K_7}
C_{\vect K_7\vect K_3}=A_{\vect K_1\vect K_2\vect K_5\vect K_6}
B_{\vect K_4\vect K_7\vect K_5\vect K_6}C_{\vect K_3\vect K_7}$.
Incorporating the other rules, we get
\begin{equation*}
+\frac{1}{2}\sum_{\vect K_5\vect K_6}\beta^*_{\vect K_1\vect K_2\vect K_5\vect K_6}
\beta^{}_{\vect K_4,-\vect K_3,\vect K_5\vect K_6}\alpha^{*}_{\vect K_3},
\end{equation*}
where we have given special names to the coefficients of the first three generators:
\begin{subequations}\label{eq:m_alpha_beta_gamma}
\begin{align}
\alpha^{}_{\vect K} &\equiv\alpha^{(2)}_{\vect K,-\vect K},\\
\beta^{}_{\vect K_1\vect K_2\vect K_3\vect K_4} &\equiv\alpha^{(4)}
_{\vect K_1\vect K_2\vect K_3\vect K_4},\\
\gamma^{}_{\vect K_1\vect K_2\vect K_3\vect K_4\vect K_5\vect K_6}&\equiv\alpha^{(6)}
_{\vect K_1\vect K_2\vect K_3\vect K_4\vect K_5\vect K_6}.
\end{align}
\end{subequations}

There are quite a few other ways to divide Fig.~\ref{fig:m_example} into trajectories. For
instance, $o_1\vect K_1A\vect K_2o_2$, $o_3\vect K_3C\vect K_7B\vect K_4o_4$,
and $A\vect K_5B\vect K_6A$, where the last trajectory is an internal one;
so $(-1)^P=+1$, $X=1$, and $(-1)^{X+P}=-1$. The expression with
the correct sign is therefore $-A_{\vect K_1\vect K_2\vect K_5\vect K_6}
B_{\vect K_7\vect K_4\vect K_5\vect K_6}C_{\vect K_3\vect K_7}$, equivalent
with the one we wrote above. Even if we \emph{reverse the direction} of any
trajectory, the final result is still the same.

\subsection{\label{subsec:m_diagrams_energy}Diagrams for $E-\mu N$}

Like in \cite{_Tan_BEC}, we use a half-solid-half-hollow diamond
to facilitate the diagrammatic calculation of $E-\mu N=\langle H-\mu \hat{N}\rangle$.
If each side is attached by a single solid line, the diamond contributes
a factor $(k^2/2-\mu)$, where $\vect K$ is the \smomentum flowing through it.
If each side is attached by two solid lines, the diamond contributes a factor
$U_{\vect K_1\vect K_2\vect K_3\vect K_4}$, where $\vect K_1\vect K_2$ leave
the diamond's solid side, and $\vect K_3\vect K_4$ enter its hollow side.

The above rules are very similar to those in \cite{_Tan_BEC}.
We now describe how to determine the sign of each diagram.

The solid lines in such a diagram can still be divided into a collection of trajectories.

For a trajectory $v_0\vect K_1v_1\cdots\vect K_Mv_M$ ($v_0=v_M$) which does not involve the
diamond, the rule is the same as in the last subsection: assign an ordered pair
of subscripts $\vect K_i\vect K_{i+1}$ to $v_i$ ($1\leq i\leq M-1$),
and assign $\vect K_1\vect K_M$ to $v_0$.

There will then be $L/2$ other, special trajectories
($L=2$ for the kinetic energy, and $L=4$ for the interaction), each of which
starts from the diamond and ends at it.

Case 1: $L=2$, and there is one special trajectory $v_0\vect K_1v_1\cdots\vect K_Mv_M$
($v_0$ and $v_M$ are the two sides of the diamond, and $\vect K_1=\vect K_M$). Assign
an ordered pair of subscripts $\vect K_i\vect K_{i+1}$ to each $v_i$ ($1\leq i\leq M-1$).
Multiply the whole expression by $(-1)$.

Case 2: $L=4$, and the two special trajectories are
$s\vect K_1\cdots\vect K_1'h$ and $s\vect K_2\cdots\vect K_2'h$,
where $s$ and $h$ are the solid and the hollow sides of the diamond, respectively.
So each trajectory is the same as the one in Case 1, and the two $(-1)$'s (overall
factors contributed by both trajectories) cancel. Assign an ordered pair
of subscripts to each vertex (other than the diamond) in the same way as above;
the diamond contributes the factor $U_{\vect K_1^{}\vect K_2^{}\vect K_1'\vect K_2'}$.

Case 3: $L=4$, and each special trajectory $o\vect K_1\cdots\vect K_Mo$
starts from and ends at the \emph{same} side of the diamond.
It can then be treated like an ordinary \emph{internal} trajectory; the pair
$\vect K_1\vect K_M$ (exactly in this order) is filled to the first two
subscript positions of $U$ if $o$ is the solid side, or to the last
two subscript positions of $U$ if $o$ is the hollow side.
The order of the fermion operators in Eqs.~\eqref{eq:m_H1} and \eqref{eq:m_H3} has been
chosen to ensure the validity of this simple rule; such order has also
affected the rule in Case 2.

\subsection{\label{subsec:m_thermodynamic}Thermodynamic orders and
a property of the wave function}

The theorem derived in \cite{_Tan_BEC} concerning the \emph{thermodynamic order} (see Appendix~\ref{a:terms})
 of any aforementioned diagram, namely
\begin{equation}\label{eq:Q}
Q=N_3-L/2,
\end{equation}
is still valid, because the prerequisites of this theorem are unaffected by the
change from a bosonic superfluid to a fermionic one;
the fermion sign does not affect these thermodynamic orders. $L$ is the number of
external points of the diagram. This theorem describes the scaling behavior of any diagram $\mathcal{D}$
in the thermodynamic limit: $\mathcal{D}\sim\Omega^Q$.

As a corollary, any diagram in the expansion of $E-\mu N$ satisfies $Q=1$ and is
thermodynamically significant.

In almost the same way as in \cite{_Tan_BEC}, we can also establish
a general property of the coefficients $\alpha^{(p)}_{\vect K_1\cdots\vect K_p}$,
namely it is zero whenever a nontrivial subset of fermions conserve spin-momentum
(that is, whenever the sum of a nontrivial subset of the subscripts is zero).
The basic idea is that any thermodynamically significant contribution to any physical observable
from a coefficient whose nontrivial subset of subscripts conserves spin-momentum
(called reducible coefficient) is only present at the ``bottleneck" vertex of a \emph{dead end}
(see Appendix~\ref{a:terms}), but that dead ends can be absorbed by redefining lower order coefficients.
Consequently, \emph{there are no dead ends in any diagrams.}
For more details, see \cite{_Tan_BEC}.

\section{\label{sec:m_fewbody}Few-Body Wave Functions}

In this section we define some few-body functions and discuss some of their properties. They are completely independent of
any many-body physics whatsoever.

But the contents listed below are \emph{restricted} to those that are \emph{directly needed}
in Sec.~\ref{sec:m_special} where we do a many-body low-density expansion.

The internal wave function $\phi_{\vect K}$ of an \emph{isolated} molecule satisfies
$(H-E_m)\sum_{\vect K}\phi_\vect K^{}c_{\vect K}^\dagger c_{-\vect K}^\dagger\ket{0}=0$.
\begin{subequations}\label{eq:m_phi}
\begin{gather}
\sum_{\vect K'}D_{\vect K\vect K'}\phi_{\vect K'}\equiv0,       \label{eq:m_phi_Schrodinger}\\
\phi_{-\vect K}\equiv-\phi_{\vect K},            \label{eq:m_phi_antisymmetry}\\
\frac{1}{2}\sum_{\vect K}\lvert\phi_{\vect K}\rvert^2\equiv1, \label{eq:m_phi_normalize}
\end{gather}
\end{subequations}
where
\begin{equation}\label{eq:m_D}
D_{\vect K\vect K'}\equiv(k^2-E_m)\delta_{\vect K\vect K'}+\frac{1}{2}U_{\vect K,-\vect K,\vect K',-\vect K'}
\end{equation}
is a hermitian matrix for the relative $D$ynamics of two fermions in opposite spin states isolated from all other fermions.
Since $E_m<0$ and the interaction is finite-ranged, the spatial representation
of $\phi$ must be finite ranged, so $\phi_{\vect k\sigma}$ depends \emph{smoothly} on $\vect k$.
Moreover, obviously $$\phi_\vect K\sim\Omega^{-1/2}.$$

Sometimes we elect to abbreviate the spin-momentum $\vect K_i$ as $i$, and $-\vect K_i$ as $\bar{i}$,
when they do not lead to confusion. $i=1,2,3,\cdots,9$. Without this notation, some equations would be
too lengthy and cumbersome.

The two-molecule zero-speed scattering wave function $\phi^{(4)}_{\vect K_1\vect K_2\vect K_3\vect K_4}$
satisfies $\bigl[H-2E_m-\epsilon^{(4)}\bigr]\sum_{1234}\phi^{(4)}_{1234}
c_{1}^\dagger c_{2}^\dagger c_{3}^\dagger c_{4}^\dagger\ket{0}=0$.
\begin{subequations}\label{eq:m_phi4}
\begin{gather}
\phi^{(4)}_{1234}=-\phi^{(4)}_{2134}=-\phi^{(4)}_{3214}=-\phi^{(4)}_{4231},\label{eq:m_phi4_antisymmetry}\\
\phi^{(4)}_{1234}=0~~~(\text{if}~\vect K_1+\cdots+\vect K_4\neq0),\label{eq:m_phi4_zeromomentum}
\end{gather}
\begin{multline}\label{eq:m_phi4_Schrodinger}
\bigl[(k_1^2+\cdots+k_4^2)/2-2E_m-\epsilon^{(4)}\bigr]\phi^{(4)}_{1234}
+\frac{1}{2}{\sum_{56}}U_{1256}\phi^{(4)}_{5634}\\
+\frac{1}{2}{\sum_{56}}U_{1356}\phi^{(4)}_{5264}
+\frac{1}{2}{\sum_{56}}U_{1456}\phi^{(4)}_{5236}
+\frac{1}{2}{\sum_{56}}U_{2356}\phi^{(4)}_{1564}\\
+\frac{1}{2}{\sum_{56}}U_{2456}\phi^{(4)}_{1536}
+\frac{1}{2}{\sum_{56}}U_{3456}\phi^{(4)}_{1256}=0,
\end{multline}
where $\epsilon^{(4)}\sim\Omega^{-1}$ is an extremely tiny positive energy, and is
always negligible \emph{unless} two of the four spin-momenta ($\vect K_1\cdots\vect K_4$) are opposite;
Ref.~\cite{_Tan_BEC} contains a simpler analog. When the distance $r$ between the
centers-of-mass of the two molecules is large [$\gg\max(l,r_m)$],
the four-fermion wave function (in the coordinate representation)
is proportional to $(1-a_m/r)$ times the internal wave functions of two isolated molecules.
Consequently, we have the infrared boundary condition
\begin{equation}\label{eq:m_phi4_smallq}
\phi^{(4)\vect q}_{\vect K_1\vect K_2}=\Bigl(\delta_{\vect q,0}-\frac{4\pi a_m}{\Omega q^2}\Bigr)\phi_{\vect K_1}\phi_{\vect K_2}
+O(q^0)
\end{equation}
\end{subequations}
when $\vect q$ is small or zero. Here
\begin{equation}\label{eq:m_phi4_shorthand}
\phi^{(4)\vect q}_{\vect K_1\vect K_2}\equiv\phi^{(4)}_{\vect K_1+\vect q/2,-\vect K_1+\vect q/2,\vect K_2-\vect q/2,-\vect K_2-\vect q/2}.
\end{equation}
The global coefficient of $\phi^{(4)}$ is set by Eq.~\eqref{eq:m_phi4_smallq}. Note that $\phi^{(4)\vect q}_{\vect K_1\vect K_2}$
has delta-function singularity at $\vect K_1=\pm\vect K_2$, but the \emph{intensity} of the delta function approaches
a finite constant when $\vect q\rightarrow0$. The term $O(q^0)$ in Eq.~\eqref{eq:m_phi4_smallq}
refers \emph{only} to the fact that the correction term approaches some function of $\vect K_1\vect K_2$ which
is \emph{independent} of $\vect q$, when $\vect q\rightarrow0$.

The following decomposition will be useful later:
\begin{equation}\label{eq:m_phi'}
\phi^{(4)}_{1234}\equiv\phi'_{1234}+\delta_{1\bar{2}}\delta_{3\bar{4}}\phi_1\phi_3
-\delta_{1\bar{3}}\delta_{2\bar{4}}\phi_1\phi_2+\delta_{1\bar{4}}\delta_{2\bar{3}}\phi_1\phi_2,
\end{equation}
where $\phi'_{\vect K_1\vect K_2\vect K_3\vect K_4}\sim\Omega^{-2}$ contains no delta-function singularity
and is also completely antisymmetric.
If $\vect K_1+\cdots+\vect K_4=0$ but $\vect K_1+\vect K_i\neq0$ (for all $2\leq i\leq4$), Eq.~\eqref{eq:m_phi4_Schrodinger} becomes
\begin{multline}\label{eq:m_phi'_Schrodinger}
\bigl[(k_1^2+k_2^2+k_3^2+k_4^2)/2-2E_m\bigr]\phi'_{1234}\\
+\frac{1}{2}{\sum_{56}}U_{1256}\phi'_{5634}
+\frac{1}{2}{\sum_{56}}U_{1356}\phi'_{5264}
+\frac{1}{2}{\sum_{56}}U_{1456}\phi'_{5236}\\
+\frac{1}{2}{\sum_{56}}U_{2356}\phi'_{1564}
+\frac{1}{2}{\sum_{56}}U_{2456}\phi'_{1536}
+\frac{1}{2}{\sum_{56}}U_{3456}\phi'_{1256}\\
-U_{12\bar{3}\bar{4}}\phi_3\phi_4
+U_{13\bar{2}\bar{4}}\phi_2\phi_4
-U_{14\bar{2}\bar{3}}\phi_2\phi_3~~~~~~\\
-U_{23\bar{1}\bar{4}}\phi_1\phi_4
+U_{24\bar{1}\bar{3}}\phi_1\phi_3
-U_{34\bar{1}\bar{2}}\phi_1\phi_2=0.
\end{multline}

A shorthand similar to Eq.~\eqref{eq:m_phi4_shorthand}:
\begin{equation}\label{eq:m_phi'_shorthand}
{\phi'}^\vect q_{\vect K_1\vect K_2}\equiv\phi'_{\vect K_1+\vect q/2,-\vect K_1+\vect q/2,\vect K_2-\vect q/2,-\vect K_2-\vect q/2}.
\end{equation}
Obviously ${\phi'}^\vect q_{12}=-(4\pi a_m/\Omega q^2)\phi_1\phi_2+O(q^0)$ as $q\rightarrow0$.

Projector onto the function space orthogonal to $\phi_\vect K$:
\begin{equation}
\proj_{\vect K_1\vect K_2}\equiv\delta_{\vect K_1\vect K_2}-\phi_{\vect K_1}\phi_{\vect K_2}^*/2.
\end{equation}
``Deviation functions":
\begin{subequations}\label{eq:m_devs}
\begin{align}
\begin{split}
\dev_{\vect K}&\equiv\Omega\lim_{\vect q\rightarrow0}\frac{1}{2}\sum_{\vect K_3\vect K_4}
   \proj_{\vect K\vect K_3}\phi^{(4)\vect q}_{\vect K_3\vect K_4}\phi^*_{\vect K_4}\\
  &=\devp_\vect K-\Omega\sum_{\vect K_3}\proj_{\vect K\vect K_3}
   \lvert\phi_{\vect K_3}\rvert^2\phi_{\vect K_3},\label{eq:m_dev}
\end{split}\\
\devp_{\vect K}&\equiv\Omega\lim_{\vect q\rightarrow0}\frac{1}{2}\sum_{\vect K_3\vect K_4}
   \proj_{\vect K\vect K_3}{\phi'}^\vect q_{\vect K_3\vect K_4}\phi^*_{\vect K_4},\label{eq:m_devp}\\
g^{}_{\vect K_1\vect K_2}&\equiv\Omega\lim_{\vect q\rightarrow0}\sum_{\vect K_3\vect K_4}
   \proj_{\vect K_1\vect K_3}\proj_{\vect K_2\vect K_4}\phi^{(4)\vect q}_{\vect K_3\vect K_4},\label{eq:m_g}\\
g'_{\vect K_1\vect K_2}&\equiv\Omega\lim_{\vect q\rightarrow0}\sum_{\vect K_3\vect K_4}
   \proj_{\vect K_1\vect K_3}\proj_{\vect K_2\vect K_4}{\phi'}^{\vect q}_{\vect K_3\vect K_4}.\label{eq:m_gp}
\end{align}
\end{subequations}
These limits exist because the divergent component in $\phi^{(4)\vect q}_{12}$
or ${\phi'}^{\vect q}_{12}$, for $\vect q\neq0$ but $\vect q\rightarrow0$, is proportional to $\phi_1\phi_2/q^2$, which is removed by the projector.
$\dev_\vect K$ and $\devp_\vect K$ are of the same thermodynamic order as $\phi_{\vect K}$,
but \emph{orthogonal} to $\phi_{\vect K}$:
$\sum_{\vect K}\phi^*_\vect K\dev_{\vect K}=0$ and $\sum_{\vect K}\phi^*_\vect K\devp_{\vect K}=0$;
likewise $\sum_1\phi_1^*g'_{12}=\sum_2\phi_2^*g'_{12}=\sum_1\phi_1^*g_{12}=\sum_2\phi_2^*g_{12}=0$.
Moreover, $\dev_{-\vect K}=-\dev_{\vect K}$, and $\devp_{-\vect K}=-\devp_{\vect K}$;
$g'_{12}=-g'_{\bar{1}2}=-g'_{1\bar{2}}=g'_{21}$; $g_{12}=-g_{\bar{1}2}=-g_{1\bar{2}}=g_{21}$;
$g'_{12}\sim\Omega^{-1}$. No evidence for the vanishing of any of these functions is found.

The identity 
\begin{equation}\label{eq:m_phi_id}
\sum_2U_{1\bar{1}2\bar{2}}\phi_2=-2(k_1^2-E_m)\phi_1
\end{equation}
and its complex conjugate (note that $U_{1234}^*=U_{3412}$) are often  used (explicitly or implicitly)
in the next section.

Two more identities:
\begin{multline}\label{eq:m_phi4_id1}
\frac{1}{2}\sum_{234}U_{1234}\phi^*_2\phi^{(4)}_{\bar{1}\bar{2}34}
+\frac{1}{2}\sum_{234}U_{\bar{1}\bar{2}34}\phi^*_2\phi^{(4)}_{1234}\\
=-\frac{2\pi a_m}{\Omega}\phi_1+\frac{1}{\Omega}\sum_{2}D_{12}\dev_2,
\end{multline}
\begin{equation}\label{eq:m_phi4_id0}
\sum_{1234}U_{\bar{1}\bar{2}34}\phi_1^*\phi_2^*\phi^{(4)}_{1234}=-4\pi a_m/\Omega.
\end{equation}
Equation~\eqref{eq:m_phi4_id1} is proved in Appendix~\ref{a:m_phi4_identity}; its
inner product with $\phi_1^*$ yields Eq.~\eqref{eq:m_phi4_id0}.
Expressing $\phi^{(4)}$ and $\dev$ in terms of $\phi'$ and $\devp$, we get (to be used in Secs.~\ref{subsec:m_20} and \ref{subsec:m_25}):
\begin{multline}\label{eq:m_phi4_id1_expanded}
\!\!\!\!\frac{1}{2}\sum_{234}U_{1234}\phi^*_2\phi'_{\bar{1}\bar{2}34}
+\frac{1}{2}\sum_{234}U_{\bar{1}\bar{2}34}\phi^*_2\phi'_{1234}+\sum_2D_{12}\lvert\phi_2\rvert^2\phi_2\\
+\lvert\phi_1\rvert^2\sum_2U_{1\bar{1}2\bar{2}}\phi_2
-\phi_1\sum_2(U_{1212}+U_{\bar{1}2\bar{1}2})\lvert\phi_2\rvert^2\\
=-\frac{2\pi a_m}{\Omega}\phi_1+\frac{1}{\Omega}\sum_{2}D_{12}\devp_2,
\end{multline}
\begin{multline}\label{eq:m_phi4_id0_expanded}
\sum\limits_{1234}U_{\bar{1}\bar{2}34}\phi^*_1\phi^*_2\phi'_{1234}
+\sum_{12}U_{1\bar{1}2\bar{2}}\lvert\phi_1\rvert^2\phi_1^*\phi_2\\
-2\sum_{12}U_{1212}\lvert\phi_1\rvert^2\lvert\phi_2\rvert^2=-4\pi a_m/\Omega.
\end{multline}
The three terms on the left side of Eq.~\eqref{eq:m_phi4_id0_expanded} are all \emph{real}.

Using a similar method as in Appendix~\ref{a:m_phi4_identity}, starting from Eq.~\eqref{eq:m_phi'_Schrodinger},
we can show the following identity:
\begin{multline}\label{eq:m_phi4_id2_expanded}
\mspace{-2mu}\frac{1}{2}\sum_{34}(U_{1234}\phi'_{\bar{1}\bar{2}34}+U_{\bar{1}\bar{2}34}\phi'_{1234}
-U_{1\bar{2}34}\phi'_{\bar{1}234}-U_{\bar{1}234}\phi'_{1\bar{2}34})\\
+U_{1\bar{1}2\bar{2}}\phi_2^2+U_{2\bar{2}1\bar{1}}\phi_1^2-(U_{1212}+U_{1\bar{2}1\bar{2}}+U_{\bar{1}2\bar{1}2}
+U_{\bar{1}\bar{2}\bar{1}\bar{2}})\phi_1\phi_2\\
=\!\!\frac{1}{\Omega}\!\sum_3\mspace{-1mu}[D_{13}(d'_3\phi_2+g'_{32})+D_{23}(\phi_1d'_3+g'_{13})]-\frac{2\pi a_m}{\Omega}\phi_1\phi_2,
\end{multline}
when $\vect K_1\neq\pm\vect K_2$. This will be used in Sec.~\ref{subsec:m_25}.

Finally, we have the $(p/2)$-molecule zero-speed scattering wave function ($p=2,4,6,\cdots$), the generalization
of Eqs.~\eqref{eq:m_phi} and \eqref{eq:m_phi4}.
\begin{multline}
\biggl[-\frac{pE_m}{2}-\epsilon^{(p)}+\sum_{i=1}^{p}\frac{k_i^2}{2}\biggr]\phi^{(p)}_{\vect K_1\cdots\vect K_p}
+\frac{1}{2}\sum_{1\leq i<j\leq p}\!U_{\vect K_i\vect K_j\vect K'\vect K''}\\
\times
\phi_{\vect K_1\cdots \vect K_{i-1}\vect K'\vect K_{i+1}\cdots\vect K_{j-1}\vect K''\vect K_{j+1}\cdots
\vect K_p}^{(p)}=0
\end{multline}
if $\vect K_1+\cdots+\vect K_p=0$; $\phi^{(p)}_{\vect K_1\cdots\vect K_p}=0$ otherwise. $\phi^{(p)}_{\vect K_1\cdots\vect K_p}$
is antisymmetric under the interchange of any two spin-momenta.
If $p=2$, $\epsilon^{(p)}=0$. If $p\geq4$, $\epsilon^{(p)}\sim\Omega^{-1}$, and
this extremely tiny energy is negligible unless the sum of a nontrivial subset of the $p$ spin-momenta vanishes.

\section{\label{sec:m_special}Diatomic molecules: low-density expansion}

At low density, the ground state
is a dilute gas of molecules, any two of which are usually - \emph{but not always} - far apart compared to
the size of a molecule. In this section we study the ground state, exploiting such a low density.

\subsection{\label{subsec:m_lowd_orders}Low-density orders}

If the \emph{leading order} contribution to a quantity $x$ scales like $n^R\Omega^Q$
in the low-density regime, we say that the \emph{low-density order} of $x$ is $R$, and may schematically write $x\sim n^R$;
more details are in Appendix~\ref{a:terms}.

It is reasonable to expect that the ground state energy
is, to the lowest order in density, simply $E=E_mN/2$. So $\mu=E_m/2+h.o.c.$ [$h.o.c.\equiv$ higher
order corrections (Appendix~\ref{a:terms})], and the
low-density order of $\mu$ is $0$, in contrast with the case of structureless
bosons \cite{_Tan_BEC}.

Since $\mu<0$, the low-density order of $k^2/2-\mu$ is 0, for any $\vect k$.
The low-density order of $U_{\vect K_1\vect K_2\vect K_3\vect K_4}$ is $0$.

The low-density order of $\alpha_{\vect K}$ is $0.5$. If we only retain the lowest
order term in the exponent in Eq.~\eqref{eq:m_psi}, we get
$n_{\vect K}=\lvert\alpha_\vect K\rvert^2/(1+\lvert\alpha_{\vect K}\rvert^2)$.
At low density, the spin-momentum distribution should
approach that of fermions in an isolated molecule times a constant, so
the \emph{width} ($\sim 1/r_m$) is roughly
 independent of $n$, and the occupation number of each $\vect K$
state should be of the order $nr_m^3\ll1$. So $\alpha_\vect K\sim n^{0.5}$.

The low-density order of $\beta_{\vect K_1\vect K_2\vect K_3\vect K_4}$ 
(where $\vect K_1+\cdots+\vect K_4=0$) is usually $1$ (see below for exceptions),
because $\lvert\beta(\vect R_1\vect R_2\vect R_3\vect R_4)\rvert^2$ 
(norm square of the Fourier transform) - for a given set of locations in a small region independent of $n$ - is approximately
associated with the probability of finding two molecules (or rather four fermions) in such a region,
and this probability is approximately proportional to $n^2$.

In general, the low-density order of $\alpha^{(p)}_{\vect K_1\cdots\vect K_p}$
is usually $p/4$.

The first exception to this estimate appears in $\beta_{\vect K_1\vect K_2\vect K_3\vect K_4}$,
since it is approximately proportional to the four-fermion wave function
describing the zero-velocity scattering of two molecules, in the low density limit.
When the two molecules are well separated, their spatial wave function should be
of the form $1-a_m/r$ times the internal wave functions of the molecules,
where $r$ is the distance between the molecules. Taking the Fourier transformation,
we see that $\beta_{\vect K_1\vect K_2\vect K_3\vect K_4}$ scales like $a_m/q^2$,
when $\vect K_1+\vect K_2=\vect q\rightarrow0$ (the term $\delta_{\vect q,0}$  is
encoded in $\alpha_\vect K$). If there were no many-body effects, this
would be an infrared divergence at small $q$. But again, just like in the case of
structureless bosons \cite{_Tan_BEC}, the divergence is regulated when $q\sim 1/\xi$, where $\xi$ is
the superfluid healing length. We will confirm in the concrete calculations that
$\xi$ still scales like $n^{-0.5}$. By reducing $\vect q=\vect K_1+\vect K_2$ from the order
$O(1)$ [for any momentum $\vect k$, by $k\sim O(1)$ we mean that $k$ has a finite and \emph{positive} lower bound
when $n\rightarrow0$]
to the order $n^{0.5}$, $\beta_{\vect K_1\vect K_2\vect K_3\vect K_4}$ must
increase from the low-density order $n^1$ to the low-density order $n^0$, in accordance with the $1/q^2$ dependence.
After this, when $q$ is further reduced, the $1/q^2$ dependence breaks down,
 and the order of magnitude of $\beta_{\vect K_1
\vect K_2\vect K_3\vect K_4}$ will not increase again.

Like in the case of structureless bosons \cite{_Tan_BEC}, we can extend this analysis to higher orders, and
arrive at a general result
\begin{equation}\label{eq:m_lowd_alpha}
\alpha^{(p)}_{\vect K_1\cdots\vect K_p}\sim n^{p/4-M+1}~~~~~~(1\leq M\leq p/2),
\end{equation}
where the $p$ spin-momenta - depending on their values - are divided into $M~$ \emph{l-clusters}.
Each l-cluster is a collection of (at least two) spin-momenta such that
\begin{enumerate}
\item[1.] their sum is of the low-density order $\sqrt{n}$ (if $M\geq2$) or is zero (if $M=1$);
\item[2.] the sum of any nontrivial subset of them is of the order
$O(1)$ [by this we mean that either the sum of spins does not vanish, or the sum of momenta is of the
order $O(1)$].
\end{enumerate}
If  $M\geq2$, the sum of spin-momenta in each l-cluster must not be zero, or
the whole $\alpha^{(p)}_{\vect K_1\cdots\vect K_p}$ vanishes (see Sec.~\ref{subsec:m_thermodynamic}).

There are still exceptions to the above result, because of the antisymmetry of
$\alpha^{(p)}_{\vect K_1\cdots\vect K_p}$. When the difference between two subscripts approaches zero, 
$\alpha^{(p)}_{\vect K_1\cdots\vect K_p}\rightarrow0$.
However, if we regard $O(n^{p/4-M+1}\Omega^{1-p/2})$ (see Appendix~\ref{a:terms}) as an \emph{upper bound}
of $\alpha^{(p)}_{\vect K_1\cdots\vect K_p}$, the Fermi statistics does not break it.

\begin{figure}
\includegraphics{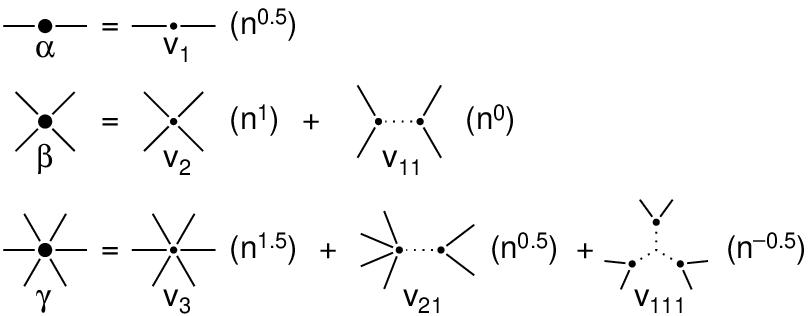}
\caption{\label{fig:m_alpha}The decomposition of $\alpha$, $\beta$, and $\gamma$
[the first three terms in the exponent in Eq.~\eqref{eq:m_psi}] into
dispersed vertices in the low-density regime. Each dispersed vertex is
denoted by $v$ with some subscript(s) indicating the number of pairs of fermions in each cluster;
the low-density order of its associated coefficients in momentum space is noted.
Dotted lines connect different clusters \emph{in} $v$. In each cluster, the sum of spins is zero and the sum
of momenta is of the order $\sqrt{n}$ (if there are two or more clusters in $v$) or
$0$ (if there is only one cluster), but for any nontrivial subset of fermions in the cluster,
either the sum of spins does not vanish, or the sum of momenta is of the order $O(1)$.}
\end{figure}

Now we decompose the $p/2$-th order (de)generator into the sum of a set of \emph{dispersed vertices},
like in the case of a dilute gas of structureless bosons \cite{_Tan_BEC}. The decomposition of
the first three generators are shown in Fig.~\ref{fig:m_alpha}; for the degenerators,
we just replace each solid circle with a hollow one. Despite the striking
similarity between Fig.~\ref{fig:m_alpha} and a similar figure in \cite{_Tan_BEC}, 
the corresponding coefficients have vastly different orders of magnitude, because of their
different \emph{thermodynamic orders}.
Nevertheless, their \emph{low-density orders} are the same. Since we have proved that
the \emph{total} thermodynamic order of any diagram in the expansion of $E-\mu N$ is $1$,
we can concentrate on the low-density orders when estimating the orders of magnitude of such
diagrams, without worrying about the thermodynamic orders of individual vertices.

In such a decomposition, $\alpha\equiv\alpha^{(2)}$ remains a single term,
denoted by $v_1$. $\beta\equiv\alpha^{(4)}$ is decomposed
to two terms, $v_{2}$ and $v_{11}$; the first one is restricted to $k>k_c$, and the second one to $k<k_c$,
where $k=\min_{\sigma_i+\sigma_j=0,1\leq i<j\leq4}\lvert\vect K_i+\vect K_j\rvert$,
and $k_c$ is a momentum scale satisfying
\begin{equation}\label{eq:m_k_c}
\begin{split}
\sqrt{na_x}\ll &k_c\ll\min(1/l,1/r_m),\\
k_c/\sqrt{na_x} &\text{ independent of } n.
\end{split}
\end{equation}
where $a_x$ is independent of $n$ and $\Omega$; we will see that $a_x$ is actually $a_m$.
Further, $\gamma\equiv\alpha^{(6)}$ is decomposed to three terms (Fig.~\ref{fig:m_alpha}), and we will see that the
second term, $v_{21}$, is most important in the low-density regime, similar to the situation for structureless
bosons \cite{_Tan_BEC}.

Such division of the spin-momentum configuration space is crucial for the practical computation. By decomposing diagrams
according to such division of the basic vertices, one will be able to analyze the orders of magnitude of diagrams and compute their
actual values much more easily. The physical result, however, is independent of such division of the spin-momentum space,
and in particular, independent of the choice of $k_c$, as long as the above inequality is satisfied.

The low-density order of the coefficients $\alpha^{[p(v)]}_{\vect K_1\cdots\vect K_{p(v)}}$
in any dispersed vertex $v$ is
\begin{equation}\label{eq:m_dispersed}
R(v)=\frac{p(v)}{4}-N_1'(v)+N_j'(v),
\end{equation}
where $p(v)$ is the number of solid lines attached to $v$, $N_{1}'(v)$ is the number of dotted
lines contained by $v$, and $N_{j}'(v)$ is the number of junctions of these dotted lines.
If $M\leq2$ [see Eq.~\eqref{eq:m_lowd_alpha}],
$N_{1}'(v)=M-1$ and $N_j'(v)=0$; but if $M\geq3$, $N_{1}'(v)=M$ and $N_{j}'(v)=1$;
see Fig.~\ref{fig:m_alpha}. Equation~\eqref{eq:m_dispersed}
 is equivalent to Eq.~\eqref{eq:m_lowd_alpha},
but will be more convenient in the next subsection.

\subsection{\label{subsec:m_power_counting}Low-density orders of diagrams}

To compute physical observables in the low density regime, it will be much more convenient to decompose
the (de)generators in the above way first.
The physical observables are then expanded in terms
of the dispersed vertices, and we obtain a new set of diagrams containing small internal circles,
solid lines, and dotted lines; for $E-\mu N$ we also have a diamond in each diagram,
and for $\langle O\rangle$ ($O$ is a product of $L$ basic fermionic operators in spin-momentum space), we have
$L$ external points; in this paper we use the term \emph{low-density expansion}
in a narrower sense, to refer to these diagrams.

To obtain the correct value of a physical observable up to a certain order in the low-density
expansion, we must take into account \emph{all} the diagrams for this observable
up to this order (called significant diagrams). \emph{The omission of any significant diagram may
invalidate the result completely.} It is therefore crucial to derive a general power-counting formula,
using which we can identify \emph{all} the significant diagrams for an observable up to any certain order.

Such formula is
\begin{equation}\label{eq:R}
R=N_3+\sum_{i=1}^{I}(P_i/4-1)+N_2'/2,
\end{equation}
where $R$ is the low-density order of any such diagram (denoted by $\mathcal{D}$ below), 
$N_3$ is its number of disconnected parts, $I$ the number of islands, $P_i$ the $P$-number of the
$i$-th island, and $N_2'$ is the number of independent loops in the skeleton diagram. The
terms are defined as follows.

If we remove all the dotted lines (if there are any) and their possible junctions (see previous subsection)
from $\mathcal{D}$, but leave all the other components unchanged, $\mathcal{D}$ is changed to $I$
disconnected parts, each of which is called an \emph{island}. Different islands are either linked
by dotted lines only, or completely disconnected.
Any small circle (other than the external points) in an island is attached by a certain number of \emph{solid} lines; the sum
of all such numbers over the whole island is called the \emph{$P$-number} of the island,
and the sum over the whole diagram is the $P$-number of the diagram.
Finally, if we reduce every entire island in $\mathcal{D}$ to a point, but leave the
dotted lines (if there are any) and their possible junctions unchanged (the dotted lines are still
attached to the islands and junctions), we get the \emph{skeleton diagram}
of $\mathcal{D}$. Obviously, if $\mathcal{D}$ contains no dotted lines, its skeleton diagram
is $N_3$ isolated points and $I=N_3$.

\begin{proof}
Only two factors contribute to the low-density order $R$ of $\mathcal{D}$:
\begin{itemize}
\item each dispersed vertex $v$
contributes $R(v)$ in accordance with Eq.~\eqref{eq:m_dispersed};
\item each independent small internal momentum $\vect q$ ($\sim\sqrt{n}$)
which flows through a dotted line contributes $1.5$, since the measure
$\sum_{\lvert\vect q\rvert\sim\sqrt{n}}=\Omega\int_{q\sim\sqrt{n}}\dif^3q/(2\pi)^3$ scales like $n^{1.5}$.
\end{itemize}
The number of independent $\vect q$'s above is clearly $N_2'$, as momentum is conserved
in any part of the diagram. So $R=P/4-N_1'+N_j'+1.5N_2'$, where $P$ is the
$P$-number of $\mathcal{D}$, $N_1'$ the total number of dotted lines, and $N_j'$ is the total
number of junctions of the dotted lines.

Now turn to the skeleton diagram $\mathcal{D'}$.
It has $I+N_j'$ vertices (islands and junctions), $N_1'$ lines (dotted lines),
$N_2'$ independent loops, and  $N_3'=N_3$ disconnected parts (the number of disconnected parts
is not changed when the islands are reduced to points). So using Eq.~\eqref{eq:any_diagram},
we get $I+N_j'-N_1'+N_2'-N_3=0$.

Using the above equation to cancel the term $-N_1'+N_j'$ in the expression of $R$, we get
Eq.~\eqref{eq:R} (note that obviously $P=\sum_{i=1}^{I}P_i$).
\end{proof}

The central result of the last subsection, Eq.~\eqref{eq:m_lowd_alpha}, is a \emph{special case}
of Eq.~\eqref{eq:R}.

To better understand $R$, we further study $P_i$.
Suppose that the contribution to $P_i$ from all the small \emph{solid} circles (other than the external points)
 in the $i$-th island is $P^{(+)}_i$, and that the contribution from all the small \emph{hollow} circles (other than the external points)
in the same island is $P^{(-)}_i$. Suppose also that the number of solid external points
in the $i$-th island is $L^{(+)}_i$, and that the number of hollow external points in the same island
is $L^{(-)}_i$. Obviously
\begin{subequations}\label{eq:P_i}
\begin{equation}
P_i=P^{(+)}_i+P^{(-)}_i,
\end{equation}
\begin{equation}
P^{(+)}_i+L^{(+)}_i=P^{(-)}_i+L^{(-)}_i>0,
\end{equation}
\begin{equation}
P^{(+)}_i, P^{(-)}_i\in\{0, 2,4,6,8,\cdots\}.
\end{equation}
\end{subequations}
The contribution to the balance of particle number from the diamond (if there is any) is cancelled.

For convenience, we shall call any island which contains two or more external points, an \emph{external island};
any island which contains no external point, an \emph{internal island}.
Any island which contains a diamond, a \emph{hamiltonian island}.
Finally, any island which contains neither a diamond nor any external point, a \emph{(vacuum) bubble island},
since it has the same topological structure as a vacuum bubble. 

An corollary of Eq.~\eqref{eq:P_i}: for any \emph{internal} island $i$ (a bubble island, or a hamiltonian island
which contains no external points), $P_i$ can only be $4,8,12,16\cdots$, and $P_i/4-1$
can only be a \emph{nonnegative integer}. It is this property that makes Eq.~\eqref{eq:R}
particularly useful in the search of all the diagrams in the low-density expansion of an observable
up to a certain order.

The bubble island whose $P$-number is $4$ has a unique topological struture - a solid circle linked to a hollow one by two
solid lines - and contributes nothing to $R$; it will be called the \emph{simplest bubble}.

An alternative (sometimes useful) form of Eq.~\eqref{eq:R} is
\begin{equation}\label{eq:R2}
R=N_3+(P_\text{ext}/4-I_\text{ext})+N_2'/2+(P_\text{int}/4-I_\text{int}),
\end{equation}
where $I_\text{ext}$ and $I_\text{int}$ are the numbers of external islands and internal ones, respectively;
$P_\text{ext}$ and $P_\text{int}$ are the total contributions to $P$ from the external islands and internal ones, respectively.
$P_\text{int}/4-I_\text{int}$ is always a nonnegative integer. 

In summary, $\mathcal{D}$ is more conveniently viewed as an ``archipelago" (islands with weak links)
when its low-density order is
analyzed. The archipelago structure is a direct consequence of the presence of two sets of well-separated length scales
in the problem: the \emph{superfluid healing length} $\propto1/\sqrt{n}$, and the \emph{few-body length scales}
$l$, $r_m$, etc.

When computing its actual value, we of course view $\mathcal{D}$ in the ``usual" way, namely as
a set of dispersed vertices with a diamond or/and external points.

\subsection{\label{subsec:m_10}$E-\mu N$ to the order $n^1$}

In any diagram in the low-density expansion of $E-\mu N$, there can not be any ``dangling" dotted line
(one whose removal would change such a diagram to two disconnected parts), since
dead ends are absent (Sec.~\ref{subsec:m_thermodynamic}). So if there is any dotted line in such
a diagram, $N_2'\geq1$ and $R\geq1.5$ [see Eq.~\eqref{eq:R}].

So if $R=1$, there can only be one island, the hamiltonian island. Moreover, the $P$-number of this
island must be $4$. So we can only find two diagrams satisfying $R=1$, in the expansion of $E-\mu N$.
They are $T_1$ and $T_2$ shown in Fig.~\ref{fig:33terms}.
\begin{subequations}\label{eq:m_10}
\begin{align}
T_1&=\sum_\vect K(k^2/2-\mu)\lvert\alpha_\vect K\rvert^2,\\
T_2&=\frac{1}{4}\sum_{\vect K\vect K'}U_{\vect K,-\vect K,\vect K',-\vect K'}^{}\alpha_{\vect K}^*
\alpha_{\vect K'}^{}.
\end{align}
\end{subequations}
We then adjust the parameters $\alpha_\vect K$ to minimize $E-\mu N$:
$\partial(E-\mu N)/\partial\alpha_\vect K^*=0$. So
\begin{equation}
k^2\alpha_{\vect K}+\frac{1}{2}\sum_{\vect K'}U_{\vect K,-\vect K,\vect K',-\vect K'}^{}\alpha_{\vect K'}
=2\mu\alpha_{\vect K}+h.o.c.,
\end{equation}
where $h.o.c.$ stands for  higher order corrections.
Since the fermions have formed bound pairs, and the density is low,
$\alpha_\vect K$ should be approximately proportional to the internal wave function of an isolated molecule,
and $2\mu\approx E_m$. Comparing the above equation with Eq.~\eqref{eq:m_phi_Schrodinger}, we find
$\alpha_{\vect K}=\eta\phi_{\vect K}+h.o.c.$.
The number of fermions is $N=\sum_{\vect K}\lvert\alpha_{\vect K}\rvert^2+h.o.c.=2\lvert\eta\rvert^2+h.o.c.$,
so $\eta=\sqrt{N/2}+h.o.c.$, where we have chosen the phase factor of $\eta$. We thus get
\begin{align}
\alpha_{\vect K}&=\sqrt{N_\up}\,\phi_{\vect K}+h.o.c.,\label{eq:m_10_alpha}\\
\mu&=E_m/2+h.o.c.,\\
E/\Omega&=nE_m/2+h.o.c.,
\end{align}
where $E$ is obtained by solving $\dif E/\dif N=\mu$, and
\begin{subequations}
\begin{alignat}{2}
N_\up&\equiv N_\down&\equiv N/2&\,,\label{eq:m_N_updown}\\
n_\up&\equiv n_\down&\equiv n/2&\equiv N/(2\Omega).\label{eq:m_n_updown}
\end{alignat}
\end{subequations}

\subsection{\label{subsec:m_1st_separation}First cluster-separation theorem}
In this subsection and the next, we discuss an elementary property of the system in the low-density regime.
It will help to further the low-density expansion.

Consider $p$ spin-momenta $\widetilde{\vect K}_1\cdots\widetilde{\vect K}_p$, where $p$ is positive and even,
$\widetilde{\vect K}_1+\cdots+\widetilde{\vect K}_p=0$, and the sum of any nontrivial subset of these spin-momenta is nonzero.
Suppose that they are divided into $M$ l-clusters ($M\geq1$), and the $\nu$-th ($1\leq\nu\leq M$) l-cluster contains $p_\nu$ spin-momenta:
$\widetilde{\vect K}_{b_\nu+1},\widetilde{\vect K}_{b_\nu+2},\cdots,\widetilde{\vect K}_{e_\nu}$,
where $\sum_{\nu=1}^Mp_\nu=p$, $e_\nu=p_1+p_2+\cdots+p_\nu$, and $b_\nu=e_\nu-p_\nu$.
Suppose also that there does not exist any l-cluster other than these $M$. 
Make the decomposition $\widetilde{\vect K}_{b_\nu+i}=\vect K_{i,\nu}+\vect q_\nu/p_\nu$ ($1\leq i\leq p_\nu$), such
that $\sum_{i=1}^{p_\nu}\vect K_{i,\nu}=0$. Consider the ground state expectation value
$\A_{\widetilde{\vect K}_1\cdots\widetilde{\vect K}_p}\equiv\langle c_{\widetilde{\vect K}_p}^{}
c_{\widetilde{\vect K}_{p-1}}^{}\cdots c_{\widetilde{\vect K}_1}^{}\rangle$.
The first cluster-separation theorem states that
there exists some coefficient $\eta$, which depends on $\vect q_1\cdots\vect q_M$ but \emph{not} on
$\vect K_{1,1}\vect K_{2,1}\cdots\vect K_{p_1,1},\cdots,\vect K_{1,M}\vect K_{2,M}\cdots\vect K_{p_M,M}$,
such that
\begin{equation}\label{eq:m_1st_separation}
F_{\widetilde{\vect K}_1\cdots\widetilde{\vect K}_p}=\eta\prod_{\nu=1}^M\phi^{(p_\nu)}_{\vect K_{1,\nu}\vect K_{2,\nu}\cdots\vect K_{p_\nu,\nu}}
\end{equation}
with a \emph{relative} error $\sim n^1$.

There are infinitely many diagrams in the low-density expansion of $\A$, but their low-density
orders have a lower bound. Consider any one of these diagrams, $\mathcal{D}$:
$N_3\geq1$, $P_\text{ext}\geq p$, and $I_\text{ext}\leq M$ (it is impossible to separate the $p$ external points
into more than $M$ islands, since spin-momentum is conserved at every internal vertex,
and the dotted lines can not carry spin or any large momentum), so we deduce from Eq.~\eqref{eq:R2}
 that $R(\mathcal{D})\geq p/4-M+1$. The low-density order of $\A$ is therefore $R\equiv p/4-M+1$ or higher.
One of the leading contributions (with low-density order $R$) to $\A$
is $\alpha^{(p)}_{\widetilde{\vect K}_1\cdots\widetilde{\vect K}_p}$.

While it is intuitively clear that the right side of Eq.~\eqref{eq:m_1st_separation}
should be the form of the \emph{lowest order} approximation of $\A$, it is not \textit{a priori}
obvious why the \emph{next-to-leading} order correction (of the order $n^{R+0.5}$) is also proportional to such
a product of zero-speed scattering wave functions.

The proof is quite involved.
In Appendix~\ref{a:1st_separation} we prove this theorem for the cases in which each l-cluster contains only two spin-momenta.
To compute $E-\mu N$ up to the order $n^{2.5}$, such an incomplete
proof will be sufficient. The complete proof will appear in a  future work.

\subsection{\label{subsec:m_2nd_separation}Second cluster-separation theorem}

There is a similar formula for the coefficient of the dispersed vertex:
\begin{equation}
\alpha^{(p)}_{\widetilde{\vect K}_1\cdots\widetilde{\vect K}_p}=\eta'\prod_{\nu=1}^M\phi^{(p_\nu)}
_{\vect K_{1,\nu}\vect K_{2,\nu}\cdots\vect K_{p_\nu,\nu}}
\end{equation}
plus a correction term whose low-density order is $R+1$ or higher, where $R\equiv p/4-M+1$,
and the $\vect q$'s, $\vect K$'s, and $\widetilde{\vect K}$'s are defined
in Sec.~\ref{subsec:m_1st_separation}. The coefficient $\eta'$ depends on $\vect q_1\vect q_2\cdots\vect q_M$
and $n$, but is \emph{independent} of the $\vect K$'s.

The nontrivial content of this theorem is that the next-to-leading order correction ($\sim n^{R+0.5}$)
to the coefficient of the dispersed vertex is factorizable in the same manner as the leading term.

In Appendix~\ref{a:2nd_separation} we give the proof of this theorem for the cases in which each l-cluster contains
only two spin-momenta; it builds upon the results \emph{already proved} in Appendix~\ref{a:1st_separation}, and is even more involved.
For the determination of $E-\mu N$ up to the order $n^{2.5}$, such special results
will suffice. The complete proof will appear later.

This partial proof will simplify the logic in the concrete calculations
of $E-\mu N$. More importantly, it enables us to accurately determine
some other interesting quantities, such as the many-body effects in the
momentum distribution of the \emph{fermions}.

The second cluster-separation theorem is more fundamental in the present formalism, since the first one
and some other similar properties, such as that of $\langle c^\dagger_{\vect K_1+\vect q/2} c^\dagger_{-\vect K_1+\vect q/2}
c^{}_{-\vect K_2+\vect q/2}c^{}_{\vect K_2+\vect q/2}\rangle$, are its corollaries.
The first one is presented for two reasons: 1) it concerns physical \emph{observables}
rather than parameters in a particular calculational framework, and 2) the proof of the second theorem
is built upon the proof of the first.
Hereafter we use the name ``cluster-separation theorem" to refer to the second theorem.

\subsection{\label{subsec:m_diagrams_25}Diagrams for $E-\mu N$ up to the order $n^{2.5}$}

In this subsection we identify all the diagrams in the low-density expansion of $E-\mu N$ up to
the order $n^{2.5}$.

Any such diagram $\mathcal{D}$ has $N_3=1$ since it is a connected one.
So $R\leq2.5$ implies $N_2'\leq3$ [Eq.~\eqref{eq:R}].

\begin{figure}
\includegraphics{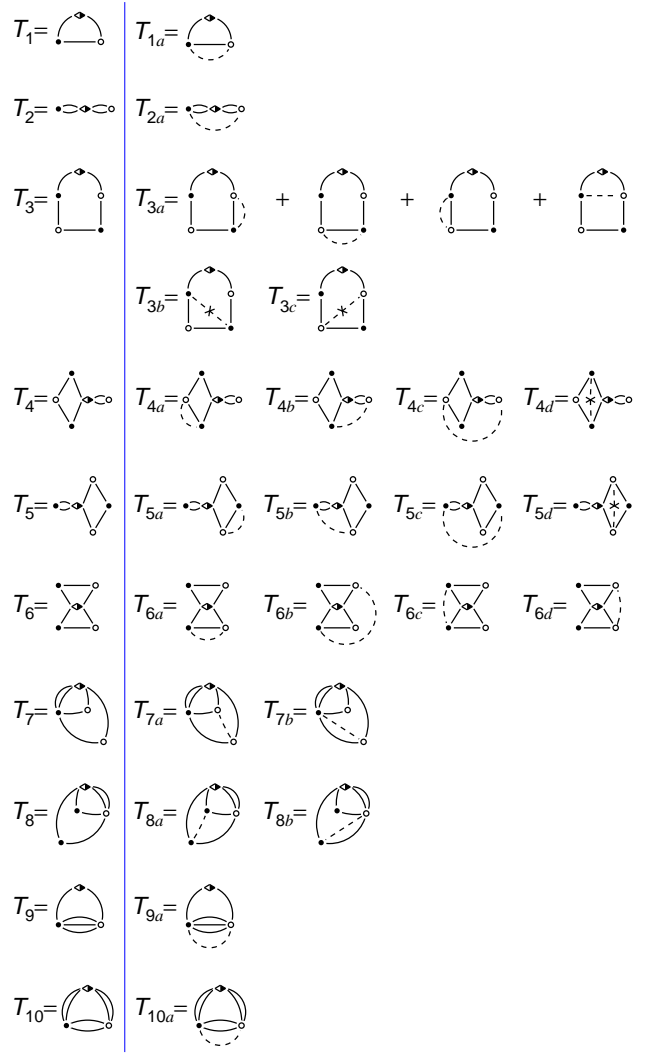}
\caption{\label{fig:33terms}Diagrams in the low-density expansion of $E-\mu N$ of molecules
(pairs of fermions) up to the order $n^{2.5}$. The composite vertices (each of which contains
a \emph{dashed} line) are defined in Fig.~\ref{fig:m_composite_vertices}; each of them
represents a geometric series. In $T_{3b}$,
$T_{3c}$, $T_{4d}$, and $T_{5d}$, a cross is marked on the dashed line, to indicate
the absence of the first term in such a series, \textit{ie.} $\beta_{\vect K_1
\vect K_2\vect K_3\vect K_4}$ (or its complex conjugate), because two of these four spin-momenta happen to be
exactly opposite.}
\end{figure}

If $N_2'=0$, there is no dotted line at all in $\mathcal{D}$. So there is only one island, the hamiltonian island,
whose $P$-number can only be $4$ ($R=1$) or $8$ ($R=2$). This is because any island in the expansion of
$E-\mu N$ is an internal one, whose $P$-number must be a multiple of $4$ (Sec.~\ref{subsec:m_power_counting}).
The two $R=1$ diagrams are $T_1$ and $T_2$ (Fig.~\ref{fig:33terms}), whose \emph{sum} however
has a \emph{higher} low-density order, $R=2$. This is the first instance of the \emph{$``R+1"$-cancellation mechanism},
discussed in the following.

Consider an arbitrary pair of diagrams in the low-density expansion of $E-\mu N$, $\mathcal{D}_1$ and $\mathcal{D}_2$,
whose hamiltonian islands are topologically the same as $T_1$
and $T_2$, respectively: if their other parts are identical, we call them an \emph{$``R+1"$-pair}. The two diagrams obviously
have the same low-density order, $R$. The second cluster-separation theorem implies that the low-density order
of their sum is at least $R+1$.

If $\mathcal{D}_1$ and $\mathcal{D}_2$ are just $T_1$ and $T_2$, the proof is simple:
substituting $\alpha_{\vect K}=\eta\phi_\vect K+O(n^{1.5})$ (Sec.~\ref{subsec:m_2nd_separation})
 into Eq.~\eqref{eq:m_10}, and noting that $\mu=E_m/2+O(n^1)$ (Appendix~\ref{a:1st_separation}),
we immediately get this result.

\begin{figure}\includegraphics{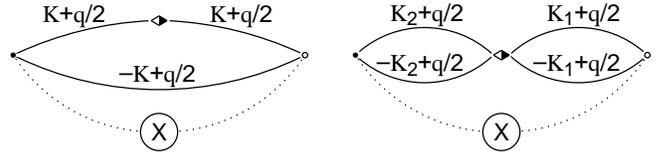}
\caption{\label{fig:m_Rplus1}A pair of diagrams $\mathcal{D}_1$ and $\mathcal{D}_2$; their sum is smaller
than each one of them by a factor of the order $n^1$. Their structures within the ``X"-circles are identical,
namely a certain number of bubble islands of certain structures, linked by dotted lines in a certain way.}
\end{figure}

\begin{figure}\includegraphics{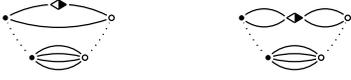}
\caption{\label{fig:m_Rplus1special}Two diagrams with $R=2.5$ each. Their sum should be $\sim n^{3.5}$,
but the proof is not yet given. However, trivially this sum is at least of the low-density order $n^3$.}
\end{figure}

If $\mathcal{D}_1$ and $\mathcal{D}_2$ contain dotted lines, they can be written in the form
[see Fig.~\ref{fig:m_Rplus1} and Eq.~\eqref{eq:m_k_c}]
\begin{subequations}\label{eq:m_Rplus1}
\begin{align}
\mathcal{D}_1&=\sum_{q<k_c;\vect K}(k^2/2-E_m/2-\deltamu+q^2/8)X_{\vect K\vect K}^\vect q,\\
\mathcal{D}_2&=\frac{1}{4}\sum_{q<k_c;\vect K_1\vect K_2}U_{\vect K_1,-\vect K_1,\vect K_2,-\vect K_2}^{}X_{\vect K_2\vect K_1}^\vect q,
\end{align}
\end{subequations}
where $\deltamu\equiv\mu-E_m/2$,
the first subscript of $X^\vect q_{\vect K_2\vect K_1}$ is associated with the creation of a pair of fermions with spin-momenta
$\vect K_2+\vect q/2$ and $-\vect K_2+\vect q/2$, and the second subscript is associated with the annihilation of a pair
with spin-momenta $\vect K_1+\vect q/2$ and $-\vect K_1+\vect q/2$. In $\mathcal{D}_1$, an average over $\vect K$ and $-\vect K$ has
been taken, since $X^\vect q_{\vect K_2\vect K_1}$ is antisymmetric for $\vect K_2\rightarrow-\vect K_2$ (and similarly for $\vect K_1$),
and thus $X_{\vect K\vect K}^\vect q$ is symmetric for $\vect K\rightarrow-\vect K$. In $\mathcal{D}_2$, $U$ has Galilean symmetry
and is independent of $\vect q$ (Sec.~\ref{subsec:m_hamiltonian}). Note finally that the \emph{signs} of $\mathcal{D}_1$ and $\mathcal{D}_2$
are correct, as is clear if one restrict every trajectory (Secs.~\ref{subsec:m_diagrams} and \ref{subsec:m_diagrams_energy})
within an island when determining the sign. If the small solid circle associated with the first subscript is contained by a dispersed
vertex $v^{}_{\underbrace{1\cdots1}_{M}}$, then from our established result we find that $X^\vect q_{\vect K_2\vect K_1}=\phi_{\vect K_2}
{X'}^{\vect q}_{\vect K_1}$ with relative error of the order $n^1$, and the $R+1$ cancellation between $\mathcal{D}_1$
and $\mathcal{D}_2$ immediately follows. If the small hollow circle associated with the second subscript is contained
by $v^\dagger_{\underbrace{1\cdots1}_M}$, then $X^\vect q_{\vect K_2\vect K_1}=\phi^*_{\vect K_1}{X''}^{\vect q}_{\vect K_2}$
with relative error of the order $n^1$, and the $R+1$ cancellation follows from the complex conjugate of Eq.~\eqref{eq:m_phi_Schrodinger}.
For $R\leq2.5$, there is only \emph{one} $``R+1"$-pair for which we have not shown the $R+1$ cancellation,
purely due to our incomplete proof of the cluster-separation theorem. This case is shown in Fig.~\ref{fig:m_Rplus1special}.
It involves the vertex $v_{21}$ and $v_{21}^\dagger$, and each diagram has $R=2.5$. However,
$\beta_{\vect K_1+\vect q/4,\vect K_2+\vect q/4,\vect K_3+\vect q/4,\vect K_4+\vect q/4,\vect K'-\vect q/2,-\vect K'-\vect q/2}$
($q\sim\sqrt{n}$, $\vect K_1+\cdots+\vect K_4=0$)
is proportional to $\phi_{\vect K'}$, at the leading order at least, since it is approximately associated with the correlation
of four fermions and a molecule which is far away (with distance $\sim\xi$) and the \emph{internal structure} of that molecule is almost
not influenced by the first four fermions. So the sum of these two diagrams is at least of the low-density order $n^3$, and can be omitted
in the computation of $E-\mu N$ up to the order $n^{2.5}$.

The counterpart of the ``R+1" mechanism in the context of \emph{structureless} bosons \cite{_Tan_BEC} is a much
more obvious fact that $q^2/2-\mu_\text{boson}$ is of the order $n^1$ when the boson momentum $q=0$ or $q\sim n^{1/2}$.

Because of the $``R+1"$-cancellation mechanism, the set of significant diagrams, up to the order $n^{2.5}$, is greatly reduced.
If the $``R+1"$-pair $\mathcal{D}_1$ and $\mathcal{D}_2$ satisfy $R\geq2$ each, then their sum is at least of the order $n^3$ and negligible.

If the hamiltonian island is $T_1$ or $T_2$, only the $N_2'=0$ or $N_2'=1$ diagrams are significant,
and the bubble islands are all the simplest ones (because of the $R+1$ mechanism).
They are $T_1$, $T_2$, $T_{1a}$ and $T_{2a}$, shown in Fig.~\ref{fig:33terms}.

If the hamiltonian island has $P=8$, then $N_2'=0$ ($R=2$) or $N_2'=1$ ($R=2.5$). If $N_2'=0$,
the hamiltonian island (the whole diagram) may contain two $v_1$'s and two $v_1^\dagger$'s
($T_3$, $T_4$, $T_5$, and $T_6$), or one $v_2$ and two $v_1^\dagger$'s ($T_7$),
or two $v_1$'s and one $v_2^\dagger$ ($T_8$), or one $v_2$ and one $v_2^\dagger$ ($T_9$ and $T_{10}$).
If $N_2'=1$, we just exhaust all the topologically distinct ways of attaching a dashed line to the hamiltonian island,
and obtain the diagrams $T_{3a}$, $T_{3b}$, $T_{3c}$, $T_{4a}$, $T_{4b}$, $T_{4c}$, $T_{4d}$,
$T_{5a}$, $T_{5b}$, $T_{5c}$, $T_{5d}$, $T_{6a}$, $T_{6b}$, $T_{6c}$, $T_{6d}$, $T_{7a}$, $T_{7b}$,
$T_{8a}$, $T_{8b}$, $T_{9a}$, and $T_{10a}$. The bubble islands in these diagrams are all the simplest ones or $R\geq3.5$
[Eq.~\eqref{eq:R}]. All these diagrams are shown in Fig.~\ref{fig:33terms}.

\begin{figure*}
\includegraphics{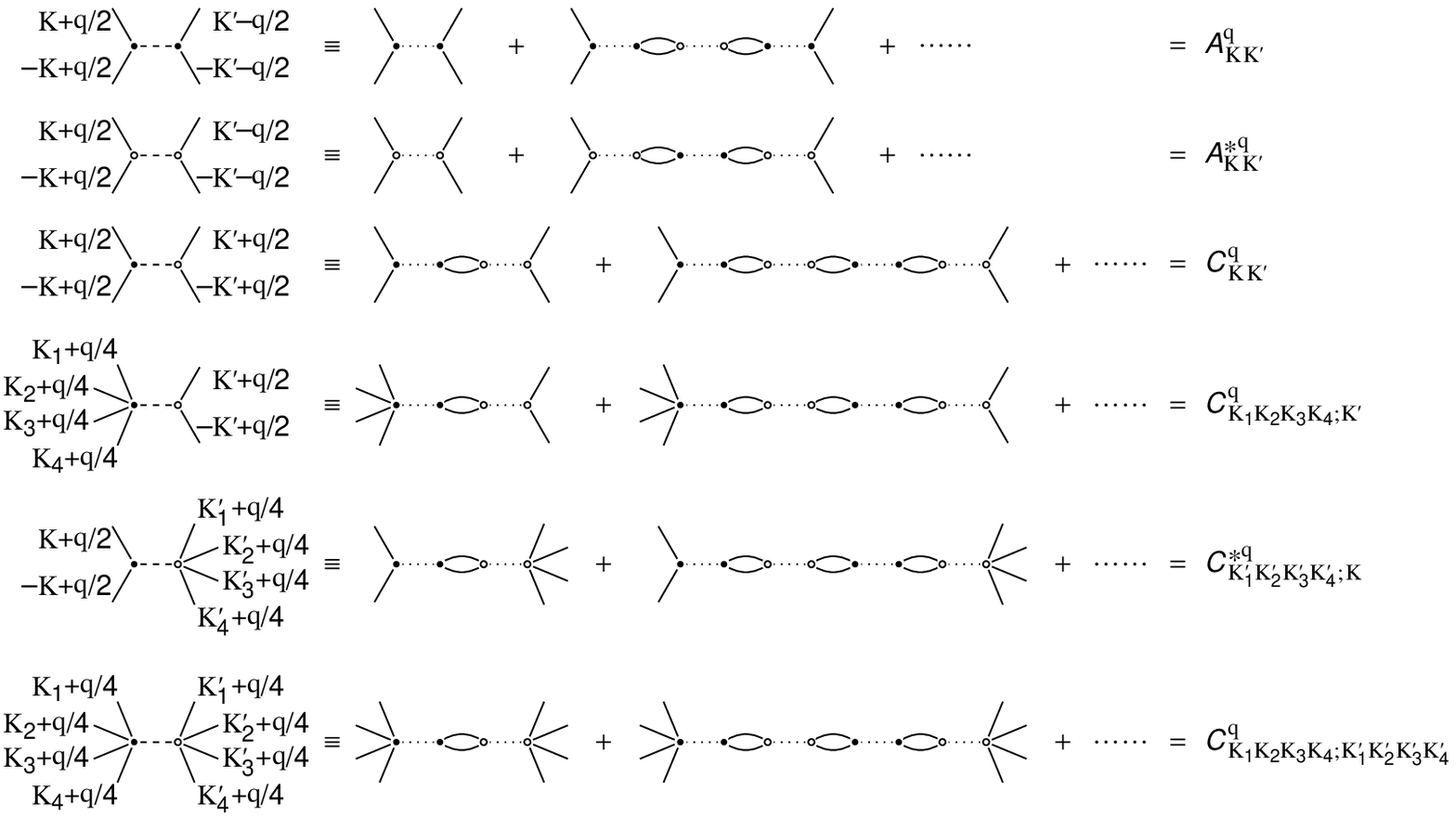}
\caption{\label{fig:m_composite_vertices}Definition of some composite vertices,
each of which is a geometric series. The spin-momenta $\vect K$'s are of the order $1$, and
the momentum $q\sim\sqrt{n}$. The dispersed vertices are defined in Fig.~\ref{fig:m_alpha}.}
\end{figure*}

Each dashed line represents a geometric series associated with a variable number of simplest bubbles.
Each end of the dashed line is attached by an arbitrary positive even number of solid lines.
The concrete meanings of the dashed lines used in Fig.~\ref{fig:33terms} are shown in Fig.~\ref{fig:m_composite_vertices}.

In $T_{3b}$, $T_{3c}$, $T_{4d}$, and $T_{5d}$, a cross is marked on the dashed line, to indicate
the absence of the first term in the geometric series, \textit{ie.} $\beta_{\vect K_1
\vect K_2\vect K_3\vect K_4}$ or its complex conjugate (recall that a diagram can not contain any dead end).

\subsection{\label{subsec:m_20}$E-\mu N$ to the order $n^{2}$}
The diagrams for $E-\mu N$ up to the order $n^2$ are $T_1$ through $T_{10}$ only. Note that $T_{1a}$ and $T_{2a}$
are an $``R+1"$-pair and their sum is of the low-density order $n^{2.5}$.
\begin{subequations}\label{eq:m_20}
{\allowdisplaybreaks
\begin{align}
T_1&=\sum_\vect K(k^2/2-\mu)\lvert\alpha_\vect K\rvert^2,\\
T_2&=\frac{1}{4}\sum_{\vect K_1\vect K_2}U_{\vect K_1,-\vect K_1,\vect K_2,-\vect K_2}^{}\alpha_{\vect K_1}^*
\alpha_{\vect K_2}^{},\\
T_3&=-\sum_\vect K(k^2/2-E_m/2)\lvert\alpha_\vect K\rvert^4,\\
T_4&=-\frac{1}{4}\sum_{\vect K_1\vect K_2}U_{\vect K_1,-\vect K_1,\vect K_2,-\vect K_2}\alpha_{\vect K_1}^*
\lvert\alpha_{\vect K_2}\rvert^2\alpha_{\vect K_2},\\
T_5&=-\frac{1}{4}\sum_{\vect K_1\vect K_2}U_{\vect K_1,-\vect K_1,\vect K_2,-\vect K_2}\lvert\alpha_{\vect K_1}^{}\rvert^2
\alpha_{\vect K_1}^*\alpha_{\vect K_2}^{},\\
T_6&=\frac{1}{2}\sum_{\vect K_1\vect K_2}U_{\vect K_1\vect K_2\vect K_1\vect K_2}\lvert\alpha_{\vect K_1}\rvert^2
\lvert\alpha_{\vect K_2}\rvert^2,\\
T_7&=-\frac{1}{4}{\sum_{\vect K\text{'s}}}'
U_{-\vect K_1,-\vect K_2,\vect K_3\vect K_4}\alpha^*_{\vect K_1}\alpha^*_{\vect K_2}
\beta_{\vect K_1\vect K_2\vect K_3\vect K_4},\\
T_8&=-\frac{1}{4}{\sum_{\vect K\text{'s}}}'U_{\vect K_1\vect K_2,-\vect K_3,-\vect K_4}\beta^*
_{\vect K_1\vect K_2\vect K_3\vect K_4}\alpha_{\vect K_3}^{}\alpha_{\vect K_4}^{},\\
T_9&=\frac{1}{6}{\sum_{\vect K\text{'s}}}'(k_1^2/2-E_m/2)\lvert\beta_{\vect K_1\vect K_2\vect K_3\vect K_4}\rvert^2,\\
T_{10}&=\frac{1}{8}{\sum_{\vect K\text{'s}}}'U_{\vect K_1\vect K_2\vect K_5\vect K_6}\beta^*_{\vect K_1\vect K_2
\vect K_3\vect K_4}\beta_{\vect K_5\vect K_6\vect K_3\vect K_4},
\end{align}
}
\end{subequations}
where $\sum'$ excludes cases in which the sum of any two subscripts of $\beta$ (or $\beta^*$) is less than $k_c$,
and in $T_3$ (and $T_9$) the term containing $\deltamu$ is omitted since it is $\sim n^3$.

The parameters $\alpha$ and $\beta$ are then  adjusted to minimize $E-\mu N$. We first write the terms containing
$\beta^*$ (\textit{ie} $T_8+T_9+T_{10}$) in the form $\sum_{\vect K\text{'s}}'\beta^*_{\vect K_1\vect K_2\vect K_3\vect K_4}
\theta_{\vect K_1\vect K_2\vect K_3\vect K_4}$; only the \emph{completely antisymmetric} component of $\theta$ contributes
to the sum, and this component must vanish since it is independent of $\beta^*$ and $\partial(E-\mu N)/\partial\beta^*=0$. We get
{\allowdisplaybreaks
\begin{multline}\label{eq:m_20_beta1}
\bigl[(k_1^2+k_2^2+k_3^2+k_4^2)/2-2E_m\bigr]\beta_{1234}\\
+\frac{1}{2}{\sum_{56}}'U_{1256}\beta_{5634}
+\frac{1}{2}{\sum_{56}}'U_{1356}\beta_{5264}
+\frac{1}{2}{\sum_{56}}'U_{1456}\beta_{5236}\\
+\frac{1}{2}{\sum_{56}}'U_{2356}\beta_{1564}
+\frac{1}{2}{\sum_{56}}'U_{2456}\beta_{1536}
+\frac{1}{2}{\sum_{56}}'U_{3456}\beta_{1256}\\
-U_{12\bar{3}\bar{4}}\alpha_3\alpha_4
+U_{13\bar{2}\bar{4}}\alpha_2\alpha_4
-U_{14\bar{2}\bar{3}}\alpha_2\alpha_3~~~~~~~\\
-U_{23\bar{1}\bar{4}}\alpha_1\alpha_4
+U_{24\bar{1}\bar{3}}\alpha_1\alpha_3
-U_{34\bar{1}\bar{2}}\alpha_1\alpha_2=0,
\end{multline}
}
The excluded regions in the spin-momentum space 
for the summation $\sum_{56}'$ contribute terms of the low-density order $n^{1.5}$,
which are smaller than the dominant terms by a factor of the order $n^{0.5}$; $\alpha_{\vect K}$'s lowest order
expression is given by Eq.~\eqref{eq:m_10_alpha}. So the comparison between Eq.~\eqref{eq:m_20_beta1} and
Eq.~\eqref{eq:m_phi'_Schrodinger} yields
\begin{equation}\label{eq:m_20_beta3}
\beta_{\vect K_1\vect K_2\vect K_3\vect K_4}=N_\up\phi'_{\vect K_1\vect K_2\vect K_3\vect K_4}+O(n^{1.5}\Omega^{-1})
\end{equation}
if $\min_{\sigma_i+\sigma_j=0,1\leq i<j\leq4}\lvert\vect K_i+\vect K_j\rvert>k_c$.

Now turn to $\partial(E-\mu N)/\partial\alpha^*_{\vect K}=0$.
Write $\delta(E-\mu N)=\sum_{\vect K}\theta'_{\vect K}\delta\alpha^*_{\vect K}
+\cdots$. Only the \emph{antisymmetric} component of $\theta'_{\vect K}$
contributes to the sum, since $\alpha_{-\vect K}=-\alpha_{\vect K}$. This component must vanish, and we get
(after a little algebra)
\begin{align}\label{eq:m_20_alpha1}
&\sum_2\widetilde{D}_{12}\alpha_{2}=\alpha_1^2\sum_2D_{21}\alpha_2^*+\sum_2D_{12}\lvert\alpha_2\rvert^2\alpha_2\notag\\
&\quad+\lvert\alpha_1\rvert^2\sum_2U_{1\bar{1}2\bar{2}}\alpha_2-\alpha_1\sum_2(U_{1212}+U_{\bar{1}2\bar{1}2})\lvert\alpha_2\rvert^2\notag\\
&\quad+\frac{1}{2}{\sum_{234}}'U_{\bar{1}\bar{2}34}\alpha_2^*\beta_{1234}+\frac{1}{2}{\sum_{234}}'U_{1234}\alpha_2^*\beta_{\bar{1}\bar{2}34}
\end{align}
plus correction terms $\sim n^2$, where $\widetilde{D}_{12}\equiv D_{12}-2\deltamu\delta_{12}$ and $\deltamu\equiv\mu-E_m/2$.
The terms on the \emph{right} side of Eq.~\eqref{eq:m_20_alpha1} are $\sim n^{1.5}$, and in this subsection
they can be simplified with the lowest order formulas [Eqs.~\eqref{eq:m_10_alpha} and \eqref{eq:m_20_beta3}]; we then solve the equation
to obtain more accurate results of $\alpha_\vect K$ and $\mu$.

The first term on the right side of Eq.~\eqref{eq:m_20_alpha1} vanishes at the order $n^{1.5}$,
and the remaining terms are further simplified with Eq.~\eqref{eq:m_phi4_id1_expanded}
(the excluded regions in the spin-momentum space for $\sum'_{234}$ can contribute terms
of the low-density order $n^2$ and are unimportant here):
\begin{equation}\label{eq:m_20_alpha2}
\sum_2\widetilde{D}_{12}\alpha_{2}=
\frac{N_\up^{3/2}}{\Omega}\Bigl(-2\pi a_m\phi_1+\sum_{2}D_{12}\devp_2\Bigr)+O(n^2).
\end{equation}
We decompose $\alpha_\vect K$ into a component \emph{parallel} to $\phi_\vect K$ and
one \emph{orthogonal} to $\phi_\vect K$:
\begin{equation}\label{eq:m_alpha_decompose}
\alpha_\vect K\equiv\eta_{\alpha}\phi_\vect K+\alpha^\perp_\vect K,~~~~~~\sum_\vect K\phi^*_\vect K\alpha^\perp_\vect K\equiv0,
\end{equation}
where $\eta_\alpha$ is elected as positive, since there is a gauge symmetry: when $\alpha^{(p)}$ in Eq.~\eqref{eq:m_psi} is
changed by a factor $\E^{\I p\theta}$ ($\theta$ is real and independent of $p$), the particle number and the energy expectation values are invariant.
Substituting Eq.~\eqref{eq:m_alpha_decompose}
into Eq.~\eqref{eq:m_20_alpha2}, and noting that $\eta_\alpha=\sqrt{N_\up}+h.o.c.$ (Sec.~\ref{subsec:m_10}), we get
\begin{gather}
-2\deltamu N_\up^{1/2}=-2\pi a_mN_\up^{3/2}/\Omega+O(n^2\Omega^{1/2}),\notag\\
\sum_2D_{12}(\alpha^\perp_2-N_\up^{3/2}\devp_2/\Omega)=2\deltamu\alpha^\perp_1+O(n^2).\label{eq:m_20_alpha3}
\end{gather}
So
\begin{equation}\label{eq:m_20_mu}
\mu=E_m/2+\pi a_m n_\up+O(n^{1.5}),
\end{equation}
and the first term on the right side of Eq.~\eqref{eq:m_20_alpha3} is $\sim n^{2.5}$ and negligible;
noting also that the spectrum of $D_{12}$ (when restricted to the subspace \emph{orthogonal} to $\phi_{\vect K}$)
has a lower bound $\lvert E_m\rvert$, we get
\begin{equation}\label{eq:m_20_alphaperp}
\alpha^\perp_\vect K=n_\up^{3/2}\Omega^{1/2}\devp_\vect K+O(n^2).
\end{equation}

Solving $\dif E/\dif N=\mu$, we get
\begin{equation}\label{eq:m_20_E}
E/\Omega=E_mn_\up+\pi a_mn_\up^2+O(n^{2.5}).
\end{equation}
As a consistency check, we may use the above results for $\alpha$ and $\beta$
to calculate $T_1+\cdots+T_{10}$, and verify that it is equal
to $E-\mu N=-\pi a_m n_\up^2\Omega+h.o.c.$.

Firstly, $T_1+T_2=-\deltamu\sum_\vect K\lvert\alpha_\vect K\rvert^2+\frac{1}{2}\sum_{12}\alpha_1^*D_{12}\alpha_2$,
but $\sum_{12}\alpha_1^*D_{12}\alpha_2=\sum_{12}\alpha^{\perp*}_1D_{12}\alpha^\perp_2\sim
n^{3}$, so $T_1+T_2=-2\pi a_m n_\up^2\Omega+h.o.c.$. 

Secondly, $T_3+\cdots+T_7=N_\up^2[t^{(1)}+t^{(2)}]+O(n^{2.5}\Omega)$,
where $t^{(1)}=-(1/2)\sum_{12}\phi_1^*D_{12}\lvert\phi_2\rvert^2\phi_2=0$, and $t^{(2)}=-(1/4)\sum_{12}U_{1\bar{1}2\bar{2}}\lvert\phi_1\rvert^2
\phi_1^*\phi_2+(1/2)\sum_{12}U_{1212}\lvert\phi_1\rvert^2\lvert\phi_2\rvert^2-(1/4)\sum\limits_{1234}U_{\bar{1}\bar{2}34}\phi^*_1\phi^*_2
\phi'_{1234}=+\pi a_m/\Omega$
[see Eq.~\eqref{eq:m_phi4_id0_expanded}], so $T_3+\cdots+T_7=+\pi a_mn_\up^2\Omega+h.o.c.$. 

Thirdly, $T_8+T_9+T_{10}
=(N_\up^2/24)\sum_{1234}{\phi'}^*_{1234}[-6U_{12\bar{3}\bar{4}}\phi_3\phi_4+2(k_1^2-E_m)\phi'_{1234}
+3\sum_{56}U_{1256}\phi'_{5634}]+O(n^{2.5}\Omega)$, but the antisymmetric component of the expression in the brackets is just
the left side of Eq.~\eqref{eq:m_phi'_Schrodinger}, so $T_8+T_9+T_{10}=O(n^{2.5}\Omega)$.

So $T_1+\cdots+T_{10}=-\pi a_mn_\up^2\Omega+h.o.c.$.

Equation~\eqref{eq:m_20_mu} is indeed the well-known equation of state at the mean-field level:
\begin{equation*}
\mu_m=E_m+4\pi a_mn_m/m_m+h.o.c.,
\end{equation*}
where $\mu_m=2\mu$ is the molecular chemical potential, $n_m= n_\up$ is the molecular density,
and $m_m=2m=2$ is the molecular mass. We have thus successfully constructed a many-body theory
which is completely compatible with the exact, nonperturbative solution to the quantum four-fermion problem,
with fairly arbitrary finite-range interparticle interaction (\emph{not} restricted to a contact interaction
for which $l\rightarrow0$).

Although common wisdom has had some success on the equation of state, a working many-body
theory is needed for the understanding of \emph{many other important features} of the system. Other predictions
of this theory will be presented in Sec.~\ref{sec:m_predictions}. Here we just point out that
Eq.~\eqref{eq:m_20_alphaperp} is directly related to the \emph{deviation} of $\langle c_{-\vect k\down}
c_{\vect k\up}\rangle$ from the internal wave function of an isolated molecule. For a two-component Fermi
gas near a Feshbach resonance, such deviation is most dramatic in the unitarity limit, in which the width of
the internal wave function of an isolated molecule in momentum space $1/r_m\rightarrow0$, but $\langle c_{-\vect k\down}
c_{\vect k\up}\rangle$ remains a finite width $\sim k_F\sim n^{1/3}$. The present theory
helps to bridge the two regimes, and describes \emph{accurately} how the physical picture of the system begins to change,
as we start from the BEC limit, going toward the unitary regime.

In the next subsection, we go beyond mean-field.

\subsection{\label{subsec:m_25}$E-\mu N$ to the order $n^{2.5}$}

$E-\mu N$ is equal to $T_1+\cdots+T_{10}$ plus the following $T$'s, up to the order $n^{2.5}$.
\begin{subequations}\label{eq:m_25}
{\allowdisplaybreaks
\begin{align}
T_{1a}+T_{2a}&=\frac{1}{2}\sum_{12,\vect q}\bigl[D_{12}+(q^2/4-2\deltamu)\delta_{12}\bigr]\Cq_{21},\\
T_{3a}&=-2\sum_{2,\vect q}(k_2^2-E_m)\lvert\alpha_2\rvert^2\Cq_{22},\\
T_{3b}+T_{3c}&=-\frac{1}{2}\biggl[\sum_{2,\vect q}(k_2^2-E_m)\alpha_2^{*2}\widetilde{A}^\vect q_{22}+c.c.\biggr],\\
T_{4a}+T_{5a}&=-\frac{1}{2}\biggl(\sum_{12,\vect q}U_{1\bar{1}2\bar{2}}\alpha_1^*\alpha_2\Cq_{22}+c.c.\biggr),\\
T_{4b}+T_{5b}&=-\frac{1}{2}\biggl(\sum_{12,\vect q}U_{1\bar{1}2\bar{2}}\lvert\alpha_2\rvert^2C^\vect q_{21}+c.c.\biggr),\\
T_{4c}+T_{5c}&=-\frac{1}{4}\biggl(\sum_{12,\vect q}U_{1\bar{1}2\bar{2}}\alpha_2^2A^{*\vect q}_{12}+c.c.\biggr),\\
T_{4d}+T_{5d}&=-\frac{1}{4}\biggl(\sum_{12,\vect q}U_{1\bar{1}2\bar{2}}\alpha_1^*\alpha_2^*\widetilde{A}^\vect q_{22}+c.c.\biggr),\\
T_{6a}&=\sum_{12,\vect q}U_{1212}\lvert\alpha_1\rvert^2\Cq_{22},\\
T_{6b}&=\sum_{12,\vect q}U_{1212}\alpha_1^*\alpha_2\Cq_{12},\\
T_{6c}+T_{6d}&=\frac{1}{2}\sum_{12,\vect q}U_{1212}\alpha_1^*\alpha_2^*A^\vect q_{12}+c.c.,\\
T_{7a}+T_{8a}&=-\frac{1}{4}\biggl({\sum_{1234,\vect q}\mspace{-8mu}}'U_{\bar{1}\bar{2}34}A^{*\vect q}_{12}\beta_{1234}+c.c.\biggr),\\
T_{7b}&=-\frac{1}{4}{\sum_{12345,\vect q}\mspace{-10mu}}'U_{\bar{1}\bar{2}34}\alpha_2^*
\gamma^\vect q_{1234;5}A^{*\vect q}_{15},\\
T_{8b}&=-\frac{1}{4}{\sum_{12345,\vect q}\mspace{-10mu}}'U_{12\bar{3}\bar{4}}\alpha_3
\gamma^{*\vect q}_{1234;5}A^\vect q_{45},
\end{align}
\begin{align}
T_{9a}&=\frac{1}{12}\,{\sum_{123478,\vect q}\mspace{-12mu}}'(k_1^2/2-E_m/2)\gamma^{*\vect q}_{1234;7}\gamma^\vect q_{1234;8}\notag\\
&\mspace{100mu}\times\bigl[(1-\beta^\dagger_\vect q\beta_\vect q/4)^{-1}\bigr]_{87},\\
T_{10a}&=\frac{1}{16}{\sum_{12345678,\vect q}\mspace{-20mu}}'U_{1256}\gamma^{*\vect q}_{1234;7}\gamma^\vect q_{5634;8}
\bigl[(1-\beta^\dagger_\vect q\beta_\vect q/4)^{-1}\bigr]_{87},
\end{align}
}
\end{subequations}
where the arabic-number \emph{sub}scripts (except those of $T$) are spin-momenta,
the summation over $\vect q$ ($\neq0$) is restricted to $q<k_c$,
corrections of order $n^3$ or higher are omitted, $c.c.\equiv\text{complex conjugate}$,
\begin{equation}
\gamma^\vect q_{\vect K_1\vect K_2\vect K_3\vect K_4;\vect K_5}\equiv\gamma^{}_{\vect K_1+\vect q/4,\cdots,\vect K_4+\vect q/4,
\vect K_5-\vect q/2,-\vect K_5-\vect q/2},
\end{equation}
$\sum'$ excludes cases in which the sum of any two spin-momenta with opposite spins in the \emph{same} 4-fermion
$l$-cluster might be smaller than $k_c$,
and the matrices
\begin{subequations}
\begin{align}
C_\vect q&\equiv\bigl(1-\beta_\vect q\beta_\vect q^\dagger/4\bigr)^{-1}\beta_\vect q\beta_\vect q^\dagger/2,\\
A_\vect q&\equiv\bigl(1-\beta_\vect q\beta_\vect q^\dagger/4\bigr)^{-1}\beta_\vect q=\beta_\vect q
(1-\beta^\dagger_\vect q\beta_\vect q/4\bigr)^{-1},\label{eq:m_Aq}\\
\widetilde{A}_\vect q&\equiv\bigl(1-\beta_\vect q\beta_\vect q^\dagger/4\bigr)^{-1}\beta_\vect q\beta_\vect q^\dagger\beta_\vect q/4
=A_\vect q-\beta_\vect q.
\end{align}
For any matrix $M_\vect q$, $(M_\vect q)_{\vect K_1\vect K_2}\equiv M^\vect q_{\vect K_1\vect K_2}$.
Equation~\eqref{eq:m_beta_matrix} defines $\beta_\vect q$.
\end{subequations}

Like in Ref.~\cite{_Tan_BEC}, we first determine $\gamma^\vect q_{1234;5}$, then $\beta^\vect q_{12}$,
and then $\beta_{1234}$, and finally determine $\alpha_1$.

We first express $\partial(E-\mu N)/\partial\gamma^{*\vect q}_{1234;7}=0$ to the leading order, using the
lowest order formulas $\alpha_\vect K=\sqrt{N_\up}\phi_\vect K+h.o.c$,
and $$\beta^\vect q_{12}=x_\vect q\phi_1\phi_2+h.o.c.,~~~x_\vect q=x_{-\vect q}$$ in accordance with the cluster-separation theorem,
where $x_\vect q$ is so-far an unknown number; the second equality follows from the symmetry $\beta^{\vect q}_{12}=\beta^{-\vect q}_{21}$
[Eq.~\eqref{eq:m_beta_matrix}]. So $A^\vect q_{12}=\sum_3\beta^\vect q_{13}[(1-r_\vect q)^{-1}]_{32}
=x_\vect q\phi_1\sum_3\phi_3[(1-r_\vect q)^{-1}]_{32}+h.o.c.$ [$r_\vect q$ is defined by Eq.~\eqref{eq:m_rq}], and
\begin{multline*}
\!\!\!\sum_8\Bigl\{-(x_\vect q\sqrt{N_\up}/4)U_{12\bar{3}\bar{4}}\phi_3\phi_4\phi_8+(1/24)(k_1^2-E_m)\gamma^\vect q_{1234;8}\\
+(1/16){\sum_{56}}'U_{1256}\gamma^\vect q_{5634;8}\Bigr\}_{1234}[(1-r_\vect q)^{-1}]_{87}=0,
\end{multline*}
where $\{\cdots\}_{1234}$ means antisymmetrization with respect to the subscripts $1234$.
Since $(1-r_\vect q)^{-1}$ is invertible,
\begin{multline}
\Bigl\{-6(2x_\vect q\sqrt{N_\up}\phi_7)U_{12\bar{3}\bar{4}}\phi_3\phi_4+2(k_1^2-E_m)\gamma^\vect q_{1234;7}\\
+3{\sum_{56}}'U_{1256}\gamma^\vect q_{5634;7}\Bigr\}_{1234}=0.
\end{multline}
Comparing this equation with Eq.~\eqref{eq:m_phi'_Schrodinger}, we find
\begin{equation}\label{eq:m_25_gamma1}
\gamma^\vect q_{1234;5}=2\sqrt{N_\up}x_\vect q\phi_5\phi'_{1234}+h.o.c..
\end{equation}

Next we determine $\beta^\vect q_{12}$ ($0<q<k_c$). We compute $\partial(E-\mu N)/\partial\beta^{*\vect q}_{12}$ up to the
order $n^1$, and then determine what value $\beta^\vect q_{12}$ should take, in order for this
partial derivative to vanish. For this purpose $\alpha_\vect K$ (in $T_{3a},\cdots,T_{10a}$),
$\gamma^\vect q_{1234;5}$ (in $T_{7b}\cdots T_{10a}$), and $\beta_{1234}$ (in $T_{7a}+T_{8a}$)
can be retained to the lowest order. \emph{After} taking the partial derivative,
we can retain $\beta^\vect q_{12}$ and $\beta^{*\vect q}_{12}$ in 
$\partial(T_{3a}+\cdots+T_{10a})/\partial\beta^{*\vect q}_{12}$ to the lowest order, since
this partial derivative is $\sim n^1$.  Note, however, that $\partial(T_{1a}+T_{2a})/\partial\beta^{*\vect q}_{12}$
is $\sim n^0$ and should be treated exactly at this point. Note also that $T_8,T_9,T_{10}$ do \emph{not}
contribute to $\partial(E-\mu N)/\partial\beta^{*\vect q}_{12}$, since in $T_8,T_9,T_{10}$, the
parameters $\beta_{1234}^*$ refer to a different region of the spin-momentum configuration space than $\beta^\vect q_{12}$ ($q<k_c$);
similarly the factor $\beta_{1234}^*$  in $T_{8a}$ is independent of the partial derivative.

$\partial(T_{3a}/2+T_{4a})/\partial\beta^{*\vect q}_{12}=0$ up to the order $n^1$, since $\sum_1\phi_1^*D_{12}=0$;
similarly $\partial(T_{3a}/2+T_{5a})/\partial\beta^{*\vect q}_{12}=0$, and $\partial[(T_{3b}+T_{3c})+(T_{4d}+T_{5d})]
/\partial\beta^{*\vect q}_{12}=0$. Using the above result of $\gamma$, writing
$A_{\vect q}=\beta_\vect q(1-r_\vect q)^{-1}$ and approximating the left factor $\beta_\vect q$ with its lowest order expression,
we easily see that $\partial(T_{8b}+T_{9a}+T_{10a})/\partial\beta^{*\vect q}_{12}=0$ up to the order $n^1$.
For the remaining terms, we use the identities
\begin{align*}
\partial A_\vect q^\dagger/\partial\beta^{*\vect q'}_{89}&\equiv X_\vect q~~
\Leftrightarrow~~\partial A^{*\vect q}_{56}/\partial\beta^{*\vect q'}_{89}=X^\vect q_{65},\\
\partial C_\vect q/\partial\beta^{*\vect q'}_{89}&=(1/2)\beta_\vect qX_\vect q,\\
\partial A_\vect q/\partial\beta^{*\vect q'}_{89}&=(1/4)\beta_\vect qX_\vect q\beta_\vect q,
\end{align*}
where the dependence of $X_\vect q$ on $\vect K_8\vect K_9\vect q'$ is suppressed for brevity. We get
\begin{gather*}
\partial(E-\mu N)/\partial\beta^{*\vect q'}_{89}=\sum_{56,\vect q}S^\vect q_{56}X^\vect q_{65}+O(n^{1.5}\Omega^{-1}),\\
S^\vect q_{56}\!\equiv\!\!\frac{1}{4}\sum_1\widetilde{D}^\vect q_{51}\beta^\vect q_{16}+N_\up \bigl[S^{(2)}_{56}+ x_\vect qS^{(1)}_5\phi_6
+ x_\vect q^2S^{(0)}\phi_5\phi_6\bigr],\\
\widetilde{D}^\vect q_{12}\equiv(k_1^2-E_m+q^2/4-2\deltamu)\delta_{12}+(1/2)U_{1\bar{1}2\bar{2}},\\
S^{(2)}_{56}\equiv-\frac{1}{4}\sum_{12}U_{\bar{5}\bar{6}12}\phi'_{5612}+\frac{1}{2}U_{5656}\phi_5\phi_6-\frac{1}{4}U_{5\bar{5}6\bar{6}}\phi_6^2,\\
\begin{split}S^{(1)}_{5}&\equiv-\frac{1}{4}\sum_1U_{5\bar{5}1\bar{1}}\lvert\phi_1\rvert^2\phi_1
  -\frac{1}{4}\lvert\phi_5\rvert^2\sum_1U_{5\bar{5}1\bar{1}}\phi_1\\
  &\quad+\phi_5\sum_1U_{1515}\norms{\phi_1}-\frac{1}{2}\sum_{234}U_{\bar{5}\bar{2}34}\phi_2^*\phi'_{5234},
\end{split}\\
\begin{split}
  S^{(0)}&\equiv-\frac{1}{16}\sum_{12}U_{1\bar{1}2\bar{2}}\norms{\phi_1}\phi_1^*\phi_2+\frac{1}{8}\sum_{12}U_{1212}\norms{\phi_1\phi_2}\\
  &\quad-\frac{1}{16}\sum_{1234}U_{34\bar{1}\bar{2}}\phi_1\phi_2{\phi'}^*_{1234}=\frac{\pi a_m}{4\Omega},
\end{split}
\end{gather*}
where $S^{(0)}$ is simplified with Eq.~\eqref{eq:m_phi4_id0_expanded} in the end; the sum of any two spin-momenta
in the function $\phi'$ is restricted to nonzero.
From the definition of $A_\vect q$, we can easily show that $X^{\vect q}_{65}\equiv-X^{\vect q}_{\bar{6}5}\equiv-X^{\vect q}_{6\bar{5}}
\equiv X^{-\vect q}_{56}$; in $S^\vect q_{56}$, only the component with this same symmetry contributes to the partial derivative,
and this component must vanish, since
\begin{equation*}
X_\vect q=(1-\beta_\vect q^\dagger\beta_\vect q/4)^{-1}(\partial \beta_\vect q^\dagger/\partial\beta^{*\vect q'}_{89})
(1-\beta_\vect q\beta_\vect q^\dagger/4)^{-1},
\end{equation*}
where the matrices $(1-\beta_\vect q^\dagger\beta_\vect q/4)^{-1}$ and  $(1-\beta_\vect q\beta_\vect q^\dagger/4)^{-1}$
are \emph{invertible}. The resultant equation is then simplified with the identities in Sec.~\ref{sec:m_fewbody}, in particular
Eqs.~\eqref{eq:m_phi4_id1_expanded}, \eqref{eq:m_phi_id}, and \eqref{eq:m_phi4_id2_expanded}. The result is
\begin{align}\label{eq:m_25_betaq1}
&\sum_1\widetilde{D}^\vect q_{51}\beta^\vect q_{16}+\sum_1\widetilde{D}^\vect q_{61}\beta^\vect q_{51}
+2\pi a_mn_\up\bigl(1+4 x_\vect q+x_\vect q^2\bigr)\phi_5\phi_6\notag\\
&-n_\up(1+2x_\vect q)\bigl[(Dd')_5\phi_6+\phi_5(Dd')_6\bigr]\notag\\
&-n_\up\sum_1D_{51}^{}g'_{16}-n_\up\sum_1D_{61}^{}g'_{51}=0,
\end{align}
with an absolute error of the low-density order $n^{1.5}$. Here $(Dd')_5\equiv\sum_1D_{51}d'_1$.
We then do an orthogonal decomposition:
\begin{subequations}
\begin{gather}
\beta^\vect q_{12}\equiv x_\vect q\phi_1\phi_2+\beta^{(1)\vect q}_1\phi_2+\phi_1\beta^{(1)\overline{\vect q}}_2
+\beta^{(2)\vect q}_{12},\label{eq:m_betaq_decompose}\\
\sum_1\phi_1^*\beta^{(1)\vect q}_1\equiv\sum_1\phi_1^*\beta^{(2)\vect q}_{12}\equiv\sum_2\phi_2^*\beta^{(2)\vect q}_{12}\equiv0,
\end{gather}
\end{subequations}
where $\overline{\vect q}\equiv-\vect q$. The third term on the right side of Eq.~\eqref{eq:m_betaq_decompose}
is related to the second by the symmetry $\beta^\vect q_{12}=\beta^{\overline{\vect q}}_{21}$. Substituting
Eq.~\eqref{eq:m_betaq_decompose} into Eq.~\eqref{eq:m_25_betaq1}, we get four equations, for the four
mutually orthogonal subspaces. The linear kernal in Eq.~\eqref{eq:m_25_betaq1} (the sum of the two $\widetilde{D}^{\vect q}$'s)
has a single eigenvalue of the order $n^1$ (ie $q^2/2-4\deltamu$), associated with the eigenvector
 $\phi_1\phi_2$, and all the other eigenvalues are at least
about $\lvert E_m\rvert$. In the subspaces orthogonal to $\phi_1\phi_2$, the difference between $\widetilde{D}^\vect q$ and $D$ is
negligible. In the subspace containing $\phi_1\phi_2$, we may approximate $\deltamu$ with its lowest order value,
$\pi a_mn_\up$ [Eq.~\eqref{eq:m_20_mu}], since at the current order, $x_\vect q$ can only be determined to the leading order. We thus get
\begin{subequations}
\begin{align}
(q^2/2+4\pi a_mn_\up)&x_\vect q+2\pi a_mn_\up(1+x_\vect q^2)=0,\\
\beta^{(1)\vect q}_{\vect K}&=n_\up(1+2x_\vect q)d'_\vect K+h.o.c.,\\
\beta^{(2)\vect q}_{\vect K_1\vect K_2}&=n_\up g'_{\vect K_1\vect K_2}+h.o.c..
\end{align}
\end{subequations}
The solution to the first equation is
\begin{equation}\label{eq:m_xq}
x_{\vect q}=-\Bigl(1+\xi^2q^2-\xi q\sqrt{2+\xi^2q^2}\Bigr),~~\xi\equiv(8\pi a_mn_\up)^{-1/2}.
\end{equation}
The solution with norm greater than $1$ is unphysical and has been
discarded, like in the case of structureless bosons (cf \cite{LeeYang1957, _Tan_BEC}). The other components of $\beta^\vect q_{12}$ remind
us that \emph{additional features} are present, in contrast with a gas of structureless bosons.

Now we turn to $\beta_{1234}$ [excluding $\beta^\vect q_{12}$ ($q<k_c$)]. Its leading order expression is
determined in Sec.~\ref{subsec:m_20}, and we now proceed to the next order ($\sim n^{1.5}$). Only one additional term contributes to
$\partial(E-\mu N)/\partial\beta^*_{1234}=0$, namely $T_{8a}$.
\begin{align}
&\biggl\{2(k_1^2-E_m)\beta_{1234}+3{\sum_{56}}'U_{1256}\beta_{5634}-6U_{12\bar{3}\bar{4}}\alpha_3\alpha_4\notag\\
&\quad-6U_{12\bar{3}\bar{4}}\sum_{0<q<k_c}A^\vect q_{34}\biggr\}_{1234}=0,
\end{align}
where $\sum'$ excludes the region occupied by $\beta_\vect q$ ($q<k_c$), so the last term is in some sense complementary
to the second. Since our accuracy goal here is $n^{1.5}$, $A^\vect q_{12}$ can be retained to the lowest order,
$\phi_1\phi_2x_\vect q/\bigl(1-x_\vect q^2\bigr)$, and $\alpha_\vect K$ should be retained to the next-to-leading order,
$\alpha_\vect K=\eta_\alpha\phi_\vect K$ (the component orthogonal to $\phi_\vect K$ is $\sim n^{1.5}$ and
is negligible here, since it is multiplied by the other $\alpha_{\vect K'}\sim n^{0.5}$).

The lowest order formula $\beta_{1234}=N_\up\phi'_{1234}$ combined with Eq.~\eqref{eq:m_phi4_smallq}
(the sum of any two spin-momenta is nonzero here) leads to $\beta_{-\vect K_3+\vect q,-\vect K_4-\vect q,\vect K_3,\vect K_4}\approx
-(4\pi a_mn_\up/q^2)\phi_3\phi_4$ and $\beta_{-\vect K_4+\vect q,-\vect K_3-\vect q,\vect K_3,\vect K_4}\approx
(4\pi a_mn_\up/q^2)\phi_3\phi_4$ for $\sqrt{a_mn_\up}\ll q\ll\min(1/l,1/r_m)$. If the term containing $A^\vect q_{34}$
followed this asymptotic formula for all $q<k_c$, it would not alter the solution $\beta_{1234}=\eta_\alpha^2\phi'_{1234}$
at the order $n^{1.5}$. It is the nonzero \emph{difference}, namely 
$$A^\vect q_{34}-\phi_3\phi_4(-4\pi a_mn_\up/q^2)=\phi_3\phi_4\biggl(\frac{x_q}{1-x_q^2}+\frac{1}{2\xi^2q^2}\biggr),$$ that constitutes
an additional driving term (in addition to the term containing $\alpha$'s). Since this term has the same structure
as the $\alpha$ term, it can be absorbed into the $\alpha$ term, thus yielding the solution
\begin{equation*}
\beta_{1234}=\biggl[\eta_\alpha^2+\sum_{0<q<k_c}\biggl(\frac{x_q}{1-x_q^2}+\frac{1}{2\xi^2q^2}\biggr)\biggr]\phi'_{1234},
\end{equation*}
accurate up to the low-density order $n^{1.5}$. Substituting Eq.~\eqref{eq:m_xq}, and noting that $k_c\gg\sqrt{a_mn_\up}$,
we get
\begin{equation}\label{eq:m_25_beta1}
\beta_{1234}=\Bigl[\eta_\alpha^2+8\Omega(a_mn_\up)^{3/2}/\sqrt{\pi}\Bigr]\phi'_{1234}.
\end{equation}

Now turn to $\partial(E-\mu N)/\partial\alpha^*_\vect K=0$. This equation is determined to the order $n^{1.5}$
in Eq.~\eqref{eq:m_20_alpha1}. Here we proceed to the order $n^2$ by including contributions from
Eq.~\eqref{eq:m_25}. 
\begin{gather*}
\sum_{2}\widetilde{D}_{12}\alpha_{2}=f^{(1.5)}_{1}+f^{(2)}_{1}+O(n^{2.5}),\\
\begin{split}
&f^{(2)}_1\equiv\sum_{0<q<k_c}\biggl\{4(k_1^2-E_m)\alpha_1C^{\vect q}_{11}+2(k_1^2-E_m)\alpha^*_1\widetilde{A}^\vect q_{11}\\
&+\sum_2U_{1\bar{1}2\bar{2}}\bigl[\alpha_2(C^\vect q_{22}+C^\vect q_{11})+\alpha_1C^\vect q_{21}+\alpha_1^*A^\vect q_{21}
\!+\!\alpha_2^*\widetilde{A}^\vect q_{22}/2\bigr]\\
&+\sum_2U_{2\bar{2}1\bar{1}}\bigl(\alpha_1C^\vect q_{12}+\alpha^*_2\widetilde{A}^\vect q_{11}/2\bigr)
+{\sum_{2345}\mspace{-2mu}}'U_{\bar{1}\bar{2}34}\gamma^\vect q_{1234;5}A^{*\vect q}_{25}/2\\
&-\sum_2U_{1212}\bigl[2\alpha_1C^\vect q_{22}+2\alpha_2C^\vect q_{12}+\alpha_2^*(A^\vect q_{12}+A^\vect q_{21})\bigr]\biggr\}_{1\bar{1}}, 
\end{split}
\end{gather*}
where $\{\cdots\}_{1\bar{1}}$ retains only the component antisymmetric under $1\leftrightarrow\bar{1}$,
and $f^{(1.5)}_1$ are the terms on the right side of Eq.~\eqref{eq:m_20_alpha1}. Our accuracy goal for this equation is $n^2$.
Each term in $f^{(1.5)}_1$ is $\sim n^{1.5}$ and is computed to the next-to-leading order, at which $\alpha_\vect K=\eta_\alpha\phi_\vect K$
(the components orthogonal to $\phi_\vect K$ are negligible here) and $\beta_{1234}$ is given by Eq.~\eqref{eq:m_25_beta1}.
Each term in $f^{(2)}_1$ is $\sim n^2$ and only computed to the leading order, at which $\alpha_\vect K=\sqrt{N_\up}\phi_\vect K$,
$C^\vect q_{12}=\phi_1\phi_2^*x_q^2/(1-x_q^2)$, $A^\vect q_{12}=\phi_1\phi_2 x_q/(1-x_q^2)$, $\widetilde{A}^\vect q_{11}
=\phi_1^2x_q^3/(1-x_q^2)$, and $\gamma^\vect q_{1234;5}=2\sqrt{N_\up}x_q\phi_5\phi'_{1234}$ [Eq.~\eqref{eq:m_25_gamma1}].
Here $x_q$ is given by Eq.~\eqref{eq:m_xq}.
\begin{align}\label{eq:m_25_f1.5}
&f^{(1.5)}_1=\biggl\{\eta_\alpha^3\Bigl[-(k_1^2-E_m)\norms{\phi_1}\phi_1+\sum_2U_{1\bar{1}2\bar{2}}\norms{\phi_2}\phi_2/2\notag\\
&\mspace{10mu}-2\phi_1\sum_2U_{1212}\norms{\phi_2}\Bigr]
+(\eta_\alpha^3+3t){\sum_{234}\mspace{-1mu}}'U_{\bar{1}\bar{2}34}\phi_2^*\phi'_{1234}\biggr\}_{1\bar{1}},
\end{align}
accurate up to the low-density order $n^2$, where
\begin{equation*}
t\equiv \bigl(8/3\sqrt{\pi}\bigr)\bigl(n_\up a_m^3\bigr)^{1/2}N_\up^{3/2}\sim n^2\Omega^{3/2},
\end{equation*}
and $\sum'$ excludes three regions in the spin-momentum configuration space:
$\lvert\vect K_2+\vect K_1\rvert<k_c$, $\lvert\vect K_3+\vect K_1\rvert<k_c$, or $\lvert\vect K_4+\vect K_1\rvert<k_c$
(the intersection of these regions is smaller by a factor $\propto k_c^3$ and negligible), \textit{ie},
\begin{align*}
&{\sum_{234}\mspace{-2mu}}'U_{\bar{1}\bar{2}34}\phi_2^*\phi'_{1234}=\sum_{234}U_{\bar{1}\bar{2}34}\phi_2^*\phi'_{1234}\\
&-\sum_{5;\lvert\vect q\rvert<k_c}U_{-\vect K_1,\vect K_1-\vect q,\vect K_5-\vect q/2,-\vect K_5-\vect q/2}\\
&\qquad\times\phi_{-\vect K_1+\vect q}^*\phi'_{\vect K_1,-\vect K_1+\vect q,\vect K_5-\vect q/2,-\vect K_5-\vect q/2}\\
&-\text{(the other two small regions)}.
\end{align*}
Since $\phi'_{1234}\sim1/q^2$ at small $q$ (Sec.~\ref{sec:m_fewbody}), the contributions of these regions to $f^{(1.5)}_1$
scales like $n^{1.5}k_c$ and is $\sim n^2$, if we let $k_c$ be $\sqrt{na_m}$ times a fixed dimensionless coefficient ($\gg1$).
So we only need to compute them at the leading order; for example, the second term on the right side
of the above equation is simplified as $+\norms{\phi_1}\sum_5U_{1\bar{1}5\bar{5}}\phi_5\sum_{0<q<k_c}4\pi a_m/\Omega q^2$.

We can thus change $\sum'$ to $\sum$ in the last term of Eq.~\eqref{eq:m_25_f1.5}, and in the same time add the following
correction term ($\sim n^2$) to the right side of Eq.~\eqref{eq:m_25_f1.5}:
\begin{align}\label{eq:m_25_f1.5_correction}
\biggl\{\norms{\phi_1}\sum_2U_{1\bar{1}2\bar{2}}\phi_2-2\phi_1\sum_2U_{1212}\norms{\phi_2}\biggr\}_{1\bar{1}}
\sum_{q<k_c}\!\frac{4\pi a_mN_\up^{\frac{3}{2}}}{\Omega q^2}
\end{align}
which has been retained to the leading order. On the other hand, the terms containing $A_\vect q$ in $f^{(2)}_1$
are also dependent of $k_c$, since $\sqrt{N_\up}\sum_{q<k_c}A^\vect q_{12}
=\phi_1\phi_2\bigl(3t-\sqrt{N_\up}\sum_{q<k_c}4\pi a_m n_\up/q^2\bigr)$. Direct calculation immediately reveals that
the sum of the terms $\propto\sum_{q<k_c}1/q^2$ in $f^{(2)}$ is opposite to Eq.~\eqref{eq:m_25_f1.5_correction}.
$f^{(1.5)}_1+f^{(2)}$ is thus \emph{independent} of $k_c$ at the order $n^2$ (the $\sum'$ in $f^{(2)}_1$
can be simplified as $\sum$ since their difference is of the order $n^{2.5}$). This independence is \emph{not} a coincidence,
since different terms in this perturbative expansion refer to complementary parts (in spin-momentum configuration space)
of an underlying \emph{seamless whole}. The division between $q<k_c$ and $q>k_c$ is just for the convenience of concrete calculations,
and the physical outcome is of course independent of $k_c$.

After straightforward calculation, we get
\begin{align*}
f^{(1.5)}_1\!+\!f^{(2)}_1\!=(\eta_\alpha^3+5t)s-\frac{16t}{5}\sum_2D_{12}\norms{\phi_2}\phi_2+O(n^{2.5}),
\end{align*}
where $s$ is exactly equal to the left side of Eq.~\eqref{eq:m_phi4_id1_expanded}. Thus
\begin{align*}
&-2\deltamu\eta_\alpha\phi_1+\sum_2\widetilde{D}_{12}\alpha^\perp_2=
-\phi_1(\eta_\alpha^3+5t)2\pi a_m/\Omega\\
&+\sum_2D_{12}\bigl[(\eta_\alpha^3+5t)\devp_2/\Omega-(16t/5)\norms{\phi_2}\phi_2\bigr]+O(n^{2.5}).
\end{align*}
The difference between $\sum_2\widetilde{D}_{12}\alpha^\perp_2$ and $\sum_2D_{12}\alpha^\perp_2$ is $\sim n^{2.5}$
and negligible. Noting $\eta_\alpha=\sqrt{N_\up}+h.o.c.$, we get
\begin{subequations}
\begin{gather}
\deltamu=\eta_\alpha^2\Bigl[1+(40/3\sqrt{\pi})\bigl(n_\up a_m^3\bigr)^{1/2}\Bigr]\pi a_m/\Omega+O(n^2),\label{eq:m_25_mu1}\\
\begin{split}
&\alpha^\perp_1=\eta_\alpha^3\Bigl[1+(40/3\sqrt{\pi})\bigl(n_\up a_m^3\bigr)^{1/2}\Bigr]\devp_1/\Omega\\
&-(128/15\sqrt{\pi})\bigl(n_\up a_m^3\bigr)^{1/2}N_\up^{3/2}\sum_2\proj_{12}\norms{\phi_2}\phi_2+O(n^{2.5}).
\end{split}
\end{gather}
\end{subequations}
where $O(n^2)$ is in the \emph{coarse-grained} sense; it does not exclude
the possibility of a correction term which may behave like $n^2[\ln(na_m^3)+\text{const.}]$.

Like in the case of structureless bosons \cite{_Tan_BEC}, we express the particle number $N$ in terms of the
wave function, in order to solve it backwards for $\eta_\alpha$. Up to the order $n^{1.5}$, there are only two diagrams
for $N=\langle\sum_1c_1^\dagger c_1\rangle$; they are the derivatives of $T_1$ and $T_{1a}$ with respect to $-\mu$.
\begin{align*}
&N\!=\!\sum_1\norms{\alpha_1}\!+\!\sum_{1,\vect q}C^\vect q_{11}=2\eta_\alpha^2+2\sum_q\frac{x_q^2}{1-x_q^2}+O(n^2\Omega)\\
&\mspace{14mu}=2\Bigl[\eta_\alpha^2+(8/3\sqrt{\pi})\bigl(n_\up a_m^3\bigr)^{1/2}N_\up\Bigr]+O(n^2\Omega).
\end{align*}
Solving the above equation for $\eta_\alpha$, substituting the result back to the formulas we obtained so far,
and solving $\dif E/\dif N=\mu$, we get our results, summarized below.
\begin{subequations}\label{eq:m_25_final}
{\allowdisplaybreaks
\begin{gather}
\mu=E_m/2+\pi a_m n_\up\Bigl[1+(32/3\sqrt{\pi})\bigl(n_\up a_m^3\bigr)^{1/2}+h.o.c.\Bigr],\label{eq:m_25_mu_final}\\
\begin{split}
E/\Omega=&E_mn_\up+\pi a_mn_\up^2\\
&\times\Bigl[1+(128/15\sqrt{\pi})\bigl(n_\up a_m^3\bigr)^{1/2}+h.o.c.\Bigr],\label{eq:m_25_E_final}
\end{split}\\
\alpha_1\equiv\eta_\alpha\phi_1+\alpha^\perp_1,~~~~~~\sum_1\phi_1^*\alpha^\perp_1\equiv0,\\
\eta_\alpha=N_\up^{1/2}\Bigl[1-(4/3\sqrt{\pi})\bigl(n_\up a_m^3\bigr)^{1/2}+h.o.c.\Bigr],\label{eq:m_25_etaalpha_final}\\
\begin{split}
&\alpha^\perp_1=N_\up^{3/2}\biggl\{\Bigl[1+(28/3\sqrt{\pi})\bigl(n_\up a_m^3\bigr)^{1/2}\Bigr]\devp_1/\Omega\\
&-(128/15\sqrt{\pi})\bigl(n_\up a_m^3\bigr)^{1/2}\sum_2\proj_{12}\norms{\phi_2}\phi_2\biggr\}+O(n^{2.5}),
\end{split}\\
\beta_{1234}=N_\up\Bigl[1+(16/3\sqrt{\pi})\bigl(n_\up a_m^3\bigr)^{1/2}\Bigr]\phi'_{1234}+O(n^2\Omega^{-1}),\label{eq:m_25_beta1234_final}\\
\beta^\vect q_{12}\equiv x_\vect q\phi_1\phi_2+\beta^{\perp\vect q}_{12},~~~\sum_{12}\phi_1^*\phi_2^*\beta^{\perp\vect q}_{12}\equiv0,\\
x_{\vect q}=-\Bigl(1+\xi^2q^2-\xi q\sqrt{2+\xi^2q^2}\Bigr)+h.o.c.,\label{eq:m_25_xq_final}\\
\beta^{\perp\vect q}_{12}=n_\up\bigl[(1+2x_\vect q)(\devp_1\phi_2+\phi_1\devp_2)+g'_{12}\bigr]+h.o.c.,\label{eq:m_25_betaqperp_final}\\
\gamma^{\vect q}_{1234;5}=2\sqrt{N_\up}x_\vect q\phi'_{1234}\phi_5+h.o.c.,\label{eq:m_25_gamma_final}
\end{gather}
}
\end{subequations}
where in Eqs.~\eqref{eq:m_25_beta1234_final} and \eqref{eq:m_25_gamma_final}, the first four spin-momenta $1234$ must be an l-cluster, thus
excluding the cases in which the sum of any two of them is $\sim\sqrt{na_m}$ or zero.
In Eqs.~\eqref{eq:m_25_xq_final}, \eqref{eq:m_25_betaqperp_final}, and \eqref{eq:m_25_gamma_final}, the momentum $q\ll\min(1/l,1/r_m)$.

As a consistency check, we calculate the sum of all the diagrams in Fig.~\ref{fig:33terms}. The calculation
is tedious but straightforward. The result is consistent with the value of $E-\mu N$ from Eqs.~\eqref{eq:m_25_mu_final}
and \eqref{eq:m_25_E_final}.


\section{\label{sec:m_predictions}Physical Observables other than the equation of state}
After a multistep process of logical deductions and analytical calculations,
we have now determined the parameters of the wave function [Eq.~\eqref{eq:m_psi}] to the leading order
\emph{beyond} mean-field, in the low-density regime. We are now in a position to determine various physical observables,
by evaluating their expectation values under the wave function.

For every observable computed below, we first use Eq.~\eqref{eq:R} and the other information in Sec.~\ref{subsec:m_power_counting}
to identify all the diagrams up to a certain order in the low-density expansion;
we also note that there are no dead ends.

\subsection{\label{subsec:m_pairing}Superfluid pairing function}
In this subsection we determine the superfluid pairing function,
and discuss the range of densities in which the result is valid.

We decompose it into a component parallel to
the internal wave function of an isolated molecule, and one orthogonal to it:
\begin{equation*}
F_\vect K\equiv\langle c_{-\vect K} c_{\vect K}\rangle\equiv F^\parallel_\vect K+F^\perp_\vect K
\equiv\eta_F\phi_\vect K+F^\perp_\vect K,
\end{equation*}
where $\sum_\vect K\phi^*_\vect K F^\perp_\vect K\equiv0$.

\begin{figure}\includegraphics{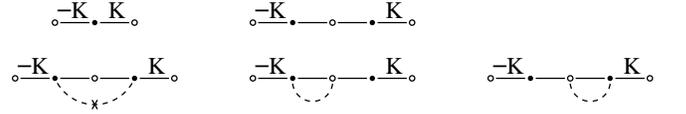}\caption{\label{fig:m_F2}Diagrams for the superfluid pairing function
$F_\vect K^{}\equiv\langle c_{-\vect K}^{} c_{\vect K}^{}\rangle$ up to the order $n^2$.}\end{figure}

All the diagrams up to the order $n^2$ are shown in Fig.~\ref{fig:m_F2}. Retaining terms to the order $n^2$, we get
\begin{equation}
F_1=\alpha_1\bigl(1-\norms{\alpha_1}\bigr)-\sum_{0<q<k_c}\bigl(\alpha_1^*\widetilde{A}^\vect q_{11}
+2\alpha_1C^\vect q_{11}\bigr)+O(n^{2.5}).
\end{equation}
Using the established results of the parameters of the many-body wave function, we get
\begin{subequations}
\begin{equation}
F^\parallel_\vect K=\biggl[1-\frac{4}{3\sqrt{\pi}}\bigl(n_\up a_m^3\bigr)^{1/2}+h.o.c.\biggr]N_\up^{1/2}\phi_\vect K,
\end{equation}
\begin{align*}
F^\perp_1=&\biggl[1\!+\!\frac{28}{3\sqrt{\pi}}\bigl(n_\up a_m^3\bigr)^{1/2}\biggr]N_\up^{3/2}\Bigl(\devp_1/\Omega-
\!\sum_{2}P^\perp_{12}\norms{\phi_2}\phi_2\Bigr)\\
&+O(n^{2.5}),
\end{align*}
and the latter equation is simplified by Eq.~\eqref{eq:m_dev}:
\begin{equation}\label{eq:m_25_F2perp}
F^\perp_{\vect K}=\biggl[1+\frac{28}{3\sqrt{\pi}}\bigl(n_\up a_m^3\bigr)^{1/2}\biggr]N_\up^{3/2}\dev_{\vect K}/\Omega+O(n^{2.5}).
\end{equation}
\end{subequations}
The accuracy in the parallel component of $F_\vect K$ is limited by that of the parallel component of $\alpha_\vect K$.
To increase the accuracy, one must determine $E-\mu N$ to higher orders.

The amplitude of the component in $F_\vect K$ parallel to $\phi_\vect K$ is mapped directly to the well-known 
condensate-depletion formula
$$\langle b_0\rangle=\biggl[1-\frac{4}{3\sqrt{\pi}}\bigl(n_b a_b^3\bigr)^{1/2}+h.o.c.\biggr]\sqrt{N_b}$$
in the limit of point-like bosons. Here $N_b$ is the total number of bosons, and $n_b$ and $a_b$ are the boson density and scattering length,
respectively.

The orthogonal \emph{deviation} of $F_\vect K$ from $\eta_F\phi_\vect K$ is proportional to $\dev_\vect K$ -
a function derived from the two-molecule zero-speed scattering wave function - at the leading and next-to-leading orders.
The coefficients are also predicted \emph{exactly}, in Eq.~\eqref{eq:m_25_F2perp}.
The next step is to determine $\dev_\vect K$ from Eq.~\eqref{eq:m_dev}.
If the interaction between fermions is zero-range with a large and positive scattering length, we should
substitute the four-fermion wave function calculated by Petrov \textit{et al} \cite{Petrov2004PRL} into Eq.~\eqref{eq:m_dev},
to determine $\dev_\vect K$. This step is elementary (involving no many-body physics at all) but tedious; it will be carried out
later, by the author himself or by some other people.

The significance of this deviation from the two-body function $\phi_\vect K$ has been pointed out in the closing paragraphs
of Sec.~\ref{subsec:m_20}. Quantitative comparison between our prediction and experiment is possible in the future.

For ultracold fermionic atoms in two internal states near a wide Feshbach resonance (on the $a>0$ side),
since our prediction is based upon the two-molecule scattering physics as studied by Petrov \textit{et al}
\cite{Petrov2004PRL}, our predictions are relevant to experiment in so far as Petrov \textit{et al}'s
results are relevant to experiment. There is already much experimental evidence supporting the value of $a_m$
 as predicted by Petrov \textit{et al} \cite{Petrov2004PRL}, most notably the cloud size of a trapped molecular Bose-Einstein condensate.
On the other hand, the two-molecule physics of Petrov \textit{et al}'s is \emph{never} exact except when the molecular density
approaches zero; our low-density theory is in a similar situation. We
are thus led to believe that the low-density results in this paper are applicable to experimentally accessible densities,
just like Petrov \textit{et al}'s \cite{Petrov2004PRL}. So the low-density results in
this paper are applicable when $0<k_Fa\lesssim 1$. The more detailed upper bound depends on the particular experiment
one wants to address, and on one's accuracy goal. Here $k_F\equiv(3\pi^2n)^{1/3}$,
and $a$ is the scattering length between fermions.

Having stated these, we must stress that our results are also valid \emph{beyond} the range of validity of Petrov \textit{et al}'s \cite{Petrov2004PRL}
in the sense that ours are valid for other interactions between fermions as well. For other interactions, 
the two-molecule scattering wave function will be different and $a_m\neq0.6a$ \cite{Petrov2004PRL,_Petrov_a_m}. But the results
in this paper still apply, provided that the conditions specified in Sec.~\ref{subsec:m_hamiltonian} are satisfied.

\subsection{Superfluid four-fermion function}
Let $\vect K_1+\vect K_2+\vect K_3+\vect K_4=0$, and
$$F^{(4)}_{\vect K_1\vect K_2\vect K_3\vect K_4}\equiv\langle c_{\vect K_4} c_{\vect K_3} c_{\vect K_2}c_{\vect K_1}\rangle.$$

Case 1: the sum of two of the four spin-momenta is zero, and the thermodynamically significant contribution to
$F^{(4)}$ equals a product of two $F^{(2)}$'s, according to Eq.~\eqref{eq:Q}. Here $F^{(2)}_\vect K\equiv F_\vect K$.
For instance, $F^{(4)}_{\vect K,-\vect K,\vect K',-\vect K'}=F_{\vect K}F_{\vect K'}$ if $\vect K\neq\pm\vect K'$.
(If $\vect K=\pm\vect K'$, $F^{(4)}_{\vect K,-\vect K,\vect K',-\vect K'}=0$.)

Case 2: for \emph{all} $i=2,3,4$, $\vect K_1$ and $\vect K_i$ either have parallel spin, or have a nonzero total momenta $\sim O(n^0)$.
There is only one diagram for $F^{(4)}$ up to the order $n^{1.5}$, namely $\beta_{\vect K_1\vect K_2\vect K_3\vect K_4}$.
\begin{align}
F^{(4)}_{\vect K_1\vect K_2\vect K_3\vect K_4}=&N_\up\biggl[1+\frac{16}{3\sqrt{\pi}}\bigl(n_\up a_m^3\bigr)^{1/2}\biggr]\phi'_{\vect K_1
\vect K_2\vect K_3\vect K_4}\notag\\
&+O(n^2\Omega^{-1}).
\end{align}

\begin{figure}\includegraphics{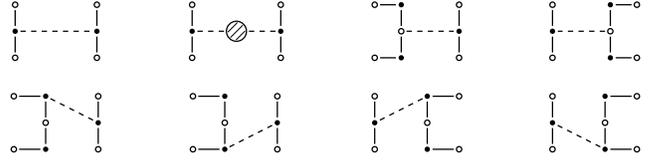}\caption{\label{fig:m_F4}Diagrams for the superfluid four-fermion function up to the low-density
order $n^1$, when
the sum of two external spin-momenta is $\sim\sqrt{n}$. The external points are distinguishable. In the second diagram,
the shaded circle stands for various internal structures (either $N_2'=1,2$, or there is one bubble island that is not simplest).}\end{figure}

Case 3: the sum of two of the four spin-momenta is a small momenta $\sim n^{1/2}$. Consider for instance
$$F^{(4)\vect q}_{\vect K_1\vect K_2}\equiv F^{(4)}_{\vect K_1+\vect q/2,-\vect K_1+\vect q/2,\vect K_2-\vect q/2,-\vect K_2-\vect q/2},$$
where $q/\sqrt{n_\up a_m}$ is a nonzero constant of order unity (in the low-density limit).
If $\vect K_1\neq\pm\vect K_2$, all the diagrams up to the low-density order $n^1$ are shown in Fig.~\ref{fig:m_F4}.
Using the part of the cluster-separation theorem that is already proved, together with the established result concerning
$\gamma^\vect q_{1234;5}$, we can easily show that the second diagram (containing the shaded circle),
whose low-density order is $0.5$, is proportional to $\phi_1\phi_2$ up to the order $n^1$. So
\begin{align*}
F^{(4)\vect q}_{12}=&\bigl(1-2\norms{\alpha_1}-2\norms{\alpha_2}\bigr)A^{\vect q}_{12}-\alpha_1^2C^\vect q_{21}-\alpha_2^2C^\vect q_{12}\\
&+\eta^\vect q\phi_1\phi_2+O(n^{1.5}\Omega^{-1}),
\end{align*}
where $\eta^\vect q\sim n^{1/2}$. We first substitute the result of $\beta^\vect q_{12}$
[Eq.~\eqref{eq:m_25_final}] to Eq.~\eqref{eq:m_Aq} to determine $A^\vect q_{12}$ accurately,
and then evaluate the other terms to the leading order. 
At $\vect K_1=\pm\vect K_2$, there is appropriate delta-function singularity in $F^{(4)\vect q}_{\vect K_1\vect K_2}$,
in accordance with case 1. Combining these subregions of the spin-momentum configuration space, we derive the final result
\begin{subequations}
\begin{gather}
F^{(4)\vect q}_{12}\equiv F^{(4)\vect q\parallel}_{12}+F^{(4)\vect q\perp}_{12},~~~\sum_{12}\phi_1^*\phi_2^*F^{(4)\vect q\perp}_{12}\equiv0,\notag\\
F^{(4)\vect q\parallel}_{12}=\biggl(\frac{x_\vect q}{1-x_\vect q^2}+h.o.c.\biggr)\phi_1\phi_2,\\
\begin{split}
F^{(4)\vect q\perp}_{12}=&n_\up\biggl[\frac{1+2x_\vect q}{1-x_\vect q^2}(\dev_1\phi_2+\phi_1\dev_2)+g_{12}\biggr]\\
&+O(n^{1.5}\Omega^{-1}).
\end{split}
\end{gather}
\end{subequations}
The physical meaning of these results will be clearer in coordinate space.

\subsection{Superfluid six-fermion function, momentum distribution, two-body reduced density matrix, etc}
The six-fermion correlation parameters $\gamma^\vect q_{1234;5}$ are an indispensible part of
the consistent beyond-mean-field theory. This is not surprising, given the fact that three-boson correlation is needed
to formulate the correct beyond-mean-field theory of structureless bosons \cite{_Tan_BEC}. Leggett (and many others)
notices that Bogoliubov wave function does \emph{not} yield the correct beyond-mean-field equation of state of structureless bosons,
since some ``bare interactions" are not renormalized in favor of the bosonic scattering length, a problem he considers
``a little disturbing" \cite{LeggettNJP}, which he then remedies by modifying the Bogoliubov wave function.

Using the pseudopotential as the effective bosonic interaction, one can escape this problem at a price:
the correct \emph{short-distance} behavior of the bosons upon collisions is omitted.
For the Fermi superfluid in the molecular Bose-Einstein condensated
state, the price of this replacement is even higher:
since all the four-fermion scattering physics is replaced by a molecular scattering length, one can no longer describe the many
physical observables that involve momentum scales much larger than $\sqrt{na_m}$, such as the many-body effects on
the fermionic momentum distribution and the superfluid pairing function.

In the approach of \cite{_Tan_BEC} and this paper, such problem is eliminated by strictly applying the power-counting formula
to identify \emph{all} the significant diagrams. When this is done, one finds that six-fermion (or three-boson) correlation terms can not be
altogether thrown away as people wish to do. In the same way as the last two subsections,
one can derive the associated nonvanishing superfluid six-fermion function:
$\langle c_{\vect K_4+\vect q/4}\cdots c_{\vect K_1+\vect q/4}c_{-\vect K_5-\vect q/2} c_{\vect K_5-\vect q/2}\rangle$.

The momentum distribution of \emph{fermions}, $\langle c^\dagger_\vect K c_{\vect K}^{}\rangle$, is another interesting quantity, for two
reasons. Firstly, it is directly measurable. Secondly, it is predicted that there are two \emph{exact} relations between the energy and the
momentum distribution of the two-component Fermi gas with large scattering length \cite{_Tan_EnergyTheorem,_Tan_Adiabatic},
valid for \emph{all} nonzero values of $k_Fa$ (including the unitary regime), and also valid for finite temperatures, few-body systems, etc.
 Performing the analytic calculation of the momentum distribution in the low-density
regime, we can directly test these earlier predictions \cite{_Tan_EnergyTheorem,_Tan_Adiabatic}.
The results obtained in this paper already enables an unambiguous determination of the fermionic momentum distribution up to the order $n^{2.5}$.
It will be presented in a subsequent work.

The two-body reduced density matrix is also observable. It remains to see if there is interesting physics in this
quantity. It will be determined analytically in the low-density regime in a subsequent work.

Another quantity to study is the fermionic Green functions (``normal" and ``anomalous").
They help us to understand the dynamic behavior of the Fermi superfluid.

The above list is still not exhaustive.

\section{\label{sec:m_conclusion}Concluding remarks}
At low densities, there is a natural coexistence of the characteristics of a conventional Bose gas and that of a Fermi gas in the present theory.
If the system is probed at length scales of the order $\xi\sim1/\sqrt{n_\up a_m}$ or longer length scales, one can hardly distinguish
it from a usual Bose gas. The equation of state, the condensate depletion, the low-momentum part of the momentum distribution of molecules
($q\sim1/\xi$), etc, all follow conventional Bogoliubov-Lee-Yang theory. If the observables that involve large momentum scale
$q\sim1/r_m$ are measured, however, one sees many features that are absent in a conventional Bose gas. The momentum distribution
of fermions, the deviation of the superfluid pairing function from being proportional to the internal wave function of an isolated molecule,
etc, are examples of such observables.

The Bose characteristics is \emph{not} put in by hand, but forced on us by the \emph{fermionic} wave function Eq.~\eqref{eq:m_psi}.
The ground state energy as derived in this paper is the absolute lower bound, as we have gone through great lengths,
making all the possible adjustments of the wave function to minimize it. The wave function itself does not seem to be improvable
any more, since all the possible correlations are included. It is the low density that makes higher correlations less important,
but their effects are not automatically ignored - instead, using the power-counting formula Eq.~\eqref{eq:R}, we have \emph{shown} that most of them,
except a few, are insignificant in the calculation of the leading order correction to the mean-field equation of state.

Will such Bose characteristics be carried over to the high density regime? It seems that various neutral isotropic superfluids 
(no matter BEC of structureless bosons, BEC of composite bosons, unitary Fermi superfluid, or BCS superfluids) have
very similar structures at length scales much longer than $\xi$ (not necessarily $\propto n^{-1/2}$). At shorter length scales,
their differences show up. However the similarity between the dilute BEC of structureless bosons and dilute BEC of composite bosons
is kept at shorter length scales, until $\max(r_m,l)$, which is much shorter than $\xi$. For a unitary Fermi superfluid at zero temperature, there is only one
length scale $\xi\sim 1/k_F\sim n^{-1/3}$, so its similarity to other superfluids must be broken at a length scale somewhat longer than $\xi$.

To formulate the quantitatively well-controlled many-body theory for the Fermi superfluid in the unitary regime, we should
probably take advantage of the above partial similarity, and do a momentum scale division at $k_c$, where $k_c$ is roughly the highest
momentum scale \emph{below which} the structure of the system is nearly indistinguishable from other superfluids. The structure
of the system at $k>k_c$ needs to be determined nonperturbatively, and smoothly matched to the low-momentum structure.

Besides this project, we have of course the shorter run subject of extending the current low-density theory to higher orders
beyond Lee-Yang. We need to first show that Wu's logarithmic term \cite{Wu1959PR}
is present in the equation of state, and then determine the leading differences
of the equations of state of molecular condensates and those of other BECs. More concretely, we need to determine the value of
the constant $C$ as defined in Ref.~\cite{_TanLevin}. We derive confidence in these plans from two sources:
the diagrammatic theory in this paper (the author has
found diagrams in the expansion of $E-\mu N$ which scale like $n^3\ln n$), and Braaten et al's Effective Field Theory calculations
\cite{Braaten2002PRL}, which support the universality of the Wu term, at least for structureless bosons.

In the same time, the fermionic momentum distribution is of high interest. Here we have some \emph{exact} theorems \cite{_Tan_EnergyTheorem,
_Tan_Adiabatic} for the two-component Fermi gas with large scattering length, valid in the low-density as well as unitary regimes.
They relate the momentum distribution to the energy, the pressure, and the rate of change of energy during a ramp of the fermionic
scattering length \cite{_Tan_EnergyTheorem,
_Tan_Adiabatic}. The confirmation of these theorems in a concrete low-density analytic calculation will push our understanding
of such a novel Fermi system to a greater depth. This has been done in Ref.~\cite{_Tan_BEC}, for the dilute \emph{Bose}
gas with large scattering length, although the theorems are only approximate in that Bose system \cite{_Tan_BEC}.

\acknowledgments
The author is very grateful to K. Levin for many helpful suggestions, and for communicating the ideas to other researchers.
The author thanks T.~L.~Ho for calling
the author's attention on testing the theorems in Refs.~\cite{_Tan_EnergyTheorem, _Tan_Adiabatic} with independent perturbative
calculations.
\appendix
\section{\label{a:m_phi4_identity}An identity satisfied by $\phi^{(4)}$}
Proof of Eq.~\eqref{eq:m_phi4_id1}. Replacing $\vect K_1\cdots\vect K_4$ in Eq.~\eqref{eq:m_phi4_Schrodinger}
by $\vect K_1+\vect q/2$, $-\vect K_1+\vect q/2$, $\vect K_2-\vect q/2$, and $-\vect K_2-\vect q/2$, respectively, using
the Galilean symmetry of $U$,  multiplying the equation by $\phi^*_{\vect K_2}$, summing over $\vect K_2$, noting the
antisymmetry of $\phi$ and $\phi^{(4)}$, and noting that dumb momenta can be freely shifted, we get
\begin{multline}\label{eq:m_phi4_longA}
\frac{q^2}{2}\sum_2\phi_2^*\phi^{(4)\vect q}_{12}+\sum_{23}D_{13}\phi_2^*\phi^{(4)\vect q}_{32}
+\sum_{23}D_{23}\phi_2^*\phi^{(4)\vect q}_{13}\\
+T^\vect q_{-\vect K_1+\vect q/2}-T^\vect q_{\vect K_1+\vect q/2}=0,
\end{multline}
$$T^\vect q_{\vect K}\equiv\sum_{\vect K_2\vect K_3\vect K_4}U_{-\vect K+\vect q,-\vect K_2-\vect q,\vect K_3\vect K_4}
\phi^*_{\vect K_2+\vect q/2}\phi^{(4)}_{\vect K\vect K_2\vect K_3\vect K_4},$$
and the infinitesimal $\epsilon^{(4)}$ is omitted in comparison with $q^2/2>0$.

Using Eq.~\eqref{eq:m_phi4_smallq}, we see that the first term of Eq.~\eqref{eq:m_phi4_longA} approaches
$-(2\pi a_m/\Omega)\sum_{2}\lvert\phi_2\rvert^2\phi_1=-(4\pi a_m/\Omega)\phi_{\vect K_1}$, when $\vect q\neq0$
but $\vect q\rightarrow0$.

Obviously $\sum_{4}D_{14}\proj_{43}=D_{13}$. So the second term
of Eq.~\eqref{eq:m_phi4_longA} is rewritten as $\sum_4D_{14}\sum_{23}\proj_{43}\phi^{(4)\vect q}_{32}\phi_2^*$.

$\sum_2D_{23}\phi_2^*=0$, and the third term vanishes.

Since $U$ and $\phi^*$ are smooth functions of momenta (when momentum is conserved by the subscripts),
$T_\vect K\equiv\lim_{\vect q\rightarrow0}T^\vect q_\vect K=\sum_{\vect K_2\vect K_3\vect K_4}U_{-\vect K,-\vect K_2,\vect K_3\vect K_4}
\phi^*_{\vect K_2}\phi^{(4)}_{\vect K\vect K_2\vect K_3\vect K_4}$. Obviously $T_{\vect k\sigma}$ depends continuously on the
momentum $\vect k$, for a given spin $\sigma$. So in the limit $\vect q\rightarrow0$,
the last two terms of Eq.~\eqref{eq:m_phi4_longA} approaches $T_{-\vect K_1}-T_{\vect K_1}$.

Combining the above results, we see that when $\vect q\rightarrow0$, Eq.~\eqref{eq:m_phi4_id1} is obtained.

\section{\label{a:terms}Definition of some technical terms}

The terms used in this paper follow convention as much as possible. However some terms in this paper
have narrower meanings than convention, some are used in a broader sense, and some are new - due to the needs of the theory.
Here we give the precise meanings of some of them (for use in this paper), in alphabetical  order.

\emph{Dead end}: an internal part of a diagram that is connected to the remaining part via a \emph{single} (de)generator.
In the present formalism, dead ends are absent because of a basic property of the coefficient $\alpha^{(p)}_{\vect K_1\cdots\vect K_p}$:
it is zero whenever the sum of a nontrivial subset of the $p$ spin-momenta is zero.

\emph{Diagram}: an arbitrary collection of  vertices and lines; each line has \emph{exactly two ends}, and they
must be attached to two vertices or the same vertex.
The number of vertices is denoted by $N_0$, the number of lines by $N_1$, the number of independent loops by $N_2$, and
the number of disconnected parts  - each part is connected, but any two different parts are disconnected - by $N_3$. There is
a simple universal identity
\begin{equation}\label{eq:any_diagram}
N_0-N_1+N_2-N_3=0,~~~\text{for any diagram}.
\end{equation}
This identity is used in the derivation of two basic power-counting formulas, Eqs.~\eqref{eq:Q} and \eqref{eq:R}.

\emph{External point}: a special vertex representing a fermionic operator (annihilation or creation)
in a product whose expectation value we are computing. Each external point is attached by a single solid line
associated with a certain spin-momentum. It is a small solid circle if it stands for a creation operator, or a hollow one
for an annihilation operator.

\emph{$h.o.c.$}: higher order correction (in the low-density expansion),
with strictly greater value of low-density order $R$, in comparision with the term immediately
preceding it.

\emph{Low-density order} $R$ of a quantity $x$: if the leading order approximation of
$x$ scales like $n^R\Omega^Q$ in the low-density regime, the low-density order of $x$ is said to be $R$.
Symbolically we may write $x\sim n^R$. If actually $x\sim n^R\ln(na_m^3)$ or things like that,
we may still write $x\sim n^R$ in the slightly broader sense. Here $n$ and $\Omega$ are the particle density and the system's volume,
respectively, and $\Omega$ is assumed to be large enough to avoid finite-size effects.
A \emph{higher} low-density order means a greater value of $R$, and usually indicates a \emph{smaller} magnitude in the low-density regime.

\emph{Nontrivial subset} of $M$ items: a nonempty proper subset, namely with $M_1$ elements such that $1\leq M_1\leq M-1$.

\emph{$O(n^R)$}: abbreviation of $O(n^R\Omega^0)$.

\emph{$O(n^R\Omega^Q)$}: a symbol describing how a quantity (in the low-density regime)
scales with $n$ and $\Omega$. The (often dimensionful)
coefficient roughly independent of $n$ and $\Omega$ is not shown in this symbol.
Things like $n^R\ln(na_m^3)\Omega^Q$ are sometimes possible, and \emph{not} excluded by this symbol.

\emph{Thermodynamic order} $Q$ of a quantity $x$: if $x$ scales like $\Omega^Q$ in the \emph{thermodynamic limit}
(in which $n$ is held constant as usual), the thermodynamic order of $x$ is said to be $Q$. Symbolically we may
write $x\sim\Omega^Q$. This notion is obviously orthogonal to the low-density order; it is also valid beyond the low-density
regime (in contrast with the low-density order).

No contradiction between $x\sim\Omega^Q$ and $x\sim n^R$ exists -
there is no contradition between ``the height of this rectangle is $Q$" and ``the width of the rectangle is $R$".
In fact we may use a plane, and map $x$ to a point with coordinates $(R,Q)$ on the plane.

\section{\label{a:1st_separation}Partial proof of the first cluster-separation theorem}
First consider $\A_\vect K\equiv\langle c_{-\vect K} c_\vect K\rangle$. The theorem states that there exists
a coefficient $\eta=\eta^{}_{0.5}n^{0.5}+\eta^{}_{1.0}n^1$, such that $\A_\vect K=\eta\phi_\vect K$,
with an absolute error of the low-density order $n^{1.5}$. The proof follows.

Since the ground state expectation value of
\begin{equation}\label{eq:Ht}
\Ht\equiv H-\mu\hat{N}
\end{equation}
is stationary under any infinitesimal change of the many-body state (the number of particles may
also be changed slightly), we can easily show an equation of motion
\begin{equation}\label{eq:motion}
\langle O\Ht-\Ht O\rangle=0
\end{equation}
for any product of fermionic operators, $O$.
Substituting Eq.~\eqref{eq:m_H3} and $O=c_{-\vect K_1}c_{\vect K_1}$ into this formula, we get
\begin{equation}\label{eq:m_separation1}
\sum_2\widetilde{D}_{12}\A_2\!
=\!\frac{1}{2}\sum_{234}U_{1234}\langle c_2^\dagger c_4^{}c_3^{}c_{\bar{1}}\rangle
+\frac{1}{2}\sum_{234}U_{2\bar{1}34}\langle c_2^\dagger c_4^{} c_3^{} c_1\!\rangle,
\end{equation}
\begin{equation}\label{eq:m_Dtilde}
\widetilde{D}_{\vect K_1\vect K_2}\equiv(k_1^2-E_m-2\deltamu)\delta_{\vect K_1\vect K_2}
+\frac{1}{2}U_{\vect K_1,-\vect K_1,\vect K_2,-\vect K_2}.
\end{equation}
where $\deltamu\equiv\mu-E_m/2$ is at least of the order $n^{0.5}$
 (Sec.~\ref{subsec:m_10}).

\begin{figure}
\includegraphics{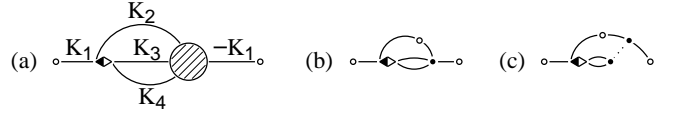}
\caption{\label{fig:m_separation1}Diagrams for
$\frac{1}{2}\!\sum\limits_{\vect K_2\vect K_3\vect K_4}\!\!\!U_{\vect K_1\vect K_2\vect K_3\vect K_4}^{}
\langle c_{\vect K_2}^\dagger c_{\vect K_4}^{}c_{\vect K_3}^{}c_{-\vect K_1}^{}\!\rangle$.
(a): the total diagram. The shaded circle stands for the ground state expectation value
$\langle c_{\vect K_2}^\dagger c_{\vect K_4}^{}c_{\vect K_3}^{}c_{-\vect K}^{}\rangle$.
(b) and (c): two individual diagrams, whose low-density orders are $1.5$ and $2$, respectively.
In (c), if the whole island is reduced to a point, the dotted line connects this same point and forms a loop.}
\end{figure}

The first term on the right side of Eq.~\eqref{eq:m_separation1}
has a diagrammatic representation in Fig.~\ref{fig:m_separation1}.
In Fig.~\ref{fig:m_separation1}(a)
the total diagram is shown; the big shaded circle stands for the expectation value of a product of
fermionic operators, which in the present case is
$\langle c_{\vect K_2}^\dagger c_{\vect K_4}^{}c_{\vect K_3}^{}c_{-\vect K}^{}\rangle$.
In Fig.~\ref{fig:m_separation1}(b)(c) we show, for illustration, two associated individual diagrams, whose low-density orders are $1.5$ and $2$,
respectively [Eq.~\eqref{eq:R}]; note that $N_2'=1$ for Fig.~\ref{fig:m_separation1}(c), since its skeleton diagram
is a point and a dotted line whose both ends are attached to such a point.

We now show that the low-density orders of the infinitely many individual diagrams associated with
Fig.~\ref{fig:m_separation1}(a) have a lower bound, $1.5$.
In Fig.~\ref{fig:m_separation1}(a), the piece on the left
of the shaded circle must be in the \emph{same} island as the solid line on the right, because
the dotted lines (if there are any) can not carry spin or any large momentum, and the separation of the
two pieces in two different islands would contradict spin-momentum conservation. Thus we have a single
external island $e$. One spin-momentum (\textit{ie} $\vect K_2$) enters the shaded circle, three leave it,
and the associated four fermion operators are normally ordered, so $\vect K_2$ must be attached to a
degenerator in the big shaded circle; so $P^{(-)}_e\geq 2$ and $P^{(+)}_e=P^{(-)}_e+L^{(-)}_e-L^{(+)}_e
=P^{(-)}_e+2\geq4$. So $P_e\geq6$. Note also that $N_3=1$. So we deduce from Eq.~\eqref{eq:R2}
that $R\geq1.5$.

The last term of Eq.~\eqref{eq:m_separation1} is similar. So the function $(\widetilde{D}\A)_
\vect K=\sum_{\vect K'}\widetilde{D}_{\vect K\vect K'}\A_{\vect K'}$ is at least of the order $n^{1.5}$.
The spectrum of $\widetilde{D}$ has a \emph{gap} equal to $\lvert E_m\rvert$; the eigenvalue associated with 
$\phi_{\vect K}$ is $-2\deltamu$, and all the other eigenvalues are at least $\lvert E_m\rvert-2\deltamu
>\lvert E_m\rvert/2\sim n^0$
(see Sec.~\ref{subsec:m_hamiltonian} for our basic assumptions about $U$). 
Now make an orthogonal decomposition: $\A_\vect K=\eta\phi_\vect K+\A'_{\vect K}$
(where $\sum_{\vect K}\phi^*_\vect K\A'_\vect K=0$). Since $\widetilde{D}$ is hermitian
and $\phi_\vect K$ is an eigenvector,
$(\widetilde{D}\A)_\vect K=-2\deltamu\eta\phi_\vect K+(\widetilde{D}\A')_\vect K$ is also a sum of two orthogonal components,
and the low-density order of each of them is at least $1.5$. 
Since the leading order contribution to $\A_{\vect K}$ is just $\alpha_\vect K$,
the low-density order of $\eta$ is $0.5$ (Sec.~\ref{subsec:m_10}). So
$\A'_\vect K$ is at least of the order $n^{1.5}$, and (as a byproduct)
\begin{equation}\label{eq:m_delta_mu}
\deltamu\equiv\mu-E_m/2~\sim n^1,
\end{equation}
which is needed in Sec.~\ref{subsec:m_diagrams_25}.

Now turn to $\A^{\vect q}_{\vect K_1\vect K_2}\equiv\langle c^{}_{-\vect K_1+\vect q/2}c^{}_{\vect K_1+\vect q/2}
c^{}_{-\vect K_2-\vect q/2}\\c^{}_{\vect K_2-\vect q/2}\rangle$, as a function of $\vect K_1\vect K_2$
(where neither $\vect K_1+\vect K_2$ nor $\vect K_1-\vect K_2$ is zero or $\sim\sqrt{n}$), for
a given nonzero momentum $\vect q$ ($q\sim\sqrt{n}$). The low-density order of this $\A$ is $0$.

\begin{figure}
\includegraphics{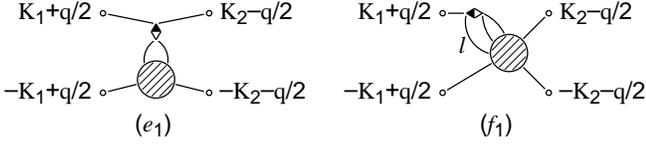}
\caption{\label{fig:m_separation2}Some residual terms in the equation of motion of
$\langle c_{-\vect K_1+\vect q/2}c^{}_{\vect K_1+\vect q/2}c^{}_{-\vect K_2-\vect q/2}c^{}_{\vect K_2-\vect q/2}\rangle$
[see Eq.~\eqref{eq:m_separation2}].
In the second diagram, ``l" marks a later ``time" than the other legs of the shaded circle, as the six fermion operators
are normally ordered. The low-density orders of these terms are at least $1$. }
\end{figure}
The cluster-separation theorem states that there exists a coefficient $\eta_\vect q$ such that
$\A_{\vect K_1\vect K_2}^{\vect q}=\eta_\vect q\phi_{\vect K_1}\phi_{\vect K_2}+{\A'}^{\vect q}_{\vect K_1\vect K_2}$,
where the low-density order of $\A'$ is $1$ or higher.
 To prove it, we substitute $O=c^{}_{-\vect K_1+\vect q/2}c^{}_{\vect K_1+\vect q/2}
c^{}_{-\vect K_2-\vect q/2}c^{}_{\vect K_2-\vect q/2}$ into Eq.~\eqref{eq:motion} and get
\begin{equation}\label{eq:m_separation2}
\sum_{\vect K_1'\vect K_2'}\Htccq_{\vect K_1\vect K_2;\vect K_1'\vect K_2'}\A_{\vect K_1'\vect K_2'}^{\vect q}
=\sum_{i=1}^4e_i+\sum_{i=1}^4f_i,
\end{equation}
where $\Htccq_{\vect K_1\vect K_2;\vect K_1'\vect K_2'}\equiv
\Htcq_{\vect K_1\vect K_1'}\delta_{\vect K_2\vect K_2'}+\delta_{\vect K_1\vect K_1'}\Htcq_{\vect K_2\vect K_2'}$,
$\Htcq_{\vect K\vect K'}\equiv(k^2-E_m-2\deltamu+q^2/4)\delta_{\vect K\vect K'}+\frac{1}{2}
U_{\vect K,-\vect K,\vect K',-\vect K'}$, and $e_i$ and $f_i$ are eight residual terms whose low-density orders
are at least $1$ (see below). In Fig.~\ref{fig:m_separation2} are shown the diagrams for $e_1$ and $f_1$.
The other $e_i$'s differ from $e_1$ in the two external points that are linked to the solid side of the diamond;
but when $\vect K_1+\vect q/2$ and $-\vect K_1+\vect q/2$ (or $\vect K_2-\vect q/2$ and $-\vect K_2-\vect q/2$)
are linked to the solid side of the diamond, we get some terms on the left side of Eq.~\eqref{eq:m_separation2}.
The other $f_i$'s differ from $f_1$ in the external point that is linked to the solid side of the diamond.

Now apply Eq.~\eqref{eq:R2}. For $e_1$, all the external points
must be in the same island, so $I_\text{ext}=1$; also $P_\text{ext}\geq4$; so $R\geq1$.
For $f_1$, $I_\text{ext}\leq2$, $P_\text{ext}^{(-)}\geq2$, $P_\text{ext}^{(+)}\geq6$, and $P_\text{ext}
=P_\text{ext}^{(+)}+P_\text{ext}^{(-)}\geq8$; so $R\geq1$.

The spectrum of $\Htcq$ has a gap: the lowest eigenvalue is of the order $n^1$ and is associated with the eigenvector
$\phi_\vect K$, and all the other eigenvalues are at least $\lvert E_m\rvert-2\deltamu+q^2/4>\lvert E_m\rvert/2$.
The two terms of $\Htccq$ act (as $\Htcq$) independently on $\vect K_1$ and $\vect K_2$, respectively. So $\Htccq$ has a single eigenvalue
of the order $n^1$, associated with the eigenvector $\phi_{\vect K_1}\phi_{\vect K_2}$; all the other eigenvalues
are at least $\lvert E_m\rvert-4\deltamu+q^2/2>\lvert E_m\rvert/2$. So the components of
$\A_{\vect K_1\vect K_2}^\vect q$ which are \emph{orthogonal} to $\phi_{\vect K_1}\phi_{\vect K_2}$ must be of the order $n^1$,
in order for $\Htccq\A$ to be of the order $n^1$.

On the left side of Eq.~\eqref{eq:m_separation2}, $\vect K_1'$ (or $\vect K_2'$) may take values close or equal
to $\pm\vect K_2$ (or $\pm\vect K_1$). But the associated contributions are not important.
When $\vect K_1'\mp\vect K_2\sim\sqrt{n}$, the associated volume in the $\vect K_1'$-space is reduced by a factor
of the order $n^{1.5}$. When $\vect K_1'\mp\vect K_2$ is exactly zero, the thermodynamic
order of $\A_{\vect K_1'\vect K_2}$ increases by $1$, but this is accompanied by the loss of an independent dumb
momentum; meanwhile
the low-density order of $\A_{\vect K_1'\vect K_2}$ becomes $1$, so the net effect on Eq.~\eqref{eq:m_separation2} is of the same
order as those of $e_i$ and $f_i$.

For more l-clusters ($M\geq3$), each of which contains two spin-momenta, we still find that
the equation of motion is dominated by the quasi-independent dynamics of individual clusters, with
small ``driving terms" [like the ones on the right side of Eq.~\eqref{eq:m_separation1} or
Eq.~\eqref{eq:m_separation2}], so that
$\A$ (at given small momenta $\vect q_l\sim\sqrt{n}$) is still proportional to a product
of single-molecule internal wave functions, with relative error of the order $n^1$.

\section{\label{a:2nd_separation}Partial proof of the second cluster-separation theorem}

Consider any diagram $\mathcal{D}$ in the low-density expansion of $\A_{\vect K}\equiv\langle c_{-\vect K}
c_{\vect K}\rangle$. Since there are only two external points [$L^{(-)}=2$, $L^{(+)}=0$], there is only
one external island, and $I_\text{ext}=1$, $P_\text{ext}^{(+)}=P_\text{ext}^{(-)}+2$.
From Eq.~\eqref{eq:R2} we deduce that $R=1/2+P^{(-)}_\text{ext}/2+(P_\text{int}/4-I_\text{int})+N_2'/2$.
If $P^{(-)}_\text{ext}=0$, the external island is of a unique structure
- a single small solid circle linked to the two external points via two solid lines; moreover, this island must be
\emph{the whole diagram}, since dead ends are absent. If $P^{(-)}_\text{ext}\geq2$, $R\geq1.5$.
So $\A_{\vect K}=\alpha_{\vect K}$ plus correction terms of the order $n^{1.5}$.
But we proved in Sec.~\ref{subsec:m_1st_separation} that there exists an $\eta$ such that
$\A_{\vect K}=\eta\phi_{\vect K}$ plus correction terms of the order $n^{1.5}$.
So $\alpha_{\vect K}=\eta\phi_{\vect K}$ plus correction terms of the order $n^{1.5}$.
The simplest case of the second cluster-separation theorem is thus proved.

\begin{figure}\includegraphics{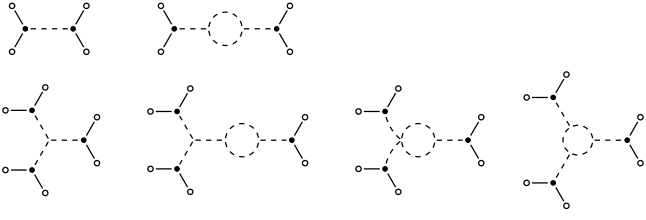}
\caption{\label{fig:m_separation3}First row: diagrams for $\A_{\vect K_1\vect K_2}^\vect q$ up to the order $n^{0.5}$.
Second row: diagrams for $\A_{\vect K_1\vect K_2\vect K_3}^{\vect q_1\vect q_2\vect q_3}$ up to the order $n^0$.
Each \emph{dashed line} represents a geometric series associated with a variable number of simplest bubbles.
The first diagram corresponds to the first row of Fig.~\ref{fig:m_composite_vertices}.}
\end{figure}
Now turn to
\begin{equation}\label{eq:m_beta_matrix}
\beta_{\vect K_1\vect K_2}^{\vect q}\equiv\beta_{\vect K_1+\vect q/2,-\vect K_1+\vect q/2,\vect K_2-\vect q/2,-\vect K_2-\vect q/2}
\end{equation}
as a function of $\vect K_1\vect K_2$ (where neither $\vect K_1+\vect K_2$ nor $\vect K_1-\vect K_2$
is zero or $\sim\sqrt{n}$), for fixed small $\vect q$ ($q\sim\sqrt{n}$). We extract its property from
$\A_{\vect K_1\vect K_2}^\vect q$ (see Sec.~\ref{subsec:m_1st_separation}). First identify all the diagrams
in the low-density expansion of $\A$ satisfying $R=0$ or $R=0.5$; using Eq.~\eqref{eq:R2}, we find that
in such a diagram, there are two external islands (each of which has $P_i=2$), the internal islands are
simplest bubbles, and $N_2'=0$ for $R=0$, $N_2'=1$ for $R=0.5$. These two sets of diagrams are shown
in Fig.~\ref{fig:m_separation3} (first row). Some diagrams have the same structure after the simplest bubbles
are suppressed, and their sum can be represented by a single diagram; each diagram in Fig.~\ref{fig:m_separation3}
is such a sum; a \emph{dashed line} represents a geometric series associated with a variable number of simplest bubbles.
See the first row ($A_{\vect K\vect K'}^{\vect q}$) of Fig.~\ref{fig:m_composite_vertices} for the content of the first diagram
in Fig.~\ref{fig:m_separation3}.
\begin{equation}\label{eq:m_beta_eq}
\A_{\vect K_1\vect K_2}^\vect q=A_{\vect K_1\vect K_2}^{\vect q}+B_{\vect K_1\vect K_2}^{\vect q}+O(n^1\Omega^{-1}).
\end{equation}
\begin{equation}\label{eq:m_A_matrix}
A_{\vect K_1\vect K_2}^{\vect q}=\beta_{\vect K_1\vect K_2}^{\vect q}+
\sum_{\vect K_3\vect K_4}\beta_{\vect K_1\vect K_3}^{\vect q}\frac{1}{2}\beta^{*\vect q}_{\vect K_4\vect K_3}\frac{1}{2}
\beta_{\vect K_4\vect K_2}^{\vect q}+\cdots,
\end{equation}
where $1/2$ is the symmetry factor of a simplest bubble, and the small momenta flowing through two
adjacent bubbles have opposite signs (and directions). In the summation over an internal spin-momentum, \textit{eg} $\vect K_3$,
$\vect K_3$ may be close
or equal to $\pm\vect K_1$ or $\pm\vect K_4$, but the corresponding contributions are negligible for our purpose, since their associated
volume in the dumb-momentum space is reduced by a factor of the order $n^{1.5}$.
Now introduce matrices
$\beta_\vect q$ and $A_\vect q$, whose matrix elements are defined by Eqs.~\eqref{eq:m_beta_matrix} and \eqref{eq:m_A_matrix},
respectively: $(\beta_\vect q)_{\vect K_1\vect K_2}\equiv\beta_{\vect K_1\vect K_2}^{\vect q}$ and similarly for $A_{\vect q}$. So
\begin{align}\label{eq:m_beta_formal_equation}
A_\vect q&=\beta_\vect q^{}+\beta_\vect q^{}\beta^\dagger_\vect q\beta_\vect q^{}/4
+\beta_\vect q^{}\beta_\vect q^\dagger\beta_\vect q^{}\beta_\vect q^\dagger\beta_\vect q^{}/16+\cdots\notag\\
&=\beta_\vect q(1-r_\vect q)^{-1},
\end{align}
where
\begin{equation}\label{eq:m_rq}
r_\vect q^{}\equiv\beta^\dagger_\vect q\beta_\vect q^{}/4
\end{equation}
is a hermitian matrix, and all its eigenvalues are nonnegative. Also the geometric series in Eq.~\eqref{eq:m_beta_formal_equation}
should be convergent, so the eigenvalues of $r_\vect q$ should be less than $1$.
Since $A_\vect q^\dagger A_{\vect q}^{}=(1-r_\vect q)^{-1}\beta_\vect q^\dagger\beta_\vect q^{}(1-r_\vect q)^{-1}
=4r_\vect q(1-r_\vect q)^{-2}$, $1+A_\vect q^\dagger A_{\vect q}^{}=[(1+r_\vect q)(1-r_\vect q)^{-1}]^2$.
But all the eigenvalues of the hermitian matrix $(1+r_\vect q)(1-r_\vect q)^{-1}$ are greater than or equal to $1$, according to
the above properties of $r_\vect q$. So $(1+r_\vect q)(1-r_\vect q)^{-1}=\sqrt{1+A_\vect q^\dagger A_{\vect q}}$.
[The square root of any \emph{positive definite hermitian} matrix $h$, $\sqrt{h}$,
is defined as a hermitian matrix with the same set of eigenvectors as $h$,
but its eigenvalues are the (positive) square roots of the corresponding eigenvalues of $h$;
this definition is unambiguous even when the spectrum of $h$ has degeneracy;
$\sqrt{h}$ depends \emph{smoothly} on $h$.]
So $1+\sqrt{1+A_\vect q^\dagger A_{\vect q}}=2(1-r_\vect q)^{-1}$, and
\begin{equation}\label{eq:m_beta_formal_solution}
\beta_\vect q^{}=2A_{\vect q}\Bigl(1+\sqrt{1+A_\vect q^\dagger A_{\vect q}}\;\Bigr)^{-1}.
\end{equation}
So $\beta_\vect q$ is a \emph{smooth} matrix function of the matrix $A_\vect q$.

$B_{\vect K_1\vect K_2}^{\vect q}$ is the $R=0.5$ correction to $\A$ (the second diagram in the first row of Fig.~\ref{fig:m_separation3}),
and $\A^\vect q_{\vect K_1\vect K_2}=\eta_\vect q\phi_{\vect K_1}\phi_{\vect K_2}+O(n/\Omega)$.
So 
\begin{equation}\label{eq:m_A_05}
A_{\vect K_1\vect K_2}^{\vect q}=\eta_\vect q\phi_{\vect K_1}\phi_{\vect K_2}+O(n^{0.5}/\Omega).
\end{equation}
Substituting this
into Eq.~\eqref{eq:m_beta_formal_solution}, we get [after using Eq.~\eqref{eq:m_phi_normalize}]
\begin{equation}\label{eq:m_beta_05}
\beta_{\vect K_1\vect K_2}^{\vect q}=\frac{2\eta_\vect q}{1+\sqrt{1+4\lvert\eta_\vect q\rvert^2}}\phi_{\vect K_1}\phi_{\vect K_2}
+O(n^{0.5}\Omega^{-1}).
\end{equation}
This is still one step from the desired result, namely $\beta_{\vect K_1\vect K_2}^{\vect q}=\eta'_{\vect q}\phi_{\vect K_1}\phi_{\vect K_2}
+O(n^1\Omega^{-1})$ for some $\eta'_{\vect q}$. So we need to consider the ``1-loop" term $B_{\vect K_1\vect K_2}^{\vect q}$,
and show that it is factorizable at the leading order ($\sim n^{0.5}$). But this term contains the dispersed vertex $v_{111}$
(or $v_{111}^\dagger$). So we need to first show that $v_{111}$,
and by the way $v_{1111}$, $v_{11111}$, etc, are factorizable at the leading order.

So we turn to $\A^{\vect q_1\vect q_2\vect q_3}_{\vect K_1\vect K_2\vect K_3}=\langle c_{-\vect K_1+\vect q_1/2}
c_{\vect K_1+\vect q_1/2}\\c_{-\vect K_2+\vect q_2/2}c_{\vect K_2+\vect q_2/2}
c_{-\vect K_3+\vect q_3/2}c_{\vect K_3+\vect q_3/2}\rangle\sim n^{-0.5}$ (where $\vect q_1+\vect q_2+\vect q_3=0$, and $q_{1,2,3}\sim\sqrt{n}$).
Using Eq.~\eqref{eq:R2}, we see again that diagrams at the lowest two orders have the same external-islands structure.
The leading expression is the first diagram in the second row of Fig.~\ref{fig:m_separation3}. It has \emph{three} dashed lines,
linked to the external islands associated with $\vect q_1\vect K_1$, $\vect q_2\vect K_2$, and $\vect q_3\vect K_3$, respectively; suppose
that the numbers of simplest bubbles on them are $s_1$, $s_2$, and $s_3$, respectively. Because the
small circles contained by a dispersed vertex must be of the same color, $s_{1,2,3}$ must be all even (for $v_{111}$)
or all odd (for $v_{111}^\dagger$).  After straightforward derivation, we get
\begin{align}\label{eq:m_gamma_eq}
&\A^{\vect q_1\vect q_2\vect q_3}_{\vect K_1\vect K_2\vect K_3}=\sum_{\vect K'_{1,2,3}}
G_{\vect K_1\vect K_1'}^{\vect q_1}G_{\vect K_2\vect K_2'}^{\vect q_2}G_{\vect K_3\vect K_3'}^{\vect q_3}
\gamma^{\vect q_1\vect q_2\vect q_3}_{\vect K_1'\vect K_2'\vect K_3'}\notag\\
&+\frac{1}{8}\sum_{\vect K'_{1,2,3}}A_{\vect K_1\vect K_1'}^{\vect q_1}A_{\vect K_2\vect K_2'}^{\vect q_2}A_{\vect K_3\vect K_3'}^{\vect q_3}
\gamma^{*\overline{\vect q}_1\overline{\vect q}_2\overline{\vect q}_3}_{\vect K_1'\vect K_2'\vect K_3'}\notag\\
&+B^{\vect q_1\vect q_2\vect q_3}_{\vect K_1\vect K_2\vect K_3}+O(n^{0.5}\Omega^{-2}).
\end{align}
Here $\gamma^{\vect q_1\vect q_2\vect q_3}_{\vect K_1\vect K_2\vect K_3}\!\equiv\gamma^{}_{\vect K_1+\vect q_1/2,-\vect K_1+\vect q_1/2,
\dots,\vect K_3+\vect q_3/2,-\vect K_3+\vect q_3/2}$,
the matrix $G_{\vect q}\equiv(1-\beta_\vect q\beta^\dagger_\vect q/4)^{-1}$, $G^\vect q_{\vect K_1\vect K_2}$ is its matrix element,
$\overline{\vect q}_i\equiv-\vect q_i$, and $B^{\vect q_1\vect q_2\vect q_3}_{\vect K_1\vect K_2\vect K_3}\sim n^{0}$ is the
next-to-leading order contribution (the last three diagrams in the second row of Fig.~\ref{fig:m_separation3}). 

The first two terms on the right side of Eq.~\eqref{eq:m_gamma_eq} are the leading-order expression of $\A$, but
the second term is proportional to $\phi_{\vect K_1}\phi_{\vect K_2}\phi_{\vect K_3}$ at the leading order,
according to Eq.~\eqref{eq:m_A_05}; since $\A$ has this same property, we deduce that the \emph{first} term on the right side
of Eq.~\eqref{eq:m_gamma_eq} is also proportional to $\phi_{\vect K_1}\phi_{\vect K_2}\phi_{\vect K_3}$ at the leading order.

The linear kernal in the first term is the direct product of three matrices:
$\mathcal{K}=G_{\vect q_1}\otimes G_{\vect q_2}\otimes G_{\vect q_3}$. Using Eq.~\eqref{eq:m_beta_05},
we find that at the leading-order, $(G_{\vect q}\phi)_\vect K=\lambda_\vect q\phi_{\vect K}$,
where $\lambda_\vect q=2/\bigl(1+\sqrt{1+4\lvert\eta_\vect q\rvert^2}\bigr)>0$,
and all the vectors orthogonal to $\phi_{\vect K}$ are unchanged under the action of $G_{\vect q}$.
So the hermitian matrix $\mathcal{K}$ is positive definite, and has $\phi_{\vect K_1}\phi_{\vect K_2}\phi_{\vect K_3}$
as one of its \emph{eigenvectors} at the leading order.

Combining the two findings in the above two paragraphs,
we deduce that $\gamma^{\vect q_1\vect q_2\vect q_3}_{\vect K_1\vect K_2\vect K_3}$ must be proportional to
$\phi_{\vect K_1}\phi_{\vect K_2}\phi_{\vect K_3}$ at the leading order.

Now turn to $\alpha^{(8)\vect q_1\vect q_2\vect q_3\vect q_4}_{\vect K_1\vect K_2\vect K_3\vect K_4}$,
namely the coefficient of the dispersed vertex $v_{1111}$. There are two diagrams, $\mathcal{D}_1$ and $\mathcal{D}_2$, for
$\A^{\vect q_1\vect q_2\vect q_3\vect q_4}_{\vect K_1\vect K_2\vect K_3\vect K_4}$ at the leading order.
The skeleton of each of them is a tree ($N_2'=0$). One of them, $\mathcal{D}_1$, contains
two $v_{111}$'s (or two $v_{111}^\dagger$'s, or $v_{111}$ and $v_{111}^\dagger$) so it is factorizable at the leading-order
according to the established properties of $v_{11}$ and $v_{111}$. The other diagram, $\mathcal{D}_2$, contains a single $v_{1111}$
(or $v_{1111}^\dagger$) together with variable numbers of $v_{11}$'s and $v_{11}^\dagger$'s.
Because both $\mathcal{D}_1$ and $\A$ are factorizable at the leading order, $\mathcal{D}_2$ is also factorizable.
Using the same logic as in the case of $v_{111}$ above, we can then deduce that $v_{1111}$ is factorizable at the
leading order.

Continuing this analysis, we see that for any $M\ll N$,
$\alpha^{(2M)\vect q_1\cdots\vect q_M}_{\vect K_1\cdots\vect K_M}$
(the coefficient of the dispersed vertex $v^{}_{\underbrace{1\cdots1}_M}$) is proportional to $\Pi_{\nu=1}^M\phi_{\vect K_\nu}$
at the leading order, for any given set of small momenta $\vect q$'s.

So we can now start the second round of logic, to strengthen the above result.

First consider $v_{11}$. In Eq.~\eqref{eq:m_beta_eq}, the term $B_{\vect K_1\vect K_2}^\vect q$ 
corresponds to the ``1-loop" ($N_2'=1$) diagram in the first row of Fig.~\ref{fig:m_separation3},
and must be factorizable (ie, proportional to $\phi_{\vect K_1}\phi_{\vect K_2}$ at any given small
$\vect q\sim\sqrt{n}$) at its \emph{own} leading order,
according to the established properties of $v_{11}$ and $v_{111}$.
But $B$ is just the next-to-leading order contribution to $\A$.
We have proved in Sec.~\ref{subsec:m_1st_separation} that $\A$ is factorizable
at the lowest \emph{two} orders. So the term $A_{\vect K_1\vect K_2}^\vect q$ must also be factorizable
at the lowest two orders, namely there exists an $\eta''_{\vect q}$, such that $A_{\vect K_1\vect K_2}^\vect q
=\eta''_{\vect q}\phi_{\vect K_1}\phi_{\vect K_2}+O(n^1\Omega^{-1})$.
Substituting this result to Eq.~\eqref{eq:m_beta_formal_solution}, we see that for
$\eta'_{\vect q}=2\eta''_{\vect q}/\bigl(1+\sqrt{1+4\lvert\eta''_\vect q\rvert^2}\bigr)$,
\begin{equation}\label{eq:m_beta_10}
\beta_{\vect K_1\vect K_2}^\vect q=\eta'_\vect q\phi_{\vect K_1}\phi_{\vect K_2}+O(n^1\Omega^{-1}).
\end{equation}

We then turn to $v_{111}$. In Eq.~\eqref{eq:m_gamma_eq}, the term $B$ corresponds to the
$N_2'=1$ diagrams in the second row of Fig.~\ref{fig:m_separation3}, which consist of the dispersed vertices
we have studied. $B$ is thus factorizable at its own leading order. But $\A$ is factorizable at the lowest two orders,
and $B$ is just the next-to-leading order contribution to $\A$. So the sum of the first two
terms on the right side of Eq.~\eqref{eq:m_gamma_eq} is factorizable at the lowest two orders.
But the second term itself is factorizable at the lowest two orders, according to the above established property
of $A_{\vect K\vect K'}^\vect q$. So the first term is factorizable at the lowest two orders. Note finally that
$G_{\vect q}$ is determined by $\beta_\vect q$, so $(G_{\vect q}\phi)_{\vect K}=\lambda'_\vect q\phi_\vect K$
with a relative error of the order $n^1$, for some $\lambda'_\vect q$. So we find that 
\begin{equation}\label{eq:m_gamma_10}
\gamma_{\vect K_1\vect K_2\vect K_3}^{\vect q_1\vect q_2\vect q_3}=\eta'_{\vect q_1\vect q_2\vect q_3}
\phi_{\vect K_1}\phi_{\vect K_2}\phi_{\vect K_3}
\end{equation}
with a relative error of the order $n^1$, for some $\eta'_{\vect q_1\vect q_2\vect q_3}$.

Extension to more clusters (each of which contains two spin-momenta) is trivial. In general, we have thus
proved that
\begin{equation}\label{eq:m_separation}
\alpha^{(2M)\vect q_1\cdots\vect q_M}_{\vect K_1\cdots\vect K_M}
=\eta'_{\vect q_1\cdots\vect q_M}\phi_{\vect K_1}\cdots\phi_{\vect K_M}
\end{equation}
with a relative error of the order $n^{1}$ in the low-density limit, for any $1\leq M\ll N$,
provided that $\vect K_\nu\pm\vect K_\rho$ are not zero or $\sim\sqrt{n}$ (for all $1\leq\nu<\rho\leq M$).


\bibliography{superfluid}
\end{document}